\documentclass[preprint,12pt]{elsarticle}

\usepackage{xcolor}
\usepackage{amssymb}
\usepackage{graphicx}
\usepackage{subcaption}
\usepackage{hyperref}
 \usepackage{float}

\newcommand{\akernummanzero}{4122}

\newcommand{\akernumdiscard}{10703}

\newcommand{\akkinummanzero}{1756}

\newcommand{\akkinumdiscard}{9886}
\newcommand{\akkinumtotalpatchpairs}{31631}

\newcommand{\akprnummanzero}{2872}

\newcommand{\akprnumdiscard}{13943}
\newcommand{\akprtotalpatchpairs}{38914}

\newcommand{\akhertwonummanzero}{13984}

\newcommand{\akhertwonumdiscard}{20103}
\newcommand{\akhertwototalpatchpairs}{41098}

\newcommand{\aktotalnumvisinspect}{147404}
\newcommand{\aktotalnumdiscard}{54635}
\newcommand{\aktotalmanzero}{22734}

\newcommand{\akfigrefHerdssample}{S1}
\newcommand{\akfigrefERdssample}{S2}
\newcommand{\akfigrefPRdssample}{S3}
\newcommand{\akfigrefKidssample}{S4}
\newcommand{\akfigrefERheatmap}{S5}
\newcommand{\akfigrefPRheatmap}{S6}
\newcommand{\akfigrefKiheatmap}{S7}

\journal{Modern Pathology}

\begin{document}

\begin{frontmatter}



\title{Predicting Ki67, ER, PR, and HER2 Statuses from H\&E-stained Breast Cancer Images}


\author[inst1]{Amir Akbarnejad}

\affiliation[inst1]{
    organization={Department Of Computing Sciences},
    addressline={University of Alberta},  
    state={AB},
    country={Canada}
}
\affiliation[inst2]{
    organization={Department of Laboratory Medicine and Pathology},
    addressline={Dalhousie University},  
    state={NS},
    country={Canada}
}
\affiliation[inst3]{
    organization={Department of Laboratory Medicine and Pathology},
    addressline={University of Alberta},  
    state={AB},
    country={Canada}
}
\author[inst1]{Nilanjan Ray}
\author[inst2]{Penny J. Barnes}
\author[inst3]{Gilbert Bigras}

\begin{abstract}
Despite the advances in machine learning and digital pathology, it is not yet clear if machine learning methods can accurately predict molecular information merely from histomorphology. In a quest to answer this question, we built a large-scale dataset (185538 images) with reliable measurements for Ki67, ER, PR, and HER2 statuses. The dataset is composed of mirrored images of H\&E and corresponding images of immunohistochemistry (IHC) assays (Ki67, ER, PR, and HER2. These images are mirrored through registration. To increase reliability, individual pairs were inspected and discarded if artifacts were present (tissue folding, bubbles, etc). Measurements for Ki67, ER and PR were determined by calculating H-Score from image analysis. HER2 measurement is based on binary classification: 0 and 1+ (IHC scores representing a negative subset) vs 3+ (IHC score positive subset). Cases with IHC equivocal score (2+) were excluded. We show that a standard ViT-based pipeline can achieve prediction performances around 90\% in terms of Area Under the Curve (AUC) when trained with a proper labeling protocol. Finally, we shed light on the ability of the trained classifiers to localize relevant regions, which encourages future work to improve the localizations. Our proposed dataset is publicly available: \href{https://ihc4bc.github.io/}{https://ihc4bc.github.io/}.    
\end{abstract}


\begin{highlights}
\item We present a publicly available dataset for predicting Ki67, ER, PR, and HER2 statuses from H\&E-stained breast cancer images. 
\item We show that a standard ViT-based pipeline trained on the dataset and with proper labeling protocol can achieve prediction performances around 90 
\item We inspect the ability of a strongly-supervised and a weakly-supervised method in localizing relevant regions. Our experiments shed light on the ability of the classifiers to localize relevant regions, and prompt future works to improve the localizations. 
\end{highlights}


\end{frontmatter}


\section{Introduction}\label{sec:intro}
Researchers have aspired to use machine learning to predict molecular information from H\&E histopathological images. Specific molecular information can be used as biomarkers to guide oncological therapy. Doing so accurately, if achievable, would be beneficial given that ancillary tests, including immunohistochemistry (IHC), are expensive, require specialized infrastructure, trained personal, control of pre-analytical conditions and participation to external proficiency programs to ensure reliable results.

As underlined in the comprehensive survey by Cifci et al \cite{kather_survey_1} (Sec. Limitations), except for some markers, the best prediction performances are around 80\% in terms of Area Under the Curve (AUC) and the unavailability of large datasets may well be the reason for not achieving higher AUCs. Indeed, the sheer amount of data used for neural network learning is probably the most influential factor for successful biomarker predictions \cite{shamai1}. Besides the small size of available datasets, a small amount of corrupted or unreliable labels may significantly reduce the prediction performance \cite{gpex}. To tackle these issues, we have built a large dataset with high-quality labels.

The contributions of our study are as follows:
\begin{itemize}
    \item We present a public dataset of images called IHC4BC representing the four frequently performed IHC assays in breast cancer pathology (ER, PR, Ki67 and HER2) matched with mirrored H\&E images to predict the statuses of the aforementioned IHC assays. This dataset is large-scale and provides high-quality labels.
    \item Using this large dataset, we show that it is possible to achieve around 90\% AUROC for the four aforementioned biomarkers.
    \item We inspect the ability of a strongly-supervised and a weakly-supervised method in localizing relevant regions. Despite reported high AUROC, our experiments shed light on the current limited ability of the classifiers to localize relevant regions, and spur future works to improve the localizations.
    
\end{itemize}

The studies done by Anand et al \cite{neerajher2} and Liu et al \cite{ki67homogen} annotate homogeneous positive or negative regions in HER2 and Ki67-stained whole-slide images (WSIs), respectively. Afterwards, the corresponding regions in H\&E slides are considered positive or negative. Consequently, machine learning is applied to distinguish between positive and negative patches extracted from those regions. The shortcoming of this approach is that in the IHC modality, many regions are associated with heterogeneous expression (especially for PR and Ki67) and all of those regions are discarded. Zeng et al \cite{prmiccai} registered pairs of H\&E and Hematoxylin \& 3,3'-Diaminobenzidine (H-DAB) IHC images to obtain corresponding H\&E and IHC patches. Afterwards, for each IHC patch they extracted the DAB signal (brown channel) and estimated the percentage of positive nuclei by computing the percentage of brown pixels. This estimate for percentage is used to train a CycleGAN-like \cite{cyclegan} virtual stainer. The label for each IHC patch (i.e. the estimate for percentage) is not a reliable estimate for the percentage of positive nuclei. Moreover, even a nuclei-by-nuclei analysis may fail due to several reasons that we will discuss in Sec. \ref{sec:visualinspection}. Shamai et al \cite{shamai2} and Couture et al (partially) \cite{cotur1} use pathologist readout Allred scores (utilized for ER and PR assessments) as ground-truth labels for training their models. This label might still not be optimally reliable, given inter-observer and intra-observer variabilities in pathologists' assessments and also by the nature of this ordinal scoring which lacks precision (i.e a score 5 encompasses any \% between 66\% and 100\% of stained nuclei) which might not provide strong-enough supervision for machine learning methods.

Given a H\&E patch, one way of obtaining genetic/molecular information is to obtain the next section (usually 4 microns) within a paraffin block representing a close-by tissue-cut stained by, e.g, IHC \cite{neerajher2, ki67homogen, prmiccai, shamai1, shamai2, cotur1}. Alternatively, one can destain a given H\&E slide and restain it to get the IHC modality on the exact same tissue-cut. Jackson et al \cite{sox10} destain/restain 12 WSI pairs to predict the expression of SOX10 from morphology. The two modalities are registered and nucleus-level labels are obtained by finding each nucleus in both modalities and for more than 3 million cells. 
Cyprys et al \cite{destainbenchmark} destain/restain 25 WSIs and benchmark many registration methods on the dataset. There are issues with destaining/restaining approach: 1) performing IHC on a destained  H\&E preparation is not a standard procedure and would need to be validated to ensure equivalent staining intensity routinely obtained from unstained sections; 2) it is hard (if not impossible) to perfectly register the two modalities at the nucleus level. For example, Jackson et al \cite{sox10} use a classifier to discard the imperfect registrations but the classifier needs some labels for training and will have some failures.

Rawat et al \cite{fingerprintsbca} train a feature-extractor for H\&E images in a self-supervised way. Afterwards, they train classifiers on roughly 1000 TMA cores to predict ER, PR, and Her2 statuses from H\&E images. They achieve slide-level AUCs of 0.89, 0.81, and 0.79 for ER, PR, and HER2 respectively. Using this sample-efficiency technique (self-supervised feature extractor) or similar ones may help improve the performance, but these techniques cannot make up for the lack of data and usually result in marginal improvement \cite{telezmultitask, contrastivepredictivecoding}. Qu et al \cite{pathwaypred} seek to predict point mutations and copy number alterations for some breast cancer genes and obtain 68-85 AUCs for different genes.
 Hohne et al \cite{brafmiccai} predict BRAF mutations and NTRK gene fusions with slide-level labels, and achieve 75-85 AUCs. He et al \cite{transcriptomstan} use spatial transcriptomics to obtain genomic information for more than 30K H\&E regions. They achieve a range of AUCs for different genes, with most AUCs reside between 70 and 80. Moreover, for predicting the continuous expression values they achieve correlation coefficients of up to 0.50 for different genes. Zeng et al \cite{prmiccai} train a CycleGAN-like \cite{cyclegan} virtual stainer for PR. Besides the issue of unreliable labels that we discussed above, the virtual stains are only evaluated using visual criteria like SSIM (structural similarity) and PSNR (peak signal-to-noise ratio) rather than clinically relevant criteria like H-Score. Shamai et al \cite{shamai2} seek to predict PD-L1 and PD-1 statuses using about 5000 TMA cores. They achieve around 90 AUC for PD-L1 and a lower AUC of 85 for PD-1. Jackson et al \cite{sox10} obtain nucleus-level labels by destaining/restaining 12 WSIs, and achieve around 90 AUCs to label each nucleus as SOX10 positive or negative. Shamai et al \cite{shamai1} seek to predict ER, PR, and HER2 status from H\&E TMA-cores. The dataset contains around 5000 TMA-cores and they achieved AUCs of up to 85 for PR, Ki67, and HER2 and up to 88 for ER. Tavolara et al \cite{niazimice} seek to predict the expression of about 20000 genes in an experimental M.tb mice infection. Some of the genes (but not all) can be predicted with above 90 correlation coefficient. The gene predictors are used as an intermediate step to accurately classify each slide as supersusceptible or not-supersusceptible. 
 All in all, there are studies that accurately predict molecular information from H\&E images, but with the best our knowledge regarding PR, Ki67, and HER2 predictions, the best performances are up to 85 in terms of AUC and slightly higher for ER.

\begin{figure}
\includegraphics[width=\textwidth]{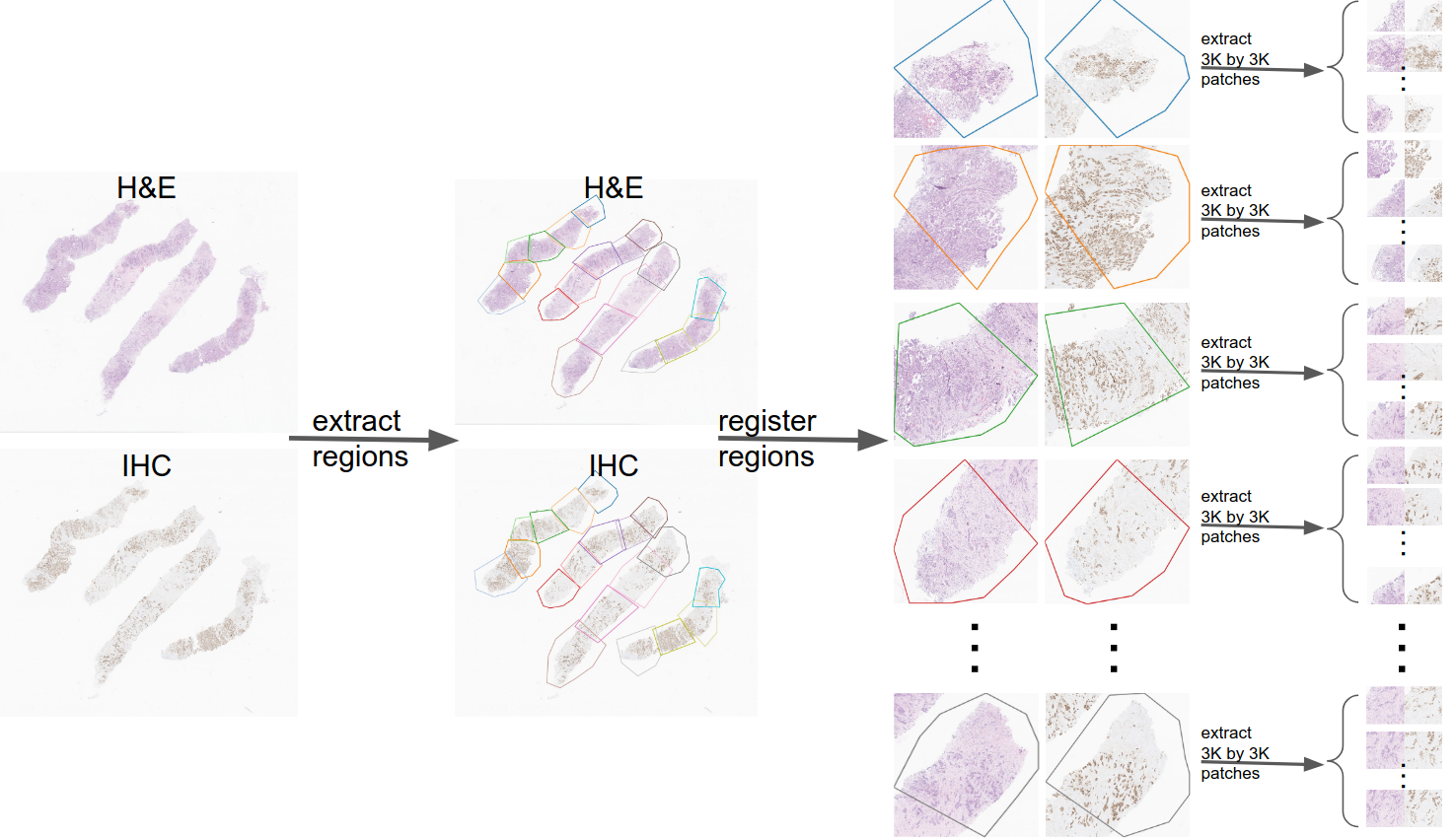}
\caption{The steps taken to obtain 3K by 3K H\&E-IHC pairs from a WSI-pair. Details are provided in Sec. \ref{sec:obtain_patchpairs}.} 
\label{fig:extract_patches}
\end{figure}

\section{Material and methods}
\subsection{Tissue and Slide Preparation}
All ER, PR, HER2 and Ki67 IHC preparations were performed on unselected sequential 50 breast biopsies collected in 2022 with strict pre-analytical and analytical controls consisting of 1) absence of cold ischemia (virtually all biopsy tissue put immediately in formalin fixative after collection), 2) minimum of 24 hours and a maximum of 48 hours in formalin, 3) utilization of on-slide controls to ensure successful IHC performance by the IHC instrument and 4) repeating confirmatory IHC procedure when finding negative tumoral ER and PR in the absence of positive internal controls (normal mammary ducts). The IHC clones utilized for ER, PR, HER2 and Ki67 were respectively SP1, PgR 636, SP3 and MIB-1. These IHCs assays were obtained using the automated platforms: Ventana Ultra (for ER, PR and HER2) and Dako Omnis (Ki67). The slides were scanned at 40X using the Aperio GT 450 - Automated, High Capacity Digital Pathology Slide Scanner.
\subsection{Obtaining Pairs of Patches}\label{sec:obtain_patchpairs}
Fig. \ref{fig:extract_patches} illustrates how we extract pairs of patches given a pair of H\&E and IHC whole-slide images. Given variability of tissue disposition on glass slide, a single rigid transformation cannot perfectly register a WSI pair. Therefore, we firstly annotated region-pairs from two given matched WSIs (in Fig. \ref{fig:extract_patches} the step labeled as "extract regions"). Afterwards, we manually registered each region-pair (in Fig. \ref{fig:extract_patches} this step is labeled as "register regions"). Finally, we traversed each region-pair with a stride of 1500 and extracted patch-pairs. The final result of this step is a set of patch-pairs each of which are 3000 by 3000 (the pairs in the right side of Fig. \ref{fig:extract_patches}).      

\begin{figure}
\centering
 \begin{subfigure}[b]{0.18000000000000002\textwidth}
  \centering
  \includegraphics[width=\textwidth]{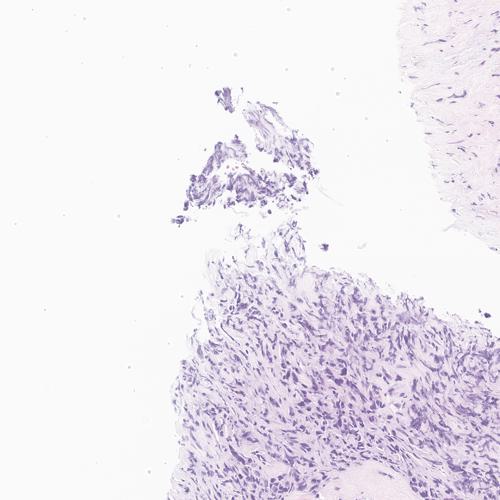}
  \end{subfigure}
\hfill
 \begin{subfigure}[b]{0.18000000000000002\textwidth}
  \centering
  \includegraphics[width=\textwidth]{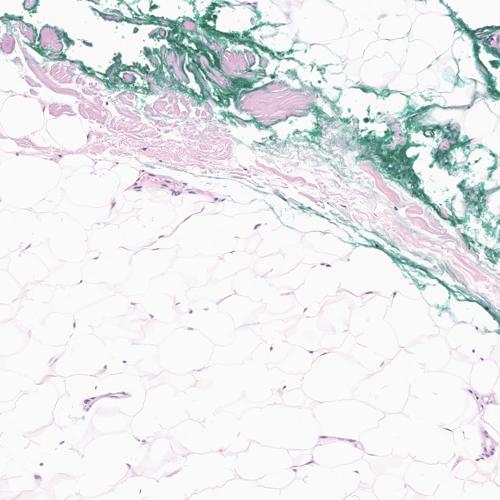}
  \end{subfigure}
\hfill
 \begin{subfigure}[b]{0.18000000000000002\textwidth}
  \centering
  \includegraphics[width=\textwidth]{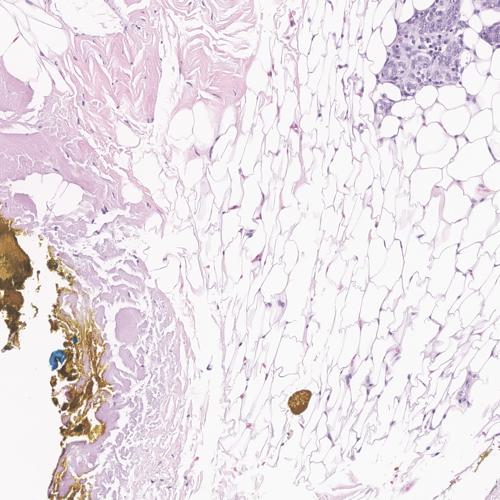}
  \end{subfigure}
\hfill
 \begin{subfigure}[b]{0.18000000000000002\textwidth}
  \centering
  \includegraphics[width=\textwidth]{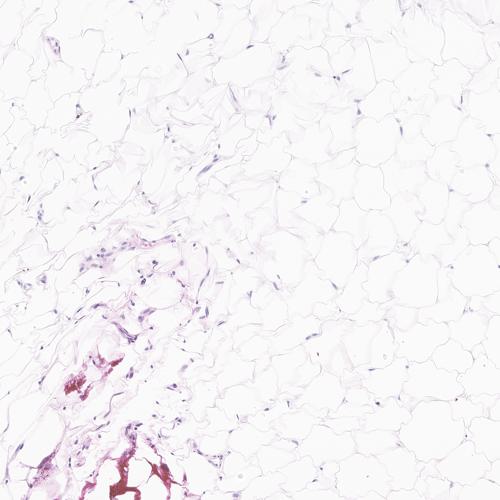}
  \end{subfigure}
\hfill
 \begin{subfigure}[b]{0.18000000000000002\textwidth}
  \centering
  \includegraphics[width=\textwidth]{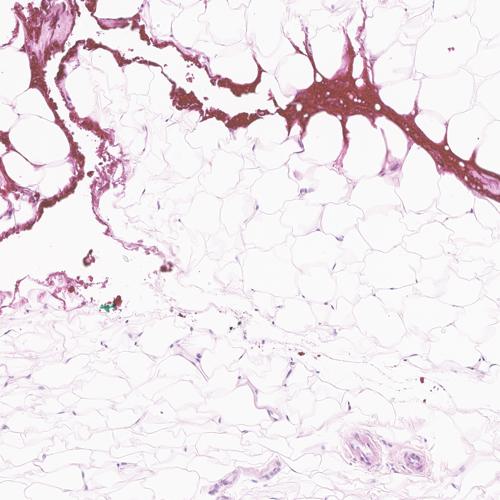}
  \end{subfigure}
\\
 \begin{subfigure}[b]{0.18000000000000002\textwidth}
  \centering
  \includegraphics[width=\textwidth]{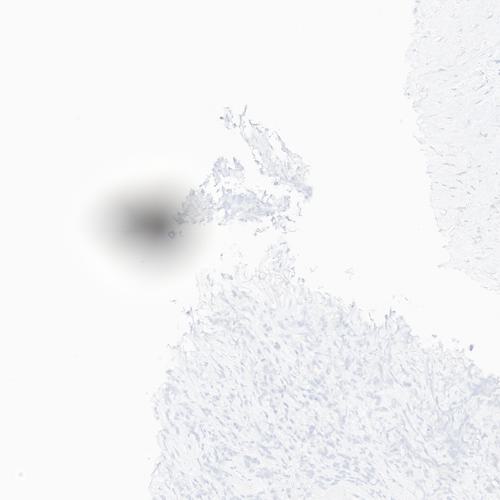}
  \end{subfigure}
\hfill
 \begin{subfigure}[b]{0.18000000000000002\textwidth}
  \centering
  \includegraphics[width=\textwidth]{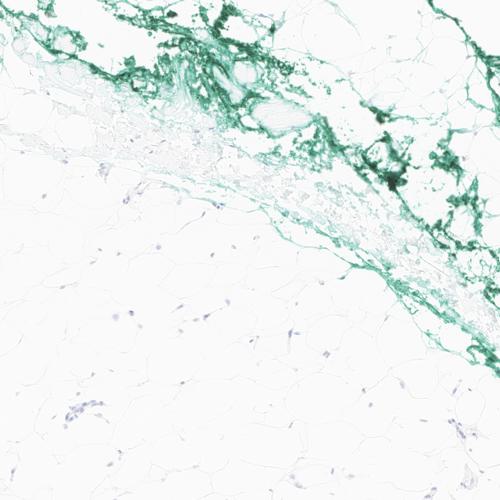}
  \end{subfigure}
\hfill
 \begin{subfigure}[b]{0.18000000000000002\textwidth}
  \centering
  \includegraphics[width=\textwidth]{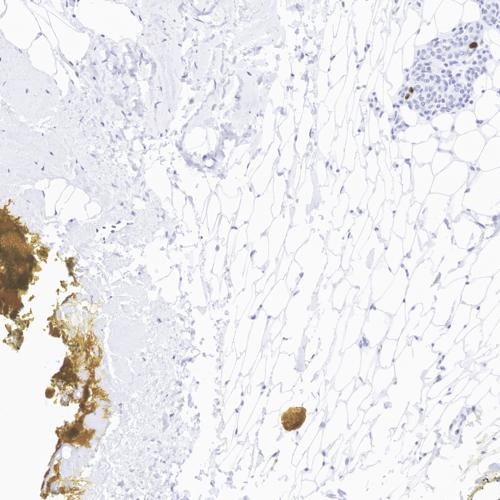}
  \end{subfigure}
\hfill
 \begin{subfigure}[b]{0.18000000000000002\textwidth}
  \centering
  \includegraphics[width=\textwidth]{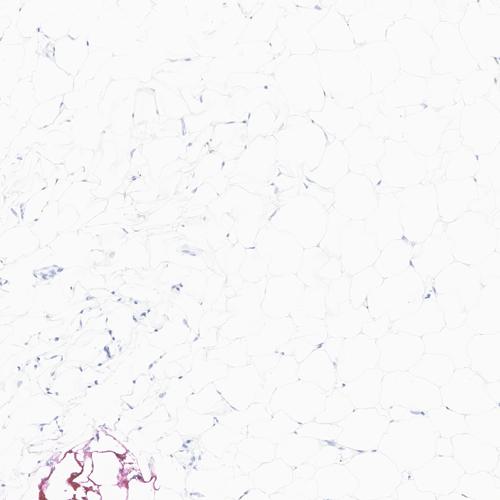}
  \end{subfigure}
\hfill
 \begin{subfigure}[b]{0.18000000000000002\textwidth}
  \centering
  \includegraphics[width=\textwidth]{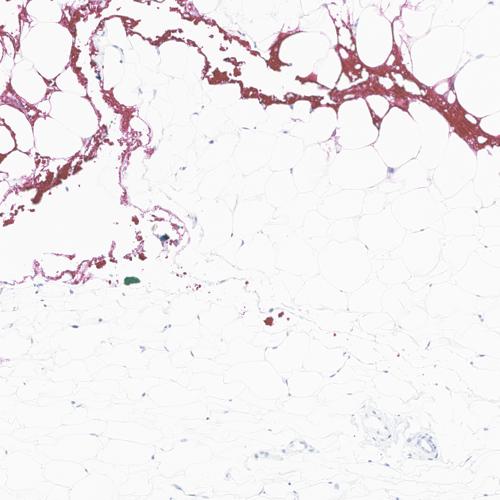}
  \end{subfigure}
\\
 \begin{subfigure}[b]{0.18000000000000002\textwidth}
  \centering
  \includegraphics[width=\textwidth]{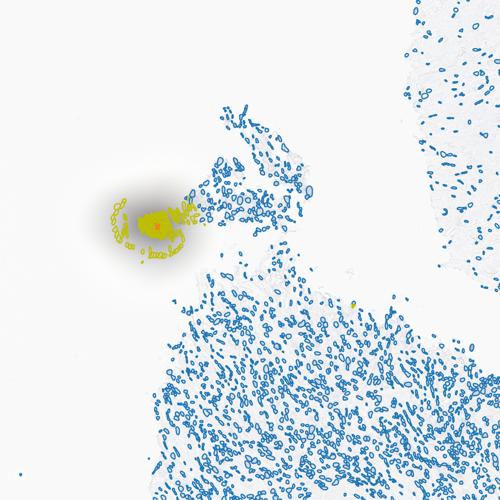}
  \end{subfigure}
\hfill
 \begin{subfigure}[b]{0.18000000000000002\textwidth}
  \centering
  \includegraphics[width=\textwidth]{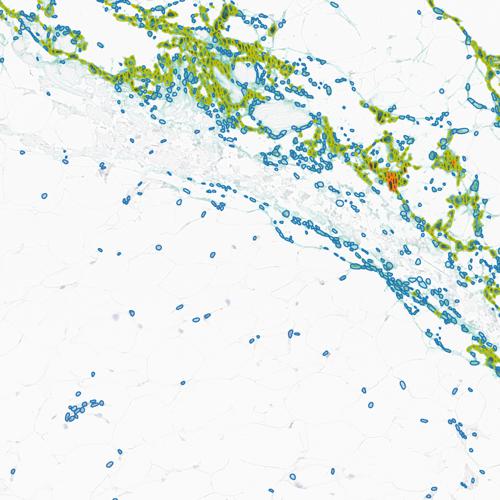}
  \end{subfigure}
\hfill
 \begin{subfigure}[b]{0.18000000000000002\textwidth}
  \centering
  \includegraphics[width=\textwidth]{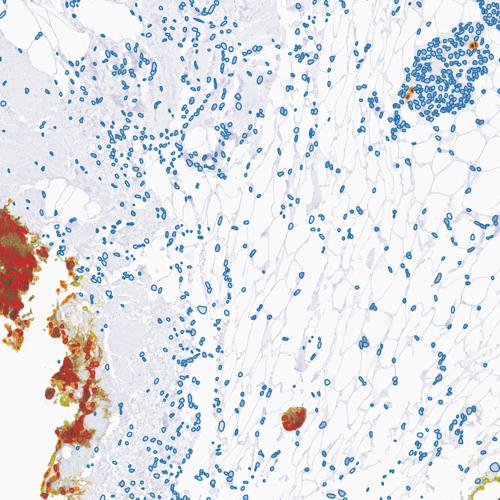}
  \end{subfigure}
\hfill
 \begin{subfigure}[b]{0.18000000000000002\textwidth}
  \centering
  \includegraphics[width=\textwidth]{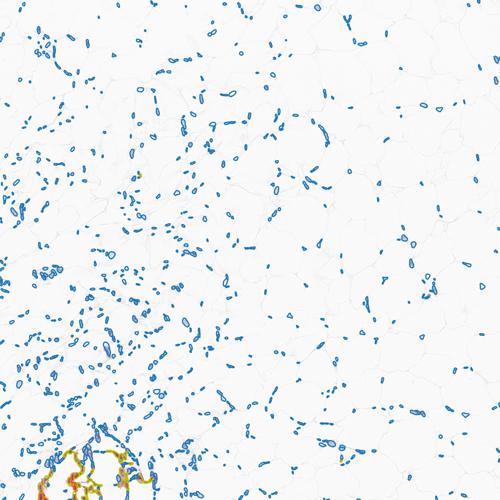}
  \end{subfigure}
\hfill
 \begin{subfigure}[b]{0.18000000000000002\textwidth}
  \centering
  \includegraphics[width=\textwidth]{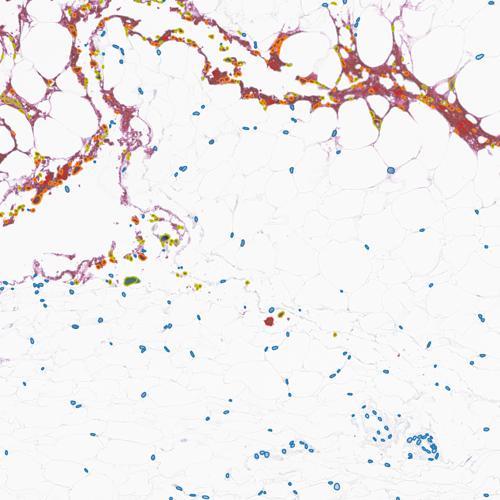}
  \end{subfigure}
  \caption{Examples of the failure of DAB-analysis, discussed in Sec. \ref{sec:visualinspection}. Each column corresponds to a H\&E-IHC pair. Row1:H\&E, Row2: IHC, Row3: The result of DAB-analysis where blue, yellow, orange, and red colors mark 0, 1+, 2+, and 3+ nuclei.}
  \label{fig:dabfailures}
\end{figure}

\subsection{H-DAB Analysis to Obtain Labels}\label{sec:hdab_basic}
Each pair contains a H\&E patch of size 3000 by 3000 and the corresponding IHC patch of size 3000 by 3000. 
We ran H-DAB analysis on the IHC image to obtain marker-information for the corresponding H\&E patch.
We used StarDist \cite{stardist} to segment the nuclei in the H-DAB patch. Afterwards, we used the conventional color-deconvolution to extract the brown DAB channel \cite{dabdeconv}. In the dataset for each IHC image we have included the average DAB channel within every and each nucleus, so different numbers like H-Score, percentage, and Allred score can be computed for each H\&E-IHC pair. The total number of nuclei (i.e. the denominator in percentage or H-score) was obtained from H\&E images, since nuclear segmentation was found more reliable when using StarDist \cite{stardist} on the hematoxylin stain of the H\&E assay compared to the hematoxylin of the H-DAB assay.  
Indeed, we noticed that in the H-DAB modality StarDist \cite{stardist} may miss many negative nuclei or may mistakenly take artifacts as negative nuclei (examples are provided in the supplementary in Fig. \akfigrefERdssample{} and its caption).
Therefore, in the dataset, we include the total number of nuclei detected in each H\&E patch, which is a more reliable estimate for the total number of nuclei when computing percentage (for Ki-67) or H-score (for ER and PR). Of notice, the H-score has a range of 0 to 300 and is computed as follows: (\% of weakly stained nuclei * 100) + (\% of moderately stained nuclei * 200) + (\% of strongly stained nuclei * 300). The determination of weak, moderate, and strong intensities was based on two fixed thresholds for DAB optical density mean established by one pathologist. A single (weak) threshold was established for the Ki67 percentage assessment. In this study, nuclei were not discriminated between tumoral and non-tumoral for H-score calculation.

\begin{figure}
\centering
 \begin{subfigure}[b]{0.18000000000000002\textwidth}
  \centering
  \includegraphics[width=\textwidth]{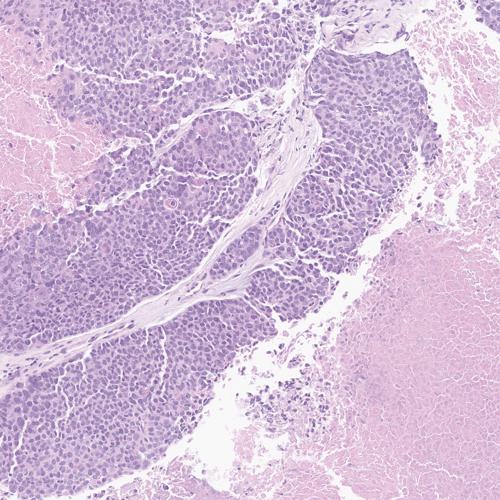}
  \end{subfigure}
\hfill
 \begin{subfigure}[b]{0.18000000000000002\textwidth}
  \centering
  \includegraphics[width=\textwidth]{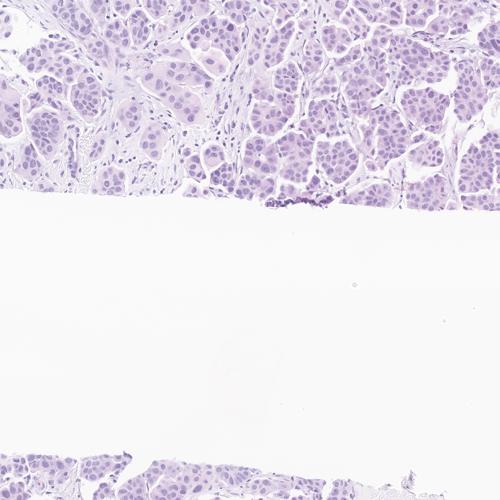}
  \end{subfigure}
\hfill
 \begin{subfigure}[b]{0.18000000000000002\textwidth}
  \centering
  \includegraphics[width=\textwidth]{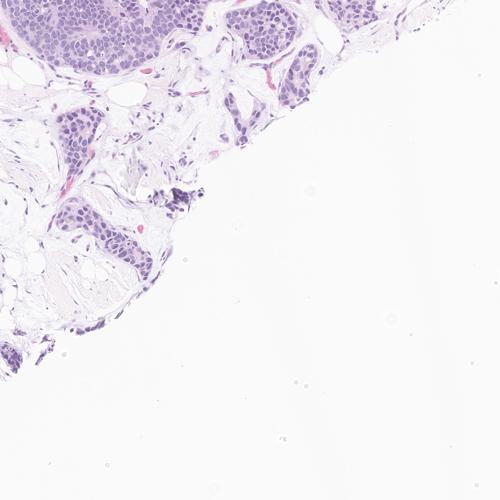}
  \end{subfigure}
\hfill
 \begin{subfigure}[b]{0.18000000000000002\textwidth}
  \centering
  \includegraphics[width=\textwidth]{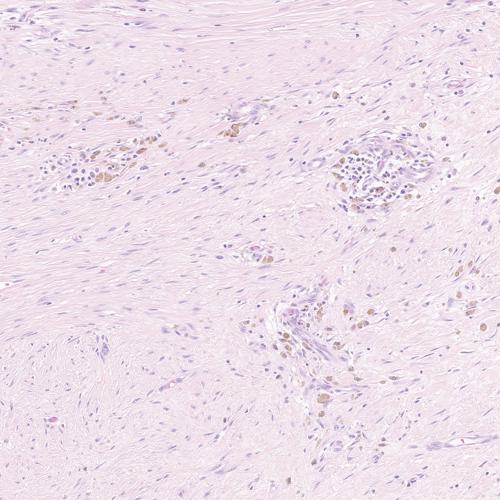}
  \end{subfigure}
\hfill
 \begin{subfigure}[b]{0.18000000000000002\textwidth}
  \centering
  \includegraphics[width=\textwidth]{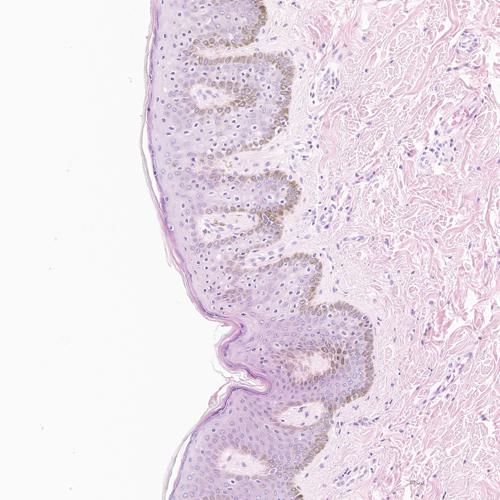}
  \end{subfigure}
\\
 \begin{subfigure}[b]{0.18000000000000002\textwidth}
  \centering
  \includegraphics[width=\textwidth]{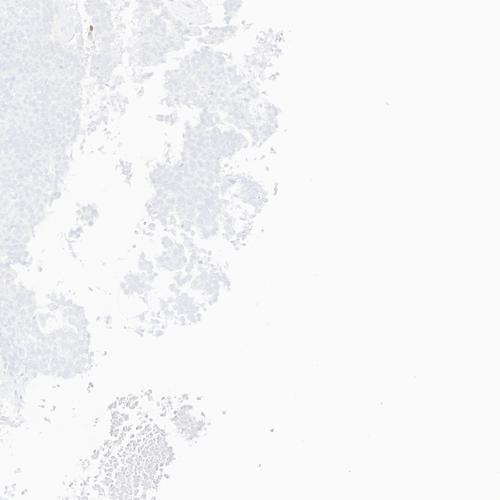}
  \end{subfigure}
\hfill
 \begin{subfigure}[b]{0.18000000000000002\textwidth}
  \centering
  \includegraphics[width=\textwidth]{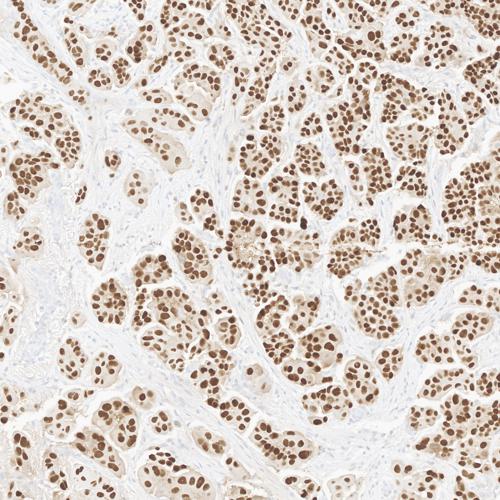}
  \end{subfigure}
\hfill
 \begin{subfigure}[b]{0.18000000000000002\textwidth}
  \centering
  \includegraphics[width=\textwidth]{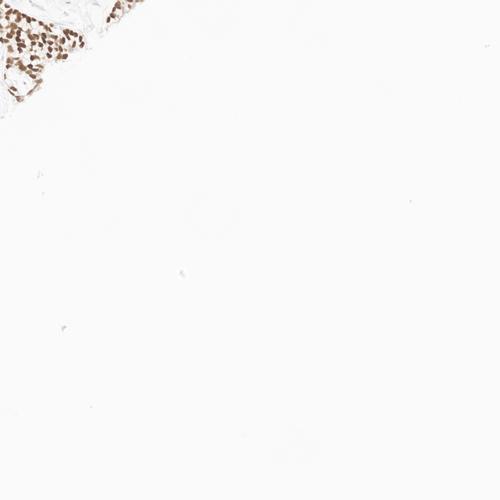}
  \end{subfigure}
\hfill
 \begin{subfigure}[b]{0.18000000000000002\textwidth}
  \centering
  \includegraphics[width=\textwidth]{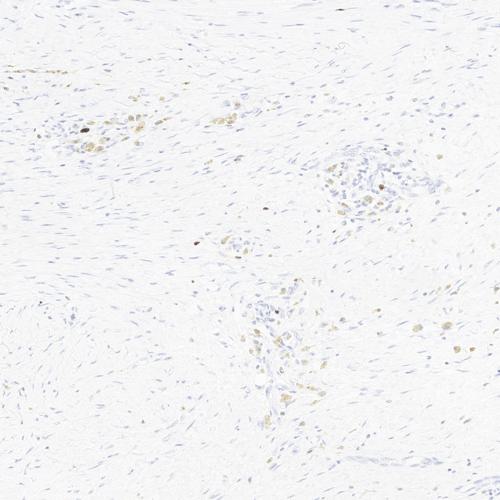}
  \end{subfigure}
\hfill
 \begin{subfigure}[b]{0.18000000000000002\textwidth}
  \centering
  \includegraphics[width=\textwidth]{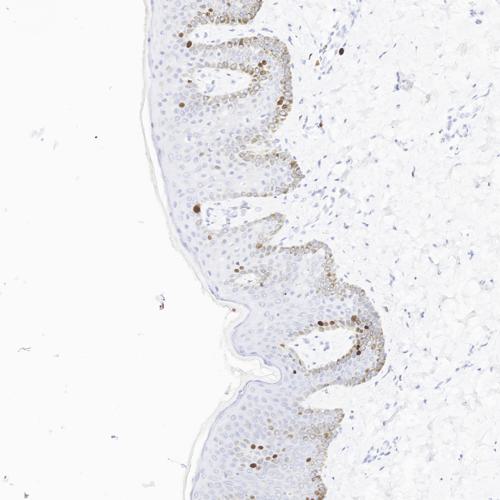}
  \end{subfigure}
  \caption{
    Examples of H\&E-IHC pairs which are discarded during the exhaustive visual inspection. Each column corresponds to a H\&E-IHC pair. Row1: H\&E, Row2: IHC. Details are provided in Sec. \ref{sec:visualinspection}
    }
  \label{fig:manualdiscard}
\end{figure}

\subsection{Visual Inspection of Pairs and DAB-Analysis}\label{sec:visualinspection}
After obtaining corresponding H\&E and H-DAB mirrored images and corresponding analytical labels, not all pairs are valid due to several reasons. So an expert pathologist (co-author 4) worked closely with a non-expert (co-author 1) to exclude all invalid pairs from the dataset. A total of \aktotalnumvisinspect{} pairs were exhaustively inspected, out of which \aktotalnumdiscard{} pairs were discarded and \aktotalmanzero{} pairs were manually labeled as zero (as we discuss below, some instances like 1\textsuperscript{st} and 2\textsuperscript{nd} columns of Fig. \ref{fig:dabfailures} were manually labeled as zero).   

Fig. \ref{fig:dabfailures} illustrates problematic images which provide wrong H-DAB analysis results. Each column illustrates one example. The rows depict the H\&E patch images, H-DAB patch images, and DAB-analysis images performed on H-DAB, respectively. First column: out of focus non-tissular artifact; second column green ink; third column yellow ink; and columns 3-5 red/brown inks. These non-tissular events are mistakenly considered as positive signal by the automatic DAB-analysis. In the case of 1\textsuperscript{st} and 2\textsuperscript{nd} columns, apart from the wrong positive signal there are no real positive nuclei elsewhere. So, in such cases, mirrors images are kept in the dataset but a manual label (H-Score) of 0 was assigned to the pair. Keeping these images (with proper H-Score) present opportunities for machine learning model to be exposed to such artifacts. On the other hand, cases like the one in the 3\textsuperscript{rd} column, besides the false positive signals, there are some real positive nuclei in the two o'clock position which made difficult to ensure a proper H-score label. Such cases were discarded from the dataset.

Another reason for discarding a pair is imperfect registration (column 3 of Fig. \ref{fig:manualdiscard}) or some tissue parts being missing in the corresponding modality (columns 1 and 2 of Fig. \ref{fig:manualdiscard}). These cases were discarded from the dataset during the exhaustive visual inspection. Other pairs were removed, as illustrated in the 4th and 5th columns of Fig. \ref{fig:manualdiscard}, when they contain brown pigments like hemosiderin and melanin identified in both modalities (H\&E and IHC), producing false DAB signal in the corresponding IHC image. During the visual inspection, if there were such signals in the H\&E image, the pair was discarded. For such WSIs not all patches (only the ones with brown color in H\&E) were excluded, to make sure that these cases are not completely discarded from the dataset. Most patches from the white slide background were discarded. For each region a few white patches (containing no tissue) were included in the dataset so a machine learning model can learn to label the white background as negative.

\subsection{Obtaining Labels for HER2 Pairs}\label{sec:labelingher2}
HER2 protein is a membrane-bound receptor and therefore the corresponding DAB signal is membranous. Thus for HER2 IHC, as a first attempt we chose not to include the 2+ cases due to poor inter-observer reproduciblity \cite{drbarneshertwo} for 2+ cases. All H\&E-HER2 pairs were inspected one-by-one. A pair was labeled as positive if any 3+ patterns were present in the IHC modality and the corresponding regions were present in the H\&E modality. In other words, a binary label was assigned to each H\&E-HER2 pair.
During the exhaustive inspection, some pairs were discarded due to the reasons that we discussed in Sec. \ref{sec:visualinspection}. For each WSI we had access to pathologist assessment result as 0, 1+, 2+, or 3+. In this study we did not include equivocal (2+) cases but instead focused on a binary approach: WSIs labeled as 0 or 1+ (high probability of non-amplified status) and WSIs labeled as 3+ (high probability of amplified status). We exhaustively inspected the pairs to confirm the veracity of this binary labeling between HER2 positive and negative cases. Some examples are provided in the supplementary in Fig. \akfigrefHerdssample{}.

\subsection{Dataset Statistics}\label{sec:datasetstats}
For ER, PR, Ki67, and HER2 59, 60, 60, and 52 (total of 231) WSI-pairs were used, respectively.
For ER, PR, Ki67, and HER2 \akhertwototalpatchpairs{}, \akprtotalpatchpairs{}, \akkinumtotalpatchpairs{}, and \akhertwototalpatchpairs{} patch-pairs were obtained, respectively, according to the procedure of Sec. \ref{sec:obtain_patchpairs}. 
Among the extracted patch-pairs, for ER, PR, Ki67, and HER2 \akernumdiscard{}, \akprnumdiscard{}, \akkinumdiscard{}, and \akhertwonumdiscard{} patch-pairs were discarded and \akernummanzero{}, \akprnummanzero{}, \akkinummanzero{}, and \akhertwonummanzero{} were manually set to 0 during the exhaustive inspection, respectively. 
Some dataset samples are illustrated in Figs. \akfigrefHerdssample{}, \akfigrefERdssample{}, \akfigrefPRdssample{}, and \akfigrefKidssample{} in the supplementary. 

This work and the authorization to publish the anonymized dataset in the public domain \href{https://ihc4bc.github.io/}{https://ihc4bc.github.io/}, received the ethical approval (HREBA.CC-19-0347) from the Health Ethics Board of Alberta.

\begin{figure}
\centering
 \begin{subfigure}[b]{0.8\textwidth}
  \centering
  \includegraphics[width=\textwidth]{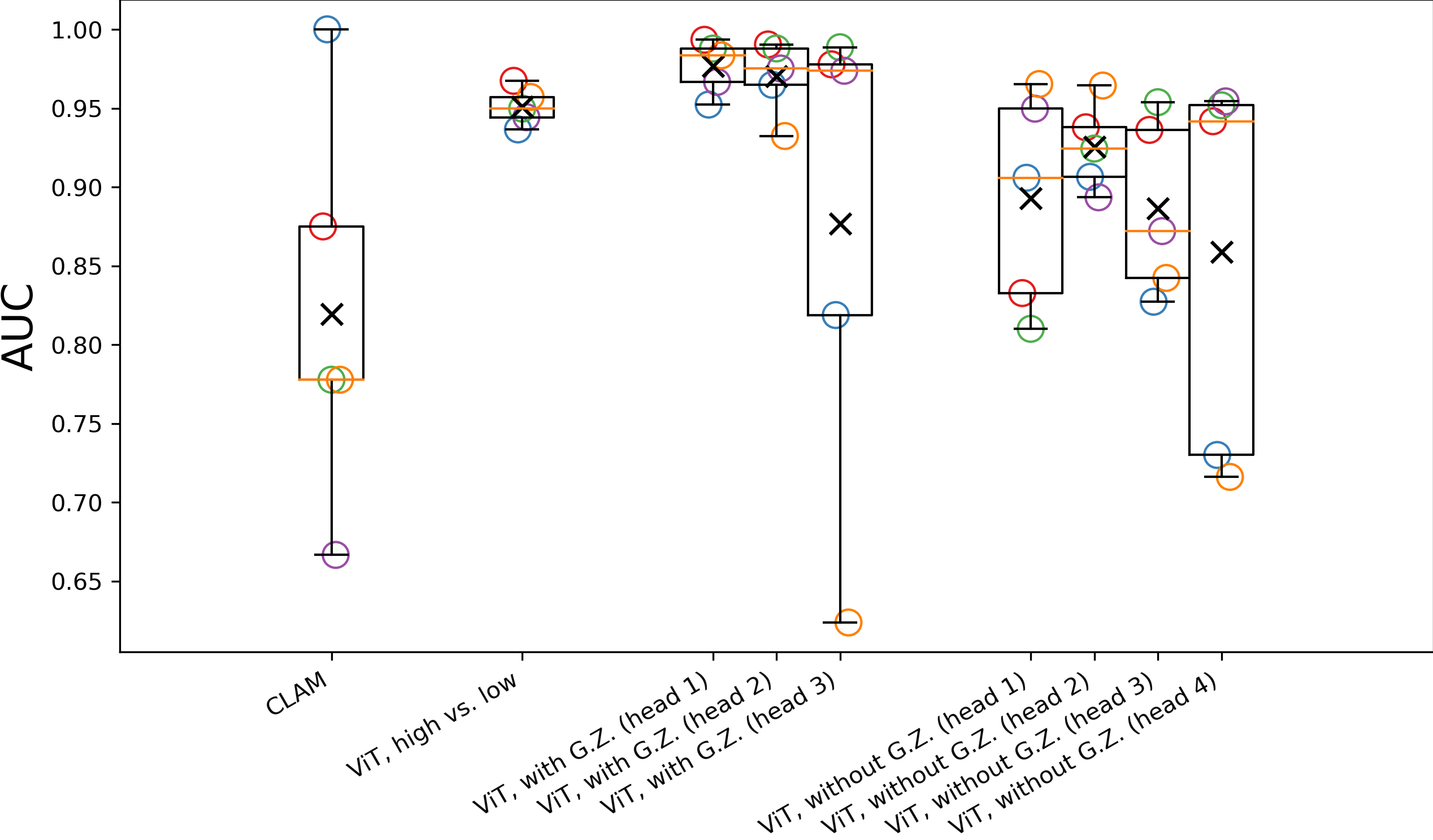}
  \end{subfigure}
\caption{
   Results for predicting Ki67 status. From left to right, 1st box-plot: CLAM \cite{clam} when predicting WSI-level Ki67-percentage below 3.82 versus above 3.82. 2nd box-plot: patch-level Ki67-percentage below 3.82 versus above 3.82. 3rd-5th box-plots: patch-level Ki67-percentage (head 1: below 5 versus above 10, head 2: below 10 versus above 15, head 3: below 15 versus above 20).
   6th-9th box-plots: patch-level Ki67-percentage (head 1: below 5 versus above 5, head 2: below 10 versus above 10, head 3: below 15 versus above 15, and head 4: below 20 versus above 20).
}
\label{fig:aucs_ki67}
\end{figure}

\begin{figure}
\centering
 \begin{subfigure}[b]{0.8\textwidth}
  \centering
  \includegraphics[width=\textwidth]{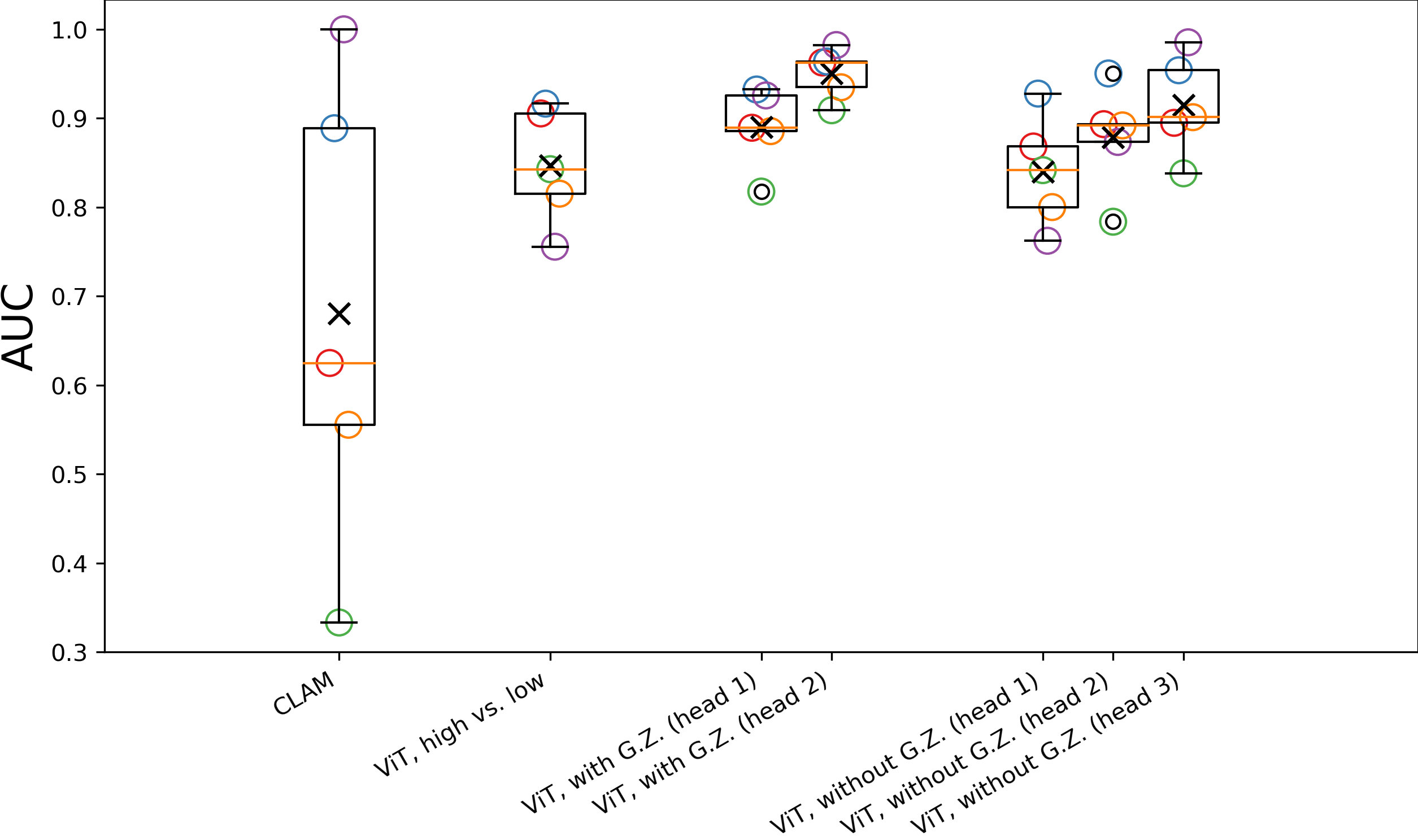}
  \end{subfigure}
\caption{
   Results for predicting ER status. From left to right, 1st box-plot: CLAM \cite{clam} when predicting WSI-level ER H-score below 42.61 versus above 42.61. 2nd box-plot: patch-level ER H-score below 42.61 versus above 42.61. 3rd-4th box-plots: patch-level ER H-score (head 1: below 30 versus above 60, head 2: below 60 versus above 90).
   5th-7th box-plots: patch-level ER-percentage (head 1: below 30 versus above 30, head 2: below 60 versus above 60, head 3: below 90 versus above 90).
}
\label{fig:aucs_er}
\end{figure}

\begin{figure}
\centering
 \begin{subfigure}[b]{0.8\textwidth}
  \centering
  \includegraphics[width=\textwidth]{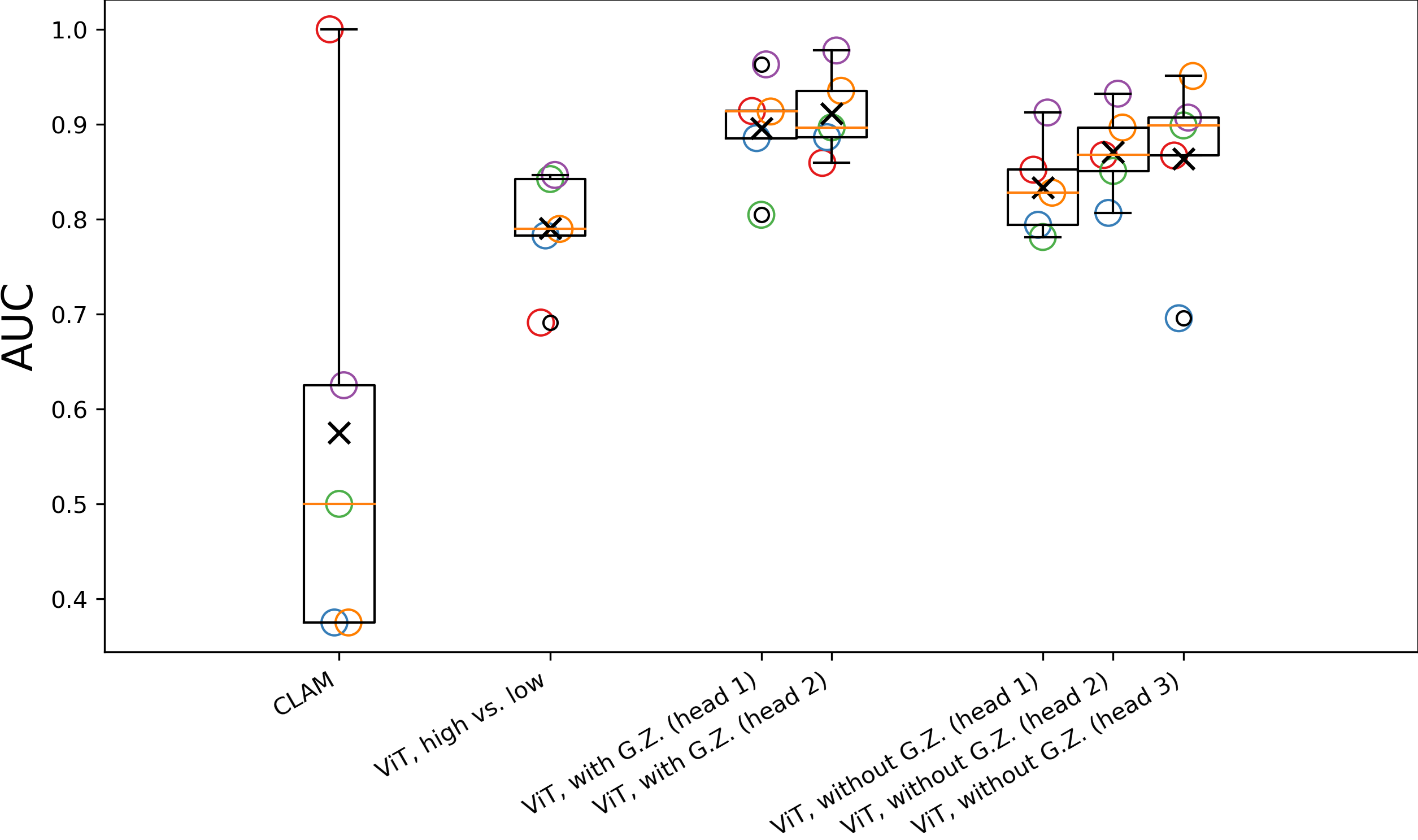}
  \end{subfigure}
\caption{
   Results for predicting PR status. From left to right, 1st box-plot: CLAM \cite{clam} when predicting WSI-level PR H-score below 7.373 versus above 7.373. 2nd box-plot: patch-level PR H-score below 7.373 versus above 7.373. 3rd-4th box-plots: patch-level PR H-score (head 1: below 30 versus above 60, head 2: below 60 versus above 90).
   5th-7th box-plots: patch-level PR-percentage (head 1: below 30 versus above 30, head 2: below 60 versus above 60, head 3: below 90 versus above 90).
}
\label{fig:aucs_pr}
\end{figure}
\begin{figure}
\centering
 \begin{subfigure}[b]{0.7\textwidth}
  \centering
  \includegraphics[width=\textwidth]{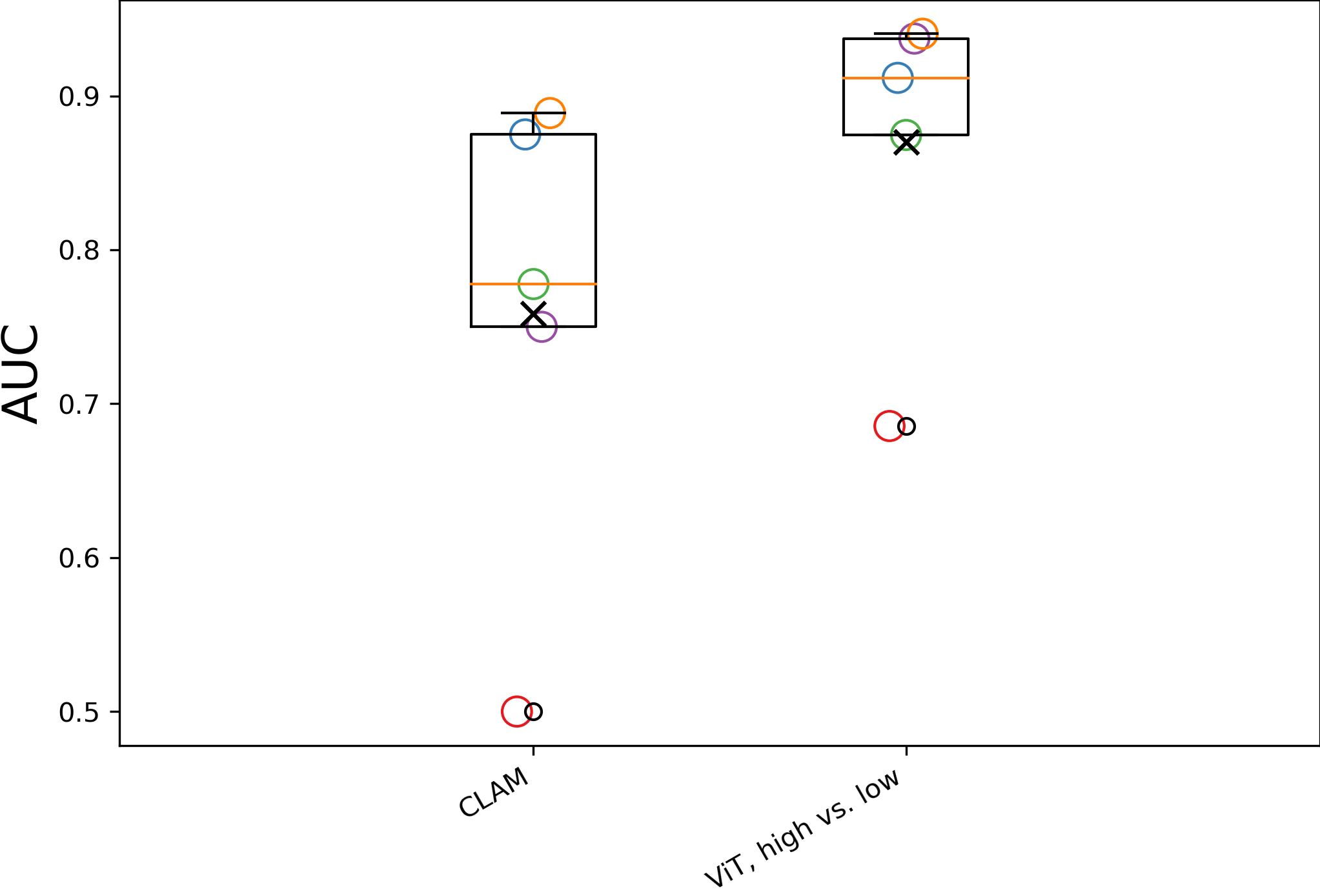}
  \end{subfigure}
\caption{
   Results for predicting HER2 status. From left to right, 1st box-plot: CLAM \cite{clam} when predicting WSI-level HER2 status positive versus negative.
   2nd box-plot: predicting whether 3+ patterns exist in a patch (positive/high) or not (negative/low). 
}
\label{fig:aucs_her2}
\end{figure}

\begin{figure}
\centering
 \begin{subfigure}[b]{0.18000000000000002\textwidth}
  \centering
  \includegraphics[width=\textwidth]{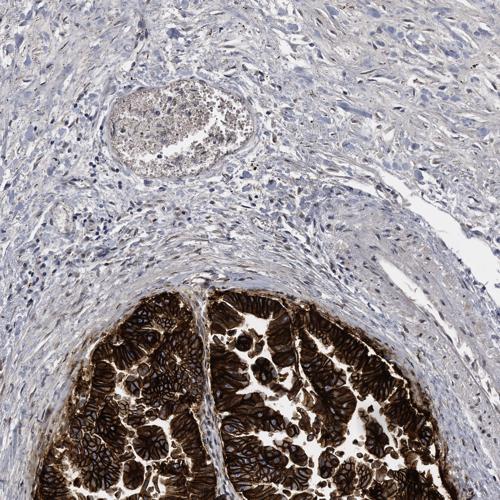}
  \end{subfigure}
\hfill
 \begin{subfigure}[b]{0.18000000000000002\textwidth}
  \centering
  \includegraphics[width=\textwidth]{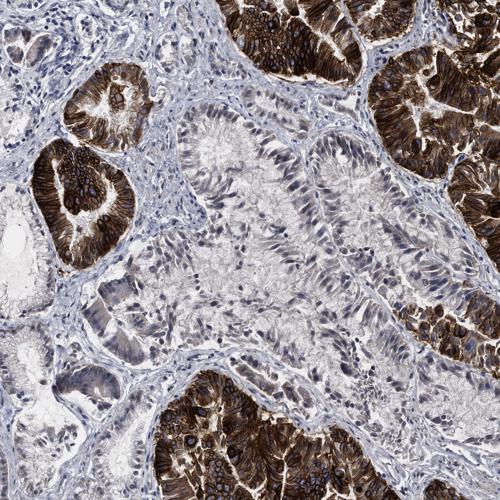}
  \end{subfigure}
\hfill
 \begin{subfigure}[b]{0.18000000000000002\textwidth}
  \centering
  \includegraphics[width=\textwidth]{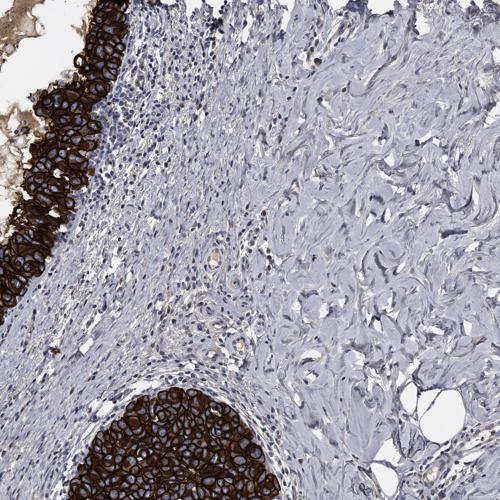}
  \end{subfigure}
\hfill
 \begin{subfigure}[b]{0.18000000000000002\textwidth}
  \centering
  \includegraphics[width=\textwidth]{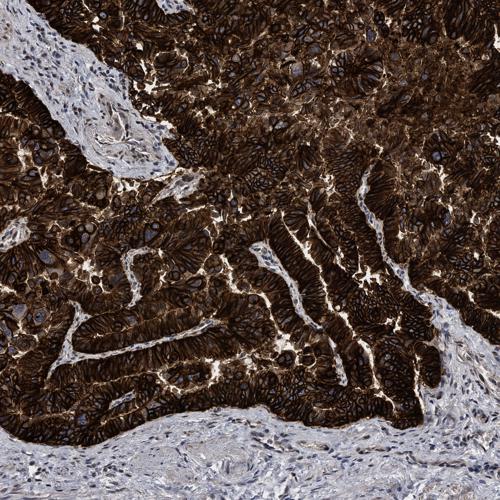}
  \end{subfigure}
\hfill
 \begin{subfigure}[b]{0.18000000000000002\textwidth}
  \centering
  \includegraphics[width=\textwidth]{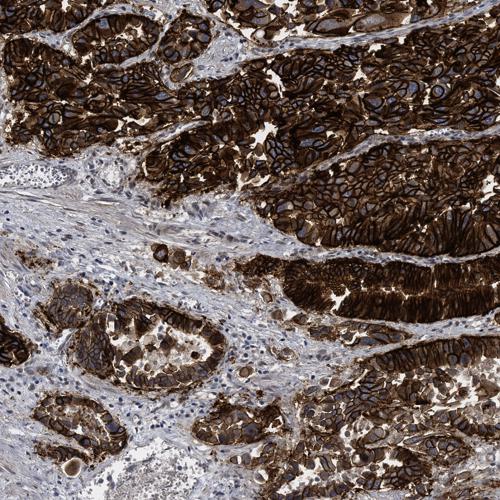}
  \end{subfigure}
\\
 \begin{subfigure}[b]{0.18000000000000002\textwidth}
  \centering
  \includegraphics[width=\textwidth]{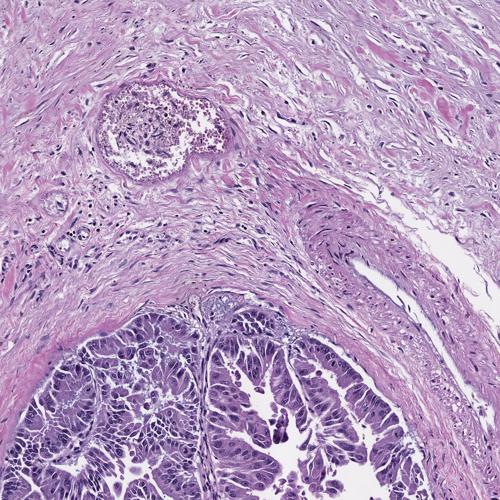}
  \end{subfigure}
\hfill
 \begin{subfigure}[b]{0.18000000000000002\textwidth}
  \centering
  \includegraphics[width=\textwidth]{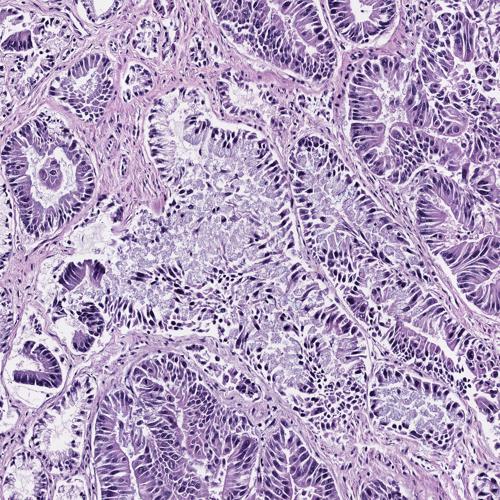}
  \end{subfigure}
\hfill
 \begin{subfigure}[b]{0.18000000000000002\textwidth}
  \centering
  \includegraphics[width=\textwidth]{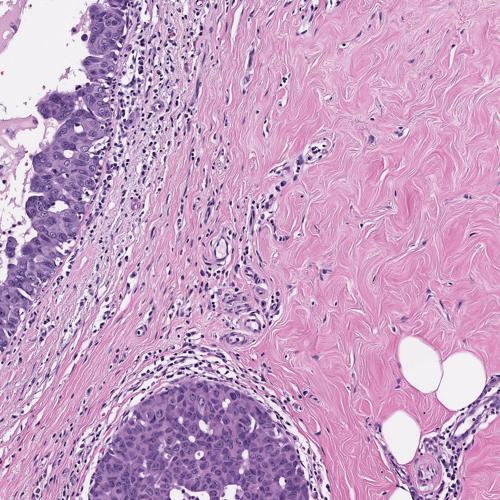}
  \end{subfigure}
\hfill
 \begin{subfigure}[b]{0.18000000000000002\textwidth}
  \centering
  \includegraphics[width=\textwidth]{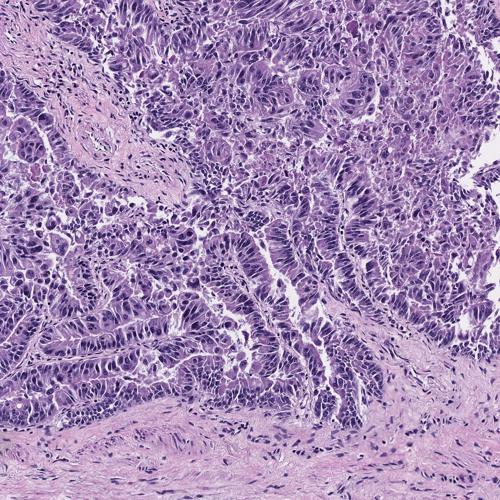}
  \end{subfigure}
\hfill
 \begin{subfigure}[b]{0.18000000000000002\textwidth}
  \centering
  \includegraphics[width=\textwidth]{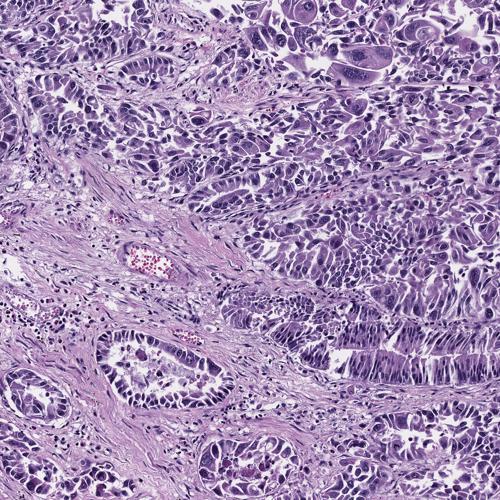}
  \end{subfigure}
\\
 \begin{subfigure}[b]{0.18000000000000002\textwidth}
  \centering
  \includegraphics[width=\textwidth]{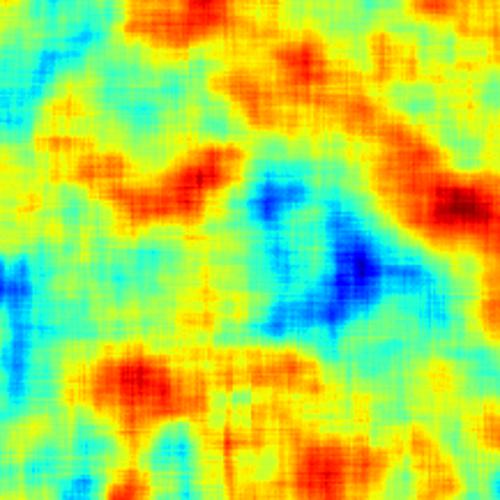}
  \end{subfigure}
\hfill
 \begin{subfigure}[b]{0.18000000000000002\textwidth}
  \centering
  \includegraphics[width=\textwidth]{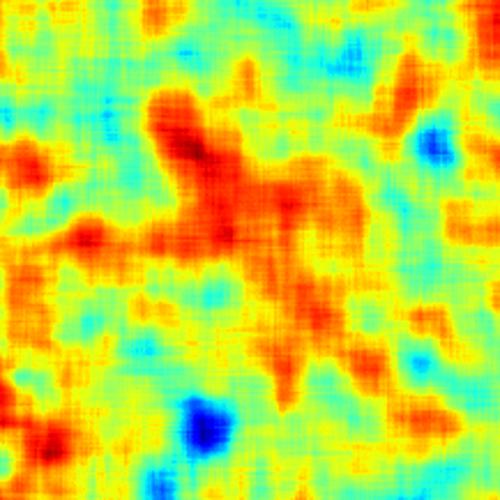}
  \end{subfigure}
\hfill
 \begin{subfigure}[b]{0.18000000000000002\textwidth}
  \centering
  \includegraphics[width=\textwidth]{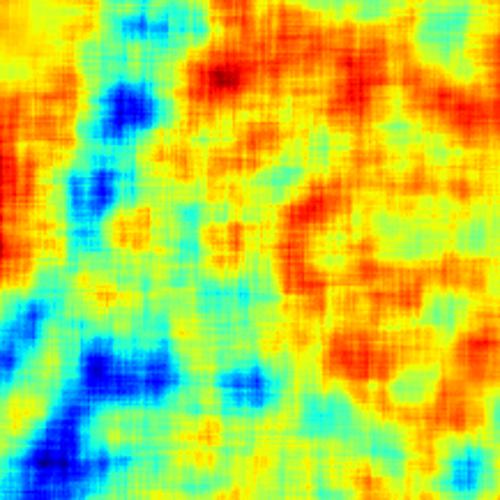}
  \end{subfigure}
\hfill
 \begin{subfigure}[b]{0.18000000000000002\textwidth}
  \centering
  \includegraphics[width=\textwidth]{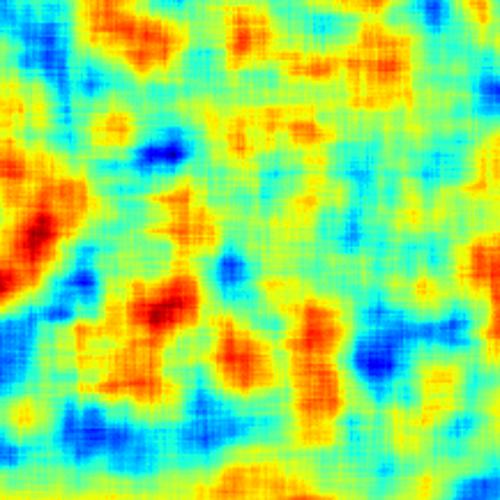}
  \end{subfigure}
\hfill
 \begin{subfigure}[b]{0.18000000000000002\textwidth}
  \centering
  \includegraphics[width=\textwidth]{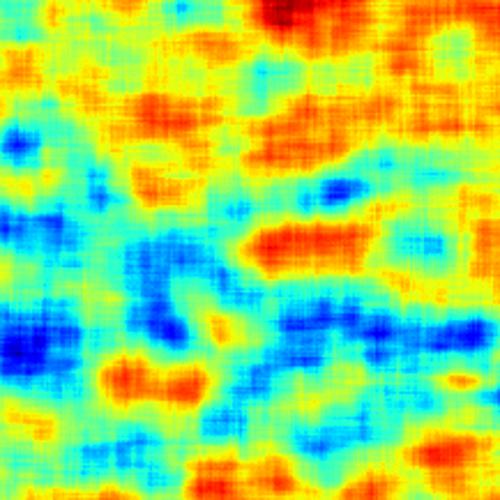}
  \end{subfigure}
\\
 \begin{subfigure}[b]{0.18000000000000002\textwidth}
  \centering
  \includegraphics[width=\textwidth]{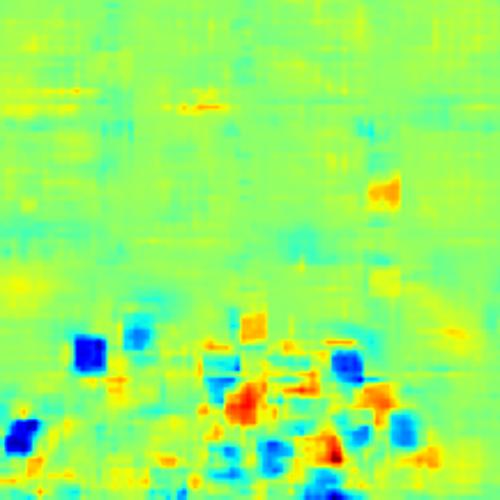}
  \end{subfigure}
\hfill
 \begin{subfigure}[b]{0.18000000000000002\textwidth}
  \centering
  \includegraphics[width=\textwidth]{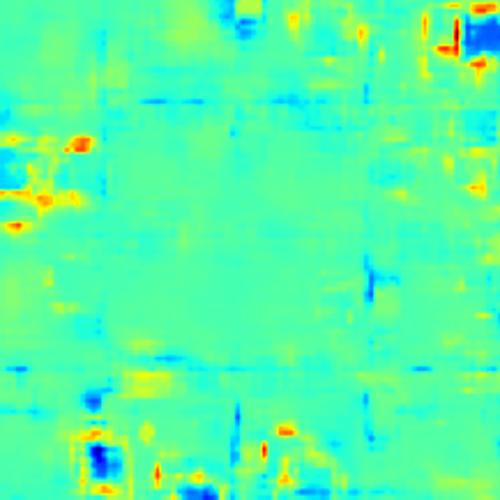}
  \end{subfigure}
\hfill
 \begin{subfigure}[b]{0.18000000000000002\textwidth}
  \centering
  \includegraphics[width=\textwidth]{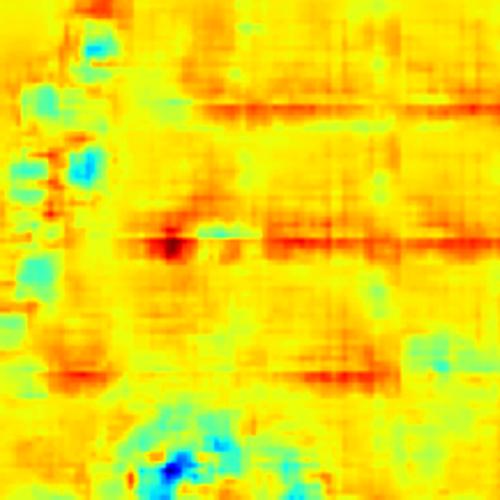}
  \end{subfigure}
\hfill
 \begin{subfigure}[b]{0.18000000000000002\textwidth}
  \centering
  \includegraphics[width=\textwidth]{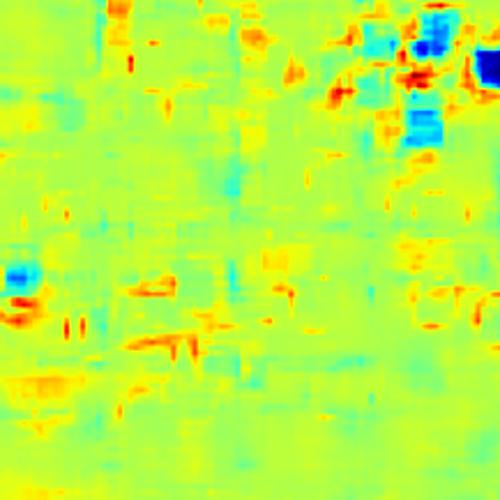}
  \end{subfigure}
\hfill
 \begin{subfigure}[b]{0.18000000000000002\textwidth}
  \centering
  \includegraphics[width=\textwidth]{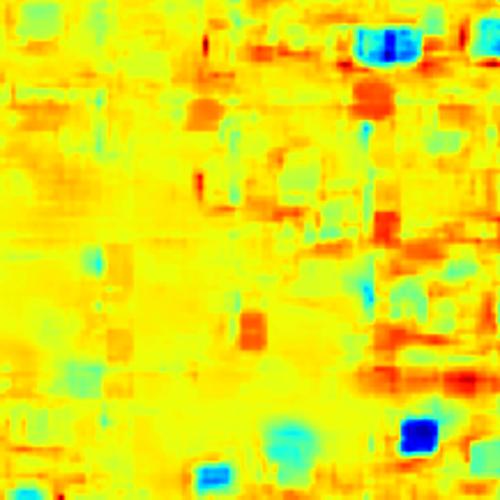}
  \end{subfigure}
  \caption{
  Localizations for HER2 predictors trained/evaluated on the split shown by the orange circle in Fig. \ref{fig:aucs_her2}. Each column corresponds to a H\&E-IHC pair. Row 1: IHC, row 2: H\&E, row 3: CLAM \cite{clam}'s attention mask, row 4: patch classifier's sensitivity to pixels. In all heatmaps we used JET color-map in which high and low values appear in red and blue, respectively.    
}
\label{fig:localizations_her2}
\end{figure}

\subsection{The Used Architectures}\label{sec:used_architectures}
We used Clustering-constrained Attention Multiple Instance Learning (CLAM) \cite{clam} whenever experimenting with weak WSI-level labels. When dealing with patch-level labels, we firstly resized each 3K by 3K patch to 1K by 1K pixels. Afterwards, we used a pipeline that processes a 1K by 1K patch as follows. A 1K by 1K patch is divided to a 4 by 4 grid whose cells are 250 by 250 pixels. Consequently, each grid cell is fed to a Resnet-18 \cite{resnet} backbone to get a volumetric map of shape $512 \times 8 \times 8$ for each cell of the grid. These volumetric maps are concatenated to form a volumetric map of shape $512 \times 32 \times 32$, which is fed to a multi-head self-attention layer (used in vision transformers \cite{vit}) followed by a classification head with a linear layer. We used a publicly available implementation of vision transformers (ViT) \cite{vit} in PyTorch \cite{vitcode}. During training we used the following data augmentations: $\pm10\%$ jitter to saturation and hue channels, and $\pm180$ degrees random rotation to images.  
\subsection{Labeling Protocols for Classifiers}\label{sec:labelingprotocol}
In this study we did not use regression to predict the continuous numbers (i.e. H-scores for ER, PR, and percentage for Ki67). Instead, we used a simple method known as cumulative logits approach \cite{cumullogits} which puts a set of thresholds like $T_1$, $T_2$, ..., $T_M$ on the continuous value where ${T_1 < T_2 < ...< T_M}$. Afterwards, for each threshold a separate classifier is considered; classifier 1 predicts whether the continuous value is below $T_1$ or above $T_1$, classifier 2 predicts whether the continuous value is below $T_2$ or above $T_2$, ..., classifier M predicts whether the continuous value is below $T_M$ or above $T_M$. Using this approach comes with the following issue. Let $\epsilon$ be a small number. The $m$-th classifier is supposed to assign a sample with number $T_m - \epsilon$ and a sample with number $T_m + \epsilon$ to different classes. In other words, the gray-zone samples whose number is close to $T_m$ may confuse the m-th classifier. To avoid this issue, we also tested a slightly modified approach as follows. Classifier 1 predicts whether the continuous value is below $T_1$ or above $T_2$, classifier 2 predicts whether the continuous value is below $T_2$ or above $T_3$, ..., classifier $M-1$ predicts whether the continuous value is below $T_{M-1}$ or above $T_M$. We refer to the former and the latter approaches as "without gray-zone" and "with gray-zone", respectively.        
\subsection{Training Setup}\label{sec:trainingsetup}
For the weakly-supervised method CLAM \cite{clam} we used the default parameter settings available in the public repository \cite{clamcode}. When training the strongly supervised patch classifiers, we always used an AdamW \cite{adamw} optimizer with the AMSGrad \cite{amsgrad} flag enabled. We trained with two learning rates, 0.0001 and 0.00001 and picked the checkpoint with the highest validation AUC. To have a fair comparison, the training cases shown to the CLAM \cite{clam} and the patch classifiers should be as close as possible. To this end, we firstly select 90 percent of WSIs for training the WSI-classifier CLAM \cite{clam} and the rest for testing. Consequently, the patches from the training/testing whole-slide images are used as training/testing sets for the patch classifiers. By doing so, the WSI-classifier CLAM \cite{clam} and the patch classifier are provided with the same set of training cases. To make a validation set for a patch classifier, for each WSI in the training set we randomly selected one of the annotated regions (like the regions shown in the third column of Fig. \ref{fig:extract_patches}) and considered its patches in the validation set. For CLAM \cite{clam} we randomly selected 5 WSIs as the validation set. Note that due to class imbalance the random testing/validation splits may become devoid of instances from some classes. In this study we did not use such splits in the experiments.       

\section{Results}
\subsection{Prediction Performances}
For each marker we created 5 splits according to the procedure of Sec. \ref{sec:trainingsetup}. When presenting the results, each split has its own unique color. For example in Fig. \ref{fig:aucs_ki67} the red circles correspond to a fixed training/testing split which is used by different methods and labeling protocols. Prediction performances in terms of AUC are provided in Figs. \ref{fig:aucs_ki67}, \ref{fig:aucs_er}, \ref{fig:aucs_pr}, and \ref{fig:aucs_her2}. In these figures the first box-plot on the left shows the performance of CLAM \cite{clam} in classifying WSIs to two classes. For HER2 (Fig. \ref{fig:aucs_her2} the box-plot on the left) WSIs from 0 and 1+ cases are labeled as CLAM \cite{clam}'s class 0 and WSIs from 3+ cases are labeled as CLAM \cite{clam}'s class 1. For Ki67, ER, and PR we firstly computed the median of the continuous values in the dataset, which are 3.82 for Ki67 percentage, 42.61 for ER H-score, and 7.373 for PR H-score. Afterwards, we used these median values to make WSI-level labels for CLAM \cite{clam}; if the WSI-level continuous number is below the median value it belongs to class 0 and otherwise to class 1. We evaluated the performance of the vision transformer-based (ViT) \cite{vit} patch classifier with the same thresholds used for CLAM \cite{clam}. The results are provided in Figs. \ref{fig:aucs_ki67}, \ref{fig:aucs_er}, \ref{fig:aucs_pr}, and \ref{fig:aucs_her2} by the box-plots labeled as "ViT, high vs. low". Note that, for example in Fig. \ref{fig:aucs_ki67} CLAM \cite{clam}'s task is to predict whether the WSI-level Ki67 percentage is below or above 3.82. But the patch classifier (the box-plot labeled "ViT, high vs. low") predicts whether the patch-level (as opposed to WSI-level) Ki67 percentage is below or above 3.82. We also experimented with the cumulative logits approach, as explained in Sec. \ref{sec:labelingprotocol} and in two settings: with gray-zone and without gray-zone. In Figs. \ref{fig:aucs_ki67}, \ref{fig:aucs_er}, and \ref{fig:aucs_pr} the corresponding box-plots are labeled as "ViT with G.Z." and "ViT without G.Z.", respectively. We used the following thresholds on Ki67 percentage: 5, 10, 15, and 20. Moreover, for ER and PR we used the following thresholds on H-score: 30, 60, and 90. 

\subsection{Localizations}\label{sec:localization}
Besides reporting the prediction performances, we inspected to what degree the weakly-supervised method CLAM \cite{clam} and the ViT-based pipeline described in Sec. \ref{sec:used_architectures} can localize relevant regions. CLAM \cite{clam} has an explicit attention mechanism, so for CLAM \cite{clam} we traversed a 1K by 1K image with a sliding window of size 256 and stride of 10, feeding each 256 by 256 patch to CLAM \cite{clam}'s attention sub-module to obtain a heatmap. If the heatmap has a large value at a pixel position, intuitively CLAM \cite{clam} has payed more attention to that pixel position.  
The ViT-based pipeline has no explicit attention mechanism, so to highlight the important pixel positions we used an attribution-based approach as follows. A 1K by 1K patch is traversed with a white patch of size 50 and stride 10, and the average change in pipeline's output is recorded in the heatmap. Intuitively, the sliding white patch hides a small region of the input image to see how it affects the pipeline's output. Note that there are more sophisticated feature-attribution methods, but they may produce discordant explanations \cite{khakzar}. In all heatmaps, we used JET color-map in which high and low values appear in red and blue, respectively.   
For instance, The localizations for HER2 results are provided in Fig. \ref{fig:localizations_her2} and for other markers in Figs. \akfigrefERheatmap{}, \akfigrefPRheatmap{}, and \akfigrefKiheatmap{} in the supplementary.
\section{Discussion} 
The potential of machine learning applied to digital pathology is evolving rapidly but predicting accurately molecular information merely from morphology remains a challenge. Breast cancer is heterogenous with a large variety of histomorphologies. A small minority of breast carcinomas have unique morphologic patterns, readily recognizable and associated with predictable molecular phenotypes. For instance, metaplastic squamous cell carcinoma is associated with triple negative status (ER, PR and HER2 negative); tubular carcinoma is associated with strong and diffuse ER positivity; mammary adenoid cystic carcinoma is typically triple negative for ER, PR, HER2 and positive for c-KIT (CD117). However the vast majority of breast cancers are not morphologically distinct (at least from a human perspective) and received the epithet "Not Otherwise Specified" or "No Special Type" (NOS or NST). In a quest to see if predictability can be improved, we built a large-scale dataset with reliable measurements for training models that predict Ki67, ER, PR, and HER2 statues from H\&E-stained images. In view of our encouraging results, we suggest that the large-scale of the dataset and the quality of labels (ground truth) are a major factor to improve prediction. 
This paper highlights the need to improve datasets both quantitatively and qualitatively. Most previous studies \cite{cotur1, shamai1, shamai2, drbarneshertwo, brafmiccai} used up to 5K training instances, compared to 20K-26K instances in this work. In addition, we managed to obtain high-quality ground-truth labels based on image analysis data extraction, instead of subjective labels provided by human readings, known for inter-observer variability.

One aspect of our methodological approach can be seen as a limitation. Indeed, when evaluating H-Score values for ER, PR and percentage for Ki67, the non-tumoral nuclei were not excluded from the automatic calculation. Since the majority of non-tumoral nuclei are negative for ER, PR and Ki67, the corresponding values are lower than expected if only tumoral cells would have been selected in the calculation. That methodological approach was initially chosen to mimic the polymerase chain reaction (PCR) assessment used in some commercial molecular signature tests which cannot entirely exclude non-tumoral cells. However since the public dataset is available, the same experiment could be repeated by excluding non-tumoral cells using a classifier.

CLAM \cite{clam}'s performance in Figs. \ref{fig:aucs_ki67}, \ref{fig:aucs_er}, \ref{fig:aucs_pr}, and \ref{fig:aucs_her2} (the 1\textsuperscript{st} bars from left) shows that weakly-supervised learning is not effective for predicting ER, PR, Ki67, and HER2 statuses.
This observation is consistent with the study done by Laleh et al \cite{benchmarkingweakly}. The best results correspond to the stongly-supervised ViT-based pipeline trained with the gray-zones explained in Sec. \ref{sec:labelingprotocol} (labled as "ViT-with G.Z." in Figs. \ref{fig:aucs_ki67}, \ref{fig:aucs_er}, \ref{fig:aucs_pr}, and labeled as "ViT, high vs. low" in Fig. \ref{fig:aucs_her2}). In this setting the prediction performances approaches or exceeds 90 in terms of AUCROC. Interestingly, when the gray-zone is not considered (labels as "ViT, without G.Z." in Figs. \ref{fig:aucs_ki67}, \ref{fig:aucs_er}, \ref{fig:aucs_pr}, and \ref{fig:aucs_her2}) we see a drastic drop in prediction performances. This shows that when the gray-zones are not considered, the small flaw of assigning different labels to patches with H-scores $T_m-\epsilon$ and $T_m+\epsilon$ (i.e. the issue elaborated upon in Sec. \ref{sec:labelingprotocol}) can significantly reduce the prediction performance. In our best results (labled as "ViT-with G.Z." in Figs. \ref{fig:aucs_ki67}, \ref{fig:aucs_er}, \ref{fig:aucs_pr}, and labeled as "ViT, high vs. low" in Fig. \ref{fig:aucs_her2}) the prediction performances approach or exceed 90\% in terms of AUC for most splits, but we see a drop of performance for some splits. Since roughly 240 WSIs are used to create the dataset, we hypothesize that in those splits some tissue-types in the test set happen to be missing in the training set. In other words, one might still need to expand the dataset to expose machine learning models to more morphological variety during training. An interesting future direction to improve molecular status prediction based on morphology is applying explainable AI \cite{gpex}\cite{reprpoint}. AI Explainability offers many opportunities of improvement by opening the "black box" nature of machine learning models. One could identify tissue variants which are missing in training phase and selectively adding those tissue variants to the dataset, or even to identify and prevent failures which are not related to diversity of images in the dataset.

Besides reporting the prediction performances, we obtained the localization maps from different methods according to the procedure of Sec. \ref{sec:localization}. These localization maps are a first step toward AI explainability. These localization maps are provided in Fig. \ref{fig:localizations_her2} (for HER2) and Figs. \akfigrefERheatmap{}, \akfigrefPRheatmap{}, and \akfigrefKiheatmap{} (respectively for ER, PR and Ki67) in the supplementary. In these figures the five columns show different (H\&E)-(H-DAB) examples: the 1\textsuperscript{st} and 2\textsuperscript{nd} rows depict respectively the H\&E and H-DAB images. Brown regions in an H-DAB image correspond to Her2/neu amplified (3+) positive and the same regions in the H\&E modality should ideally be highlighted by machine learning methods. The 3\textsuperscript{rd} and 4\textsuperscript{th} rows respectively illustrate the CLAM \cite{clam}'s and the ViT-based pipline's heatmaps. According to these heatmaps, although both pipelines have achieved around 90 AUCs neither of them can successfully localize the relevant regions. For example, according to the heatmap in 3\textsuperscript{rd} column and 3\textsuperscript{rd} row, CLAM \cite{clam} mistakenly pays attention to the negative region in the two o'clock position. This observation is consistent with the experiments done by Laleh et al. \cite{benchmarkingweakly} in which weakly-supervised methods mistakenly highlight tissue borders or other artefact when trained to predict molecular information.
Another observation is that the heatmaps related to the ViT-based pipeline (i.e. the heatmaps in the 4\textsuperscript{th} row) are often homogeneous. This might be due to the fact that the pipeline is by-design good at making use of contextual information. More precisely, although a CNN normally takes in or "sees" a relatively small patch, the ViT-based pipeline explained in Sec. \ref{sec:used_architectures} can see a 1K by 1K patch in its entirety. All in all, although in Figs. \ref{fig:localizations_her2}, \akfigrefERheatmap{}, \akfigrefPRheatmap{}, and \akfigrefKiheatmap{} the selected checkpoints can achieve around or above 90\% AUCs, none of them can successfully localize relevant tissue regions. Intuitively, during training the pipelines are never asked to decide if each region is positive or negative, and they merely need to output a label for a patch or WSI. This motivates evaluating machine learning methods based on criteria that measure how good a method is at localizing relevant regions, or even training models to optimize those criteria. 
One way of implementing such a criterion is to compute the "closeness" of a method's localization map to the brown DAB channel of the corresponding H-DAB image provided in our dataset. Since the H\&E-stained patch the corresponding H-DAB patch might be misaligned, the measure for "closeness" should be invariant to rigid body movements in either of the images.

In conclusion, this work demonstrates that larger datasets combined with high-quality labels can improve the accuracy to predict molecular status of the classical 4 IHCs (ER, PR, HER2 and Ki67) performed in human breast cancer assessment. It also demonstrates that this dataset improvement is probably not the only limiting factor to achieve accuracy improvement. Using existing machine learning models shows obvious discrepancy between the concept of human and machine learning models "attentions". We suggest that AI explainability will be the key to make significant progresses in the field. Our dataset was made public with the expectation to trigger new proposals from the scientific AI community to improve prediction.

\section{Acknowledgements}
This work was supported by Mitacs accelerate. The experiments of this study were enabled in part by the Digital Research Alliance of Canada. We would like to thank DynaLIFE Medical Laboratory (Edmonton, Canada) for their support in the slide scanning process.  



 \bibliographystyle{elsarticle-num} 
 \bibliography{cas-refs}
  \newpage

\renewcommand\thesection{S\arabic{section}}
\renewcommand{\thefigure}{S\arabic{figure}}

\newcommand{\akfigrefkiaucs}{4}
\newcommand{\akfigreferaucs}{5}
\newcommand{\akfigrefpraucs}{6}
\newcommand{\akciteclam}{[21]}




\setcounter{figure}{0} 





\centering{
\Large{
Supplementary Material For \\
Predicting Ki67, ER, PR, and HER2 Statuses from H\&E-stained Breast-cancer Images
}
}
\vspace{2cm}
\begin{figure}[H]
\centering
 \begin{subfigure}[b]{0.18000000000000002\textwidth}
  \centering
  \includegraphics[width=\textwidth]{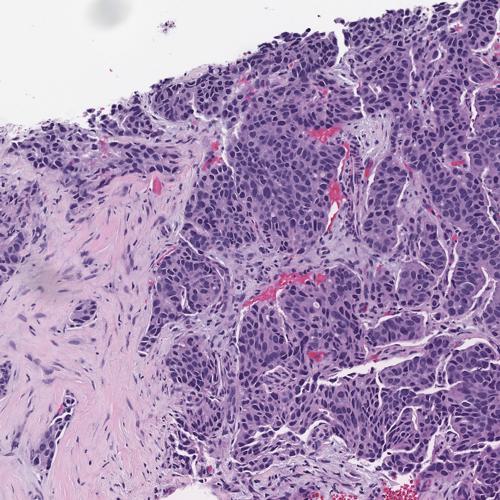}
  \end{subfigure}
\hfill
 \begin{subfigure}[b]{0.18000000000000002\textwidth}
  \centering
  \includegraphics[width=\textwidth]{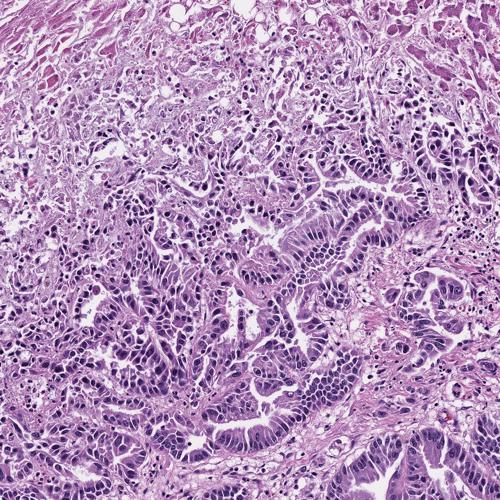}
  \end{subfigure}
\hfill
 \begin{subfigure}[b]{0.18000000000000002\textwidth}
  \centering
  \includegraphics[width=\textwidth]{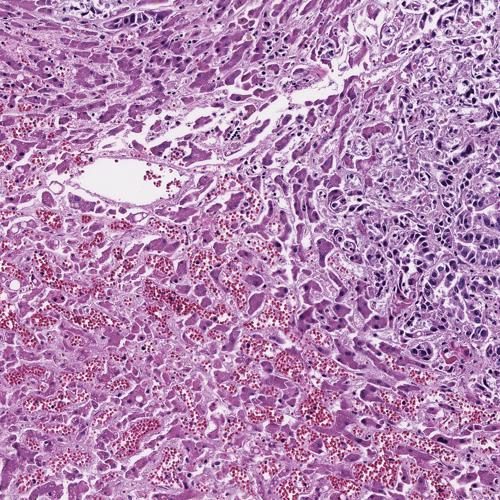}
  \end{subfigure}
\hfill
 \begin{subfigure}[b]{0.18000000000000002\textwidth}
  \centering
  \includegraphics[width=\textwidth]{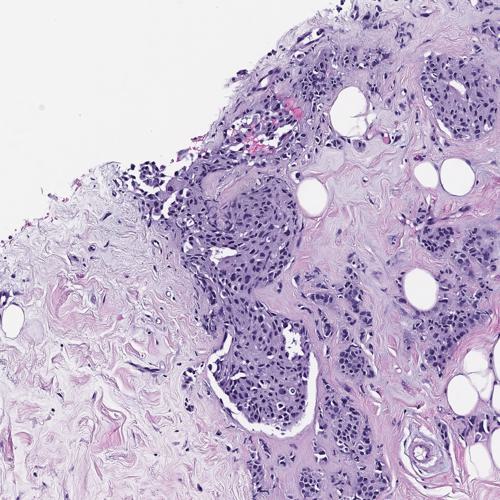}
  \end{subfigure}
\hfill
 \begin{subfigure}[b]{0.18000000000000002\textwidth}
  \centering
  \includegraphics[width=\textwidth]{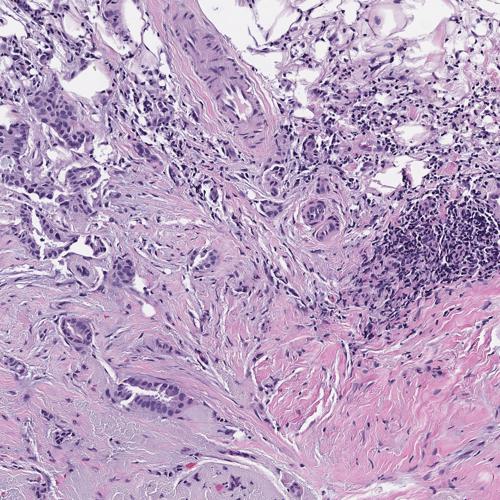}
  \end{subfigure}
\\
 \begin{subfigure}[b]{0.18000000000000002\textwidth}
  \centering
  \includegraphics[width=\textwidth]{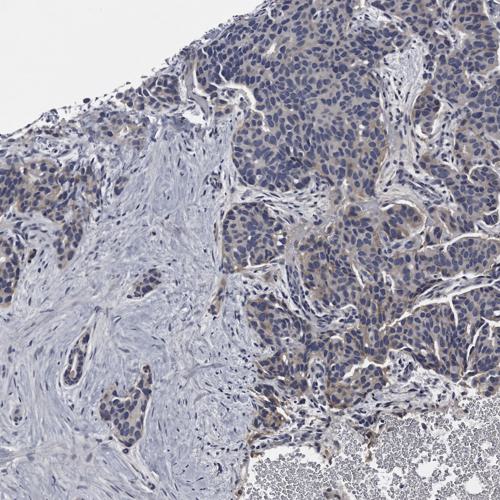}
  \end{subfigure}
\hfill
 \begin{subfigure}[b]{0.18000000000000002\textwidth}
  \centering
  \includegraphics[width=\textwidth]{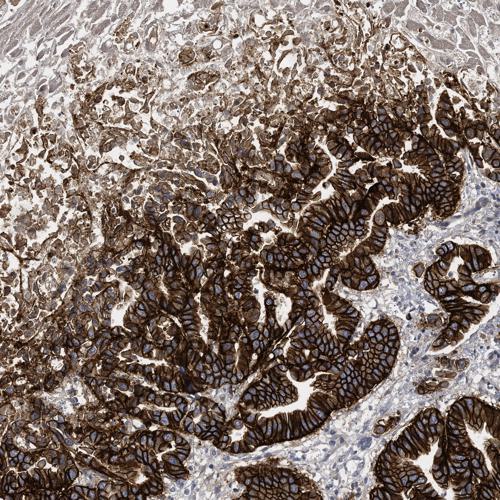}
  \end{subfigure}
\hfill
 \begin{subfigure}[b]{0.18000000000000002\textwidth}
  \centering
  \includegraphics[width=\textwidth]{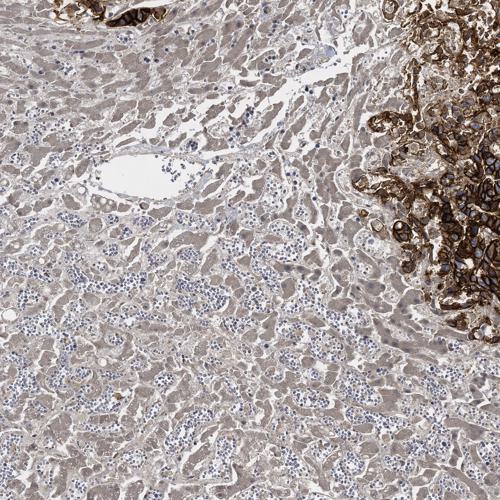}
  \end{subfigure}
\hfill
 \begin{subfigure}[b]{0.18000000000000002\textwidth}
  \centering
  \includegraphics[width=\textwidth]{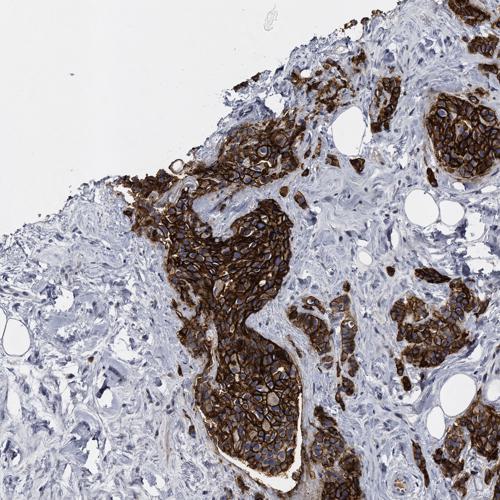}
  \end{subfigure}
\hfill
 \begin{subfigure}[b]{0.18000000000000002\textwidth}
  \centering
  \includegraphics[width=\textwidth]{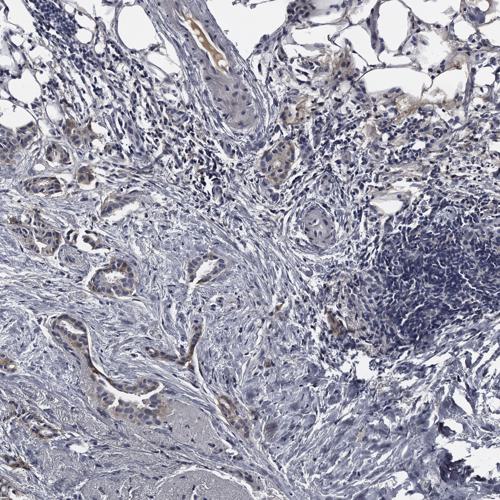}
  \end{subfigure}
  \caption{
    Examples of HER2 pairs in the dataset. Each column corresponds to a H\&E-IHC pair. Row1: H\&E, Row2: IHC.
    In columns 2, 3, and 4 some 3+ patterns exist in the H-DAB modality and the corresponding regions exist in the H\&E modality. So those pairs are labeled as positive during the exhaustive inspection. On the other hand, the pairs in the 1st and 5th columns are labeled as negative. 
    }
  \label{fig:dsintro_Her2}
\end{figure}
\begin{figure}
\centering
 \begin{subfigure}[b]{0.18000000000000002\textwidth}
  \centering
  \includegraphics[width=\textwidth]{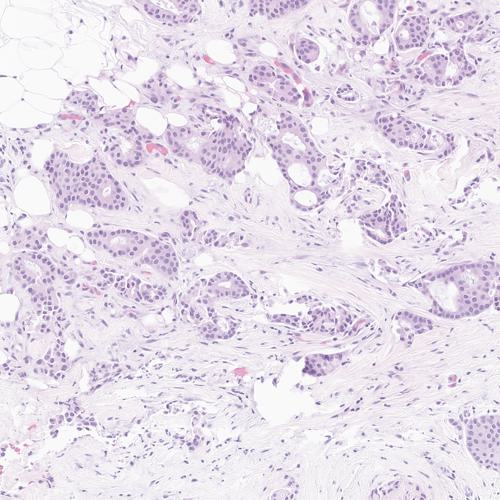}
  \end{subfigure}
\hfill
 \begin{subfigure}[b]{0.18000000000000002\textwidth}
  \centering
  \includegraphics[width=\textwidth]{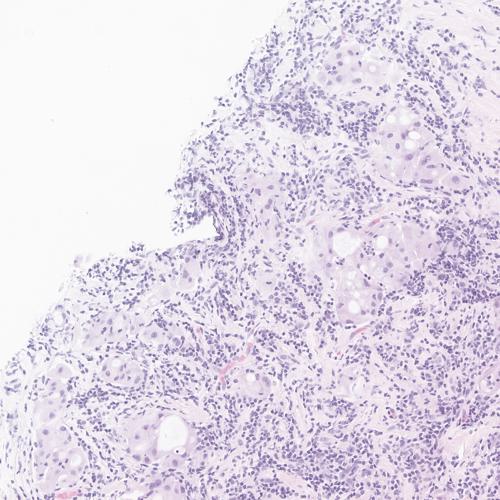}
  \end{subfigure}
\hfill
 \begin{subfigure}[b]{0.18000000000000002\textwidth}
  \centering
  \includegraphics[width=\textwidth]{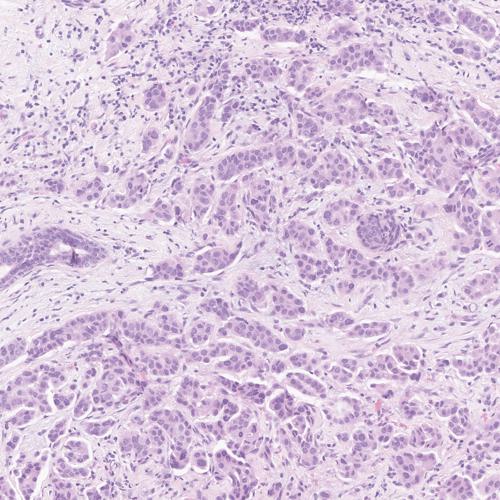}
  \end{subfigure}
\hfill
 \begin{subfigure}[b]{0.18000000000000002\textwidth}
  \centering
  \includegraphics[width=\textwidth]{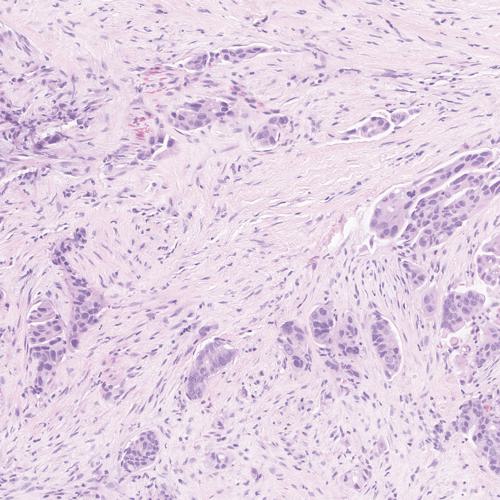}
  \end{subfigure}
\hfill
 \begin{subfigure}[b]{0.18000000000000002\textwidth}
  \centering
  \includegraphics[width=\textwidth]{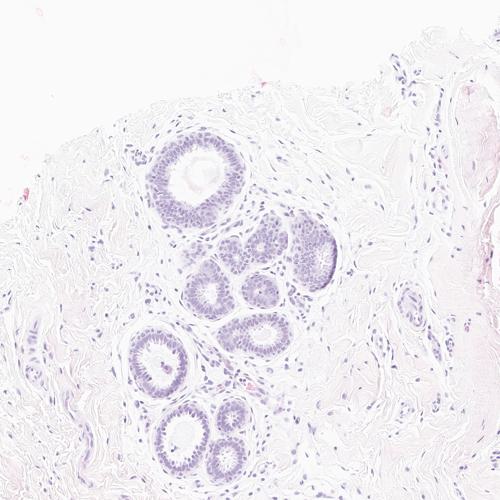}
  \end{subfigure}
\\
 \begin{subfigure}[b]{0.18000000000000002\textwidth}
  \centering
  \includegraphics[width=\textwidth]{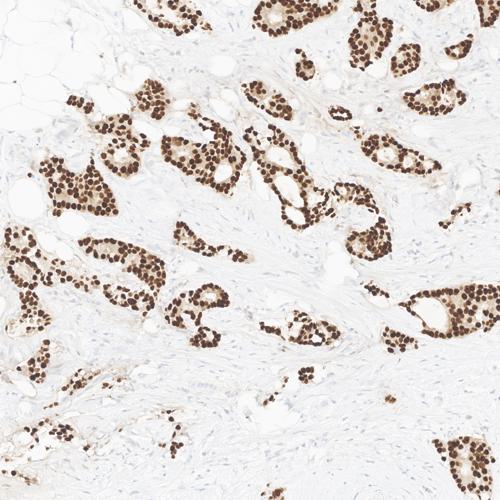}
  \end{subfigure}
\hfill
 \begin{subfigure}[b]{0.18000000000000002\textwidth}
  \centering
  \includegraphics[width=\textwidth]{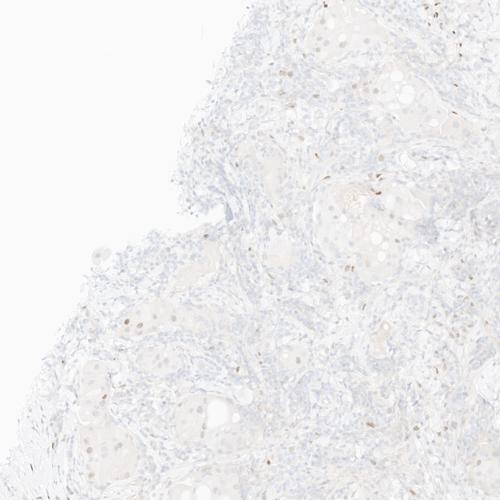}
  \end{subfigure}
\hfill
 \begin{subfigure}[b]{0.18000000000000002\textwidth}
  \centering
  \includegraphics[width=\textwidth]{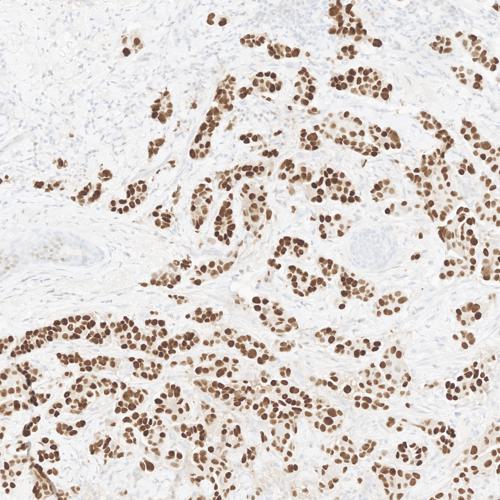}
  \end{subfigure}
\hfill
 \begin{subfigure}[b]{0.18000000000000002\textwidth}
  \centering
  \includegraphics[width=\textwidth]{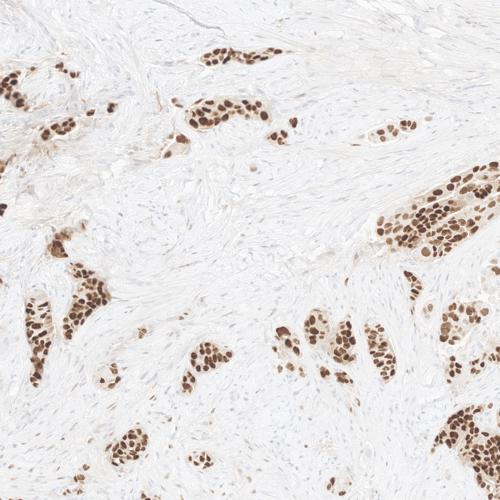}
  \end{subfigure}
\hfill
 \begin{subfigure}[b]{0.18000000000000002\textwidth}
  \centering
  \includegraphics[width=\textwidth]{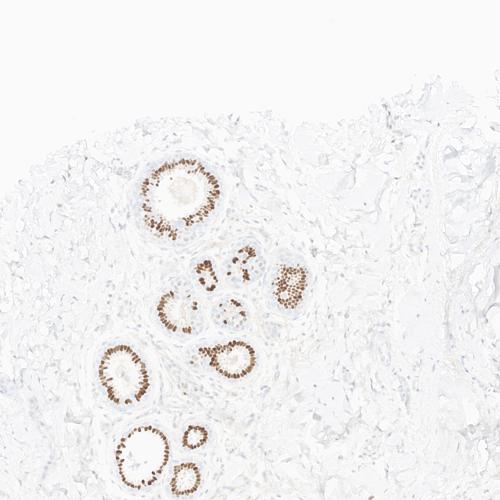}
  \end{subfigure}
\\
 \begin{subfigure}[b]{0.18000000000000002\textwidth}
  \centering
  \includegraphics[width=\textwidth]{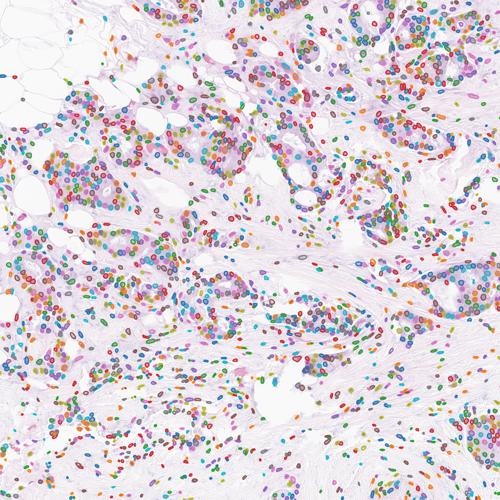}
  \end{subfigure}
\hfill
 \begin{subfigure}[b]{0.18000000000000002\textwidth}
  \centering
  \includegraphics[width=\textwidth]{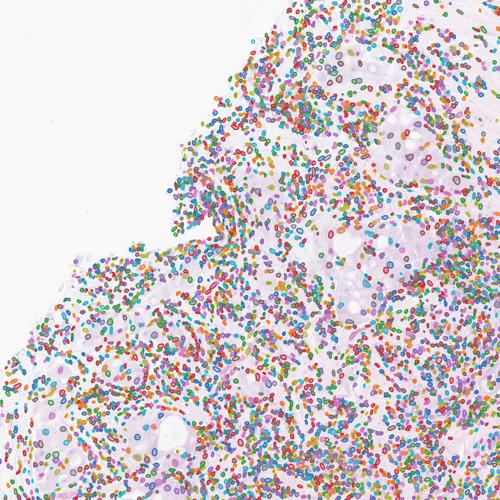}
  \end{subfigure}
\hfill
 \begin{subfigure}[b]{0.18000000000000002\textwidth}
  \centering
  \includegraphics[width=\textwidth]{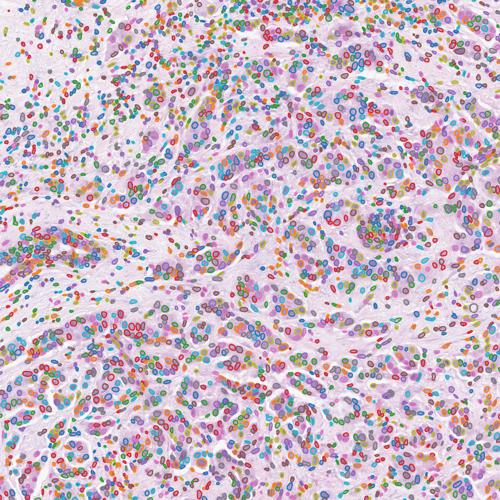}
  \end{subfigure}
\hfill
 \begin{subfigure}[b]{0.18000000000000002\textwidth}
  \centering
  \includegraphics[width=\textwidth]{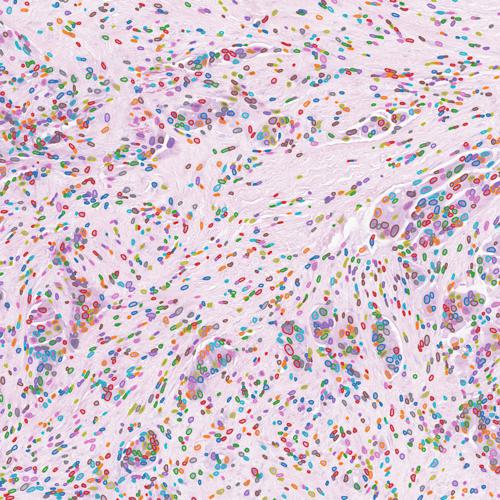}
  \end{subfigure}
\hfill
 \begin{subfigure}[b]{0.18000000000000002\textwidth}
  \centering
  \includegraphics[width=\textwidth]{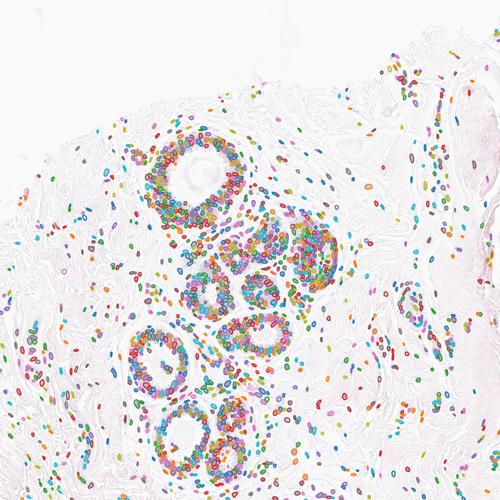}
  \end{subfigure}
\\
 \begin{subfigure}[b]{0.18000000000000002\textwidth}
  \centering
  \includegraphics[width=\textwidth]{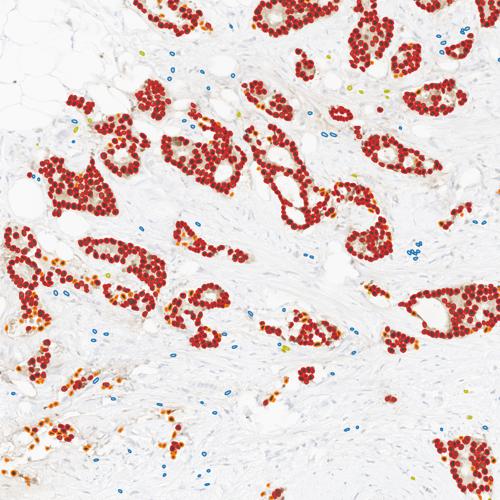}
  \end{subfigure}
\hfill
 \begin{subfigure}[b]{0.18000000000000002\textwidth}
  \centering
  \includegraphics[width=\textwidth]{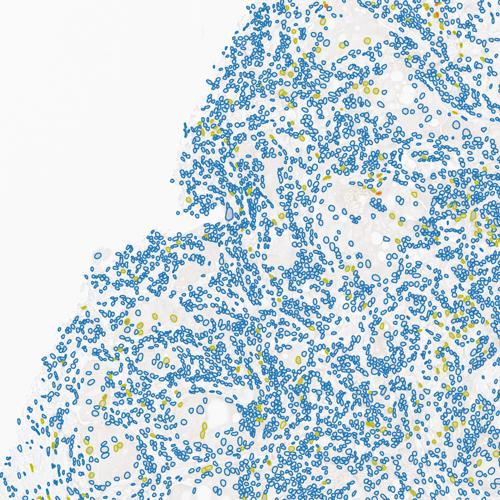}
  \end{subfigure}
\hfill
 \begin{subfigure}[b]{0.18000000000000002\textwidth}
  \centering
  \includegraphics[width=\textwidth]{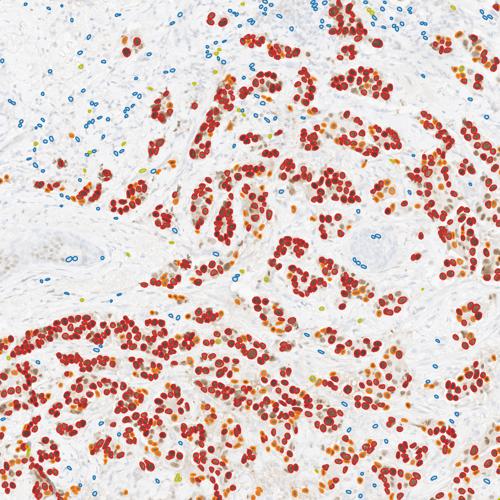}
  \end{subfigure}
\hfill
 \begin{subfigure}[b]{0.18000000000000002\textwidth}
  \centering
  \includegraphics[width=\textwidth]{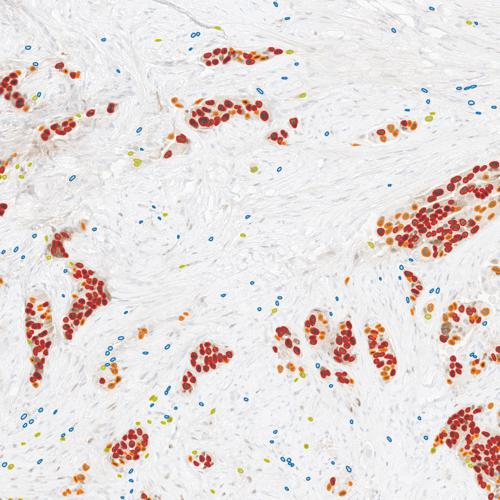}
  \end{subfigure}
\hfill
 \begin{subfigure}[b]{0.18000000000000002\textwidth}
  \centering
  \includegraphics[width=\textwidth]{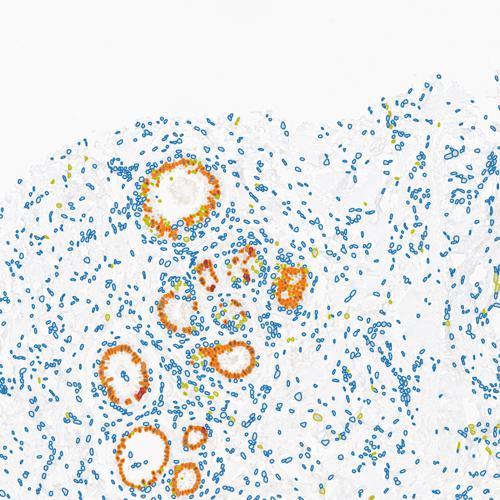}
  \end{subfigure}
  \caption{
   Examples of  H\&E-ER pairs in the dataset. Each column corresponds to a H\&E-IHC pair. Row 1: H\&E, Row 2: IHC, Row 3:nuclei sementations on H\&E, Row 4: the result of DAB-analysis on IHC where blue, yellow, orange, and red colors mark 0, 1+, 2+, and 3+ nuclei, respectively. 
   In the first and fourth columns there are several nuclei which are detected in the H\&E modality (row 3) but they are missed in the IHC modality (row 4). Moreover, in the fifth column in the IHC modality (row 4) the number of nuclei are overestimated. 
  }
  \label{fig:dsintro_ER}
\end{figure}

\begin{figure}
\centering
 \begin{subfigure}[b]{0.18000000000000002\textwidth}
  \centering
  \includegraphics[width=\textwidth]{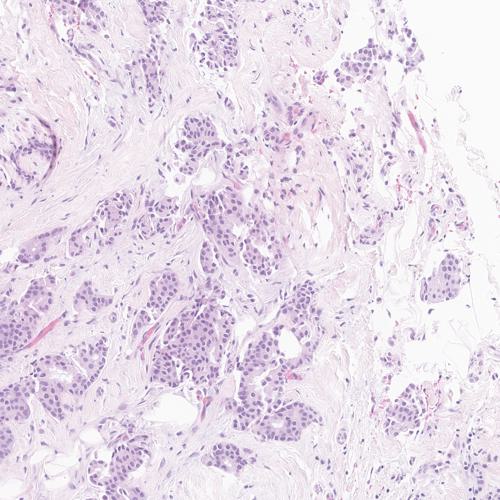}
  \end{subfigure}
\hfill
 \begin{subfigure}[b]{0.18000000000000002\textwidth}
  \centering
  \includegraphics[width=\textwidth]{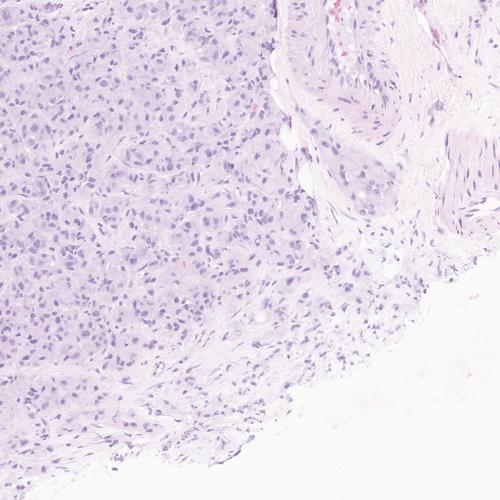}
  \end{subfigure}
\hfill
 \begin{subfigure}[b]{0.18000000000000002\textwidth}
  \centering
  \includegraphics[width=\textwidth]{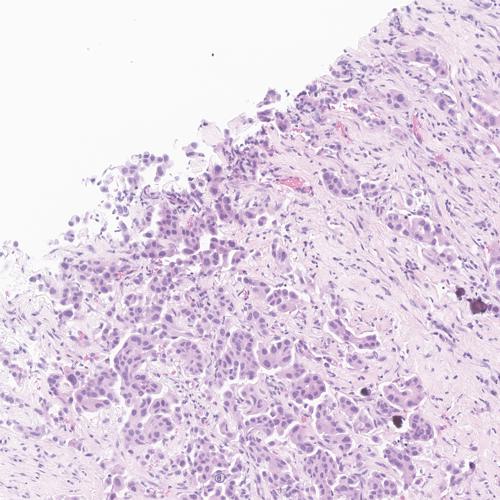}
  \end{subfigure}
\hfill
 \begin{subfigure}[b]{0.18000000000000002\textwidth}
  \centering
  \includegraphics[width=\textwidth]{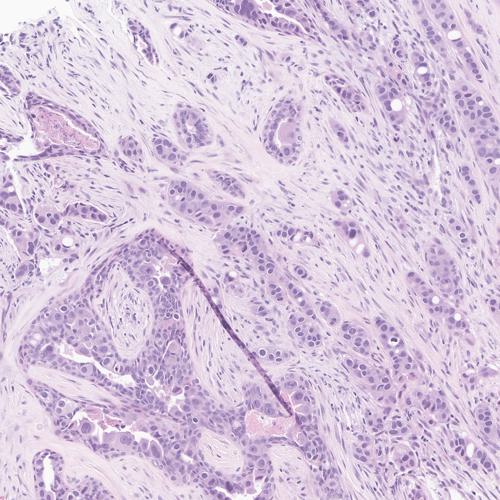}
  \end{subfigure}
\hfill
 \begin{subfigure}[b]{0.18000000000000002\textwidth}
  \centering
  \includegraphics[width=\textwidth]{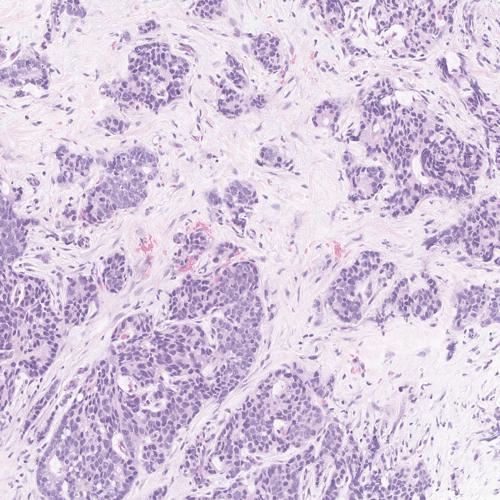}
  \end{subfigure}
\\
 \begin{subfigure}[b]{0.18000000000000002\textwidth}
  \centering
  \includegraphics[width=\textwidth]{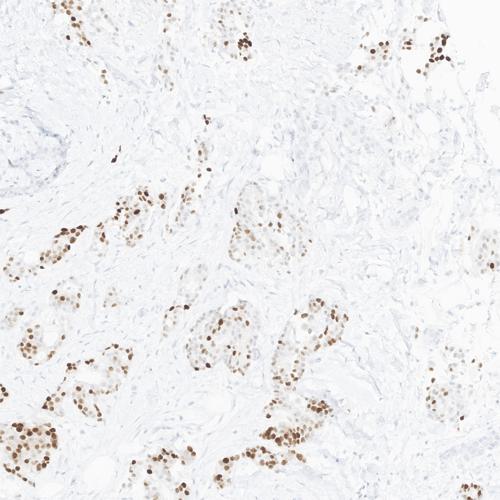}
  \end{subfigure}
\hfill
 \begin{subfigure}[b]{0.18000000000000002\textwidth}
  \centering
  \includegraphics[width=\textwidth]{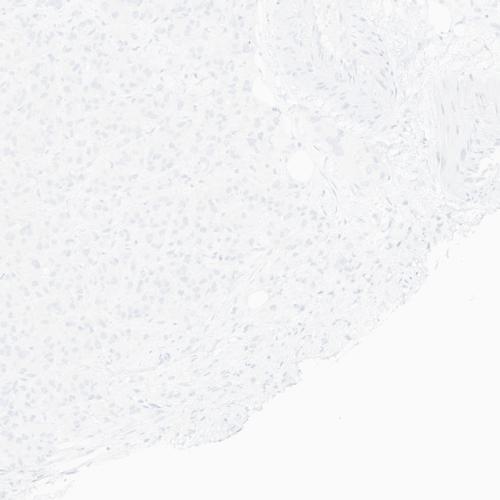}
  \end{subfigure}
\hfill
 \begin{subfigure}[b]{0.18000000000000002\textwidth}
  \centering
  \includegraphics[width=\textwidth]{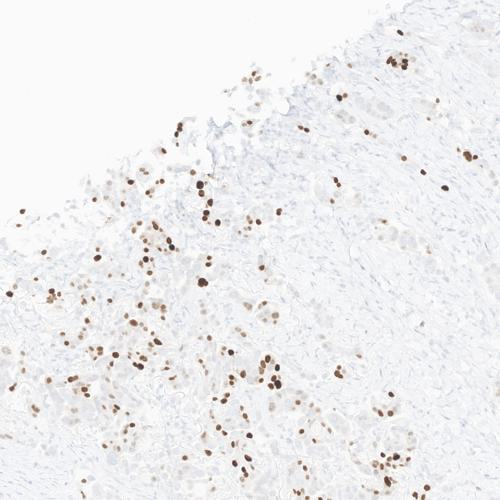}
  \end{subfigure}
\hfill
 \begin{subfigure}[b]{0.18000000000000002\textwidth}
  \centering
  \includegraphics[width=\textwidth]{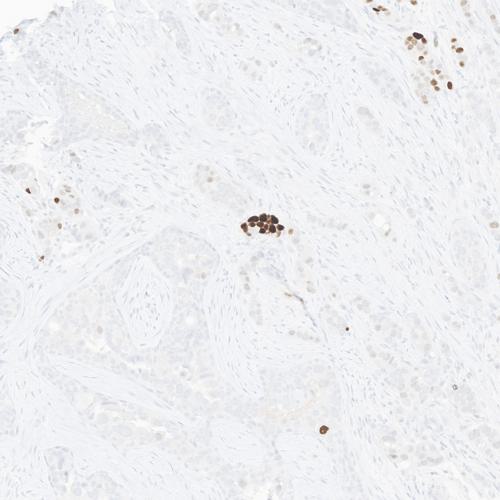}
  \end{subfigure}
\hfill
 \begin{subfigure}[b]{0.18000000000000002\textwidth}
  \centering
  \includegraphics[width=\textwidth]{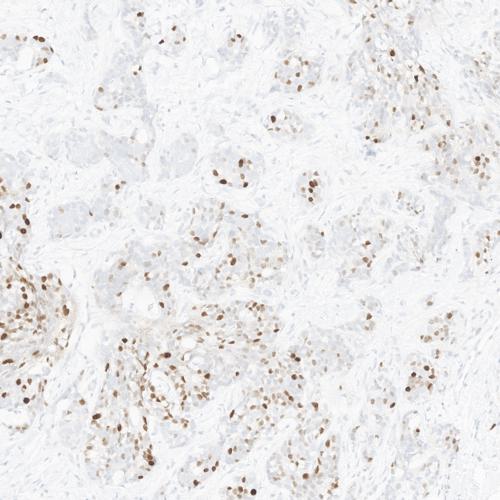}
  \end{subfigure}
\\
 \begin{subfigure}[b]{0.18000000000000002\textwidth}
  \centering
  \includegraphics[width=\textwidth]{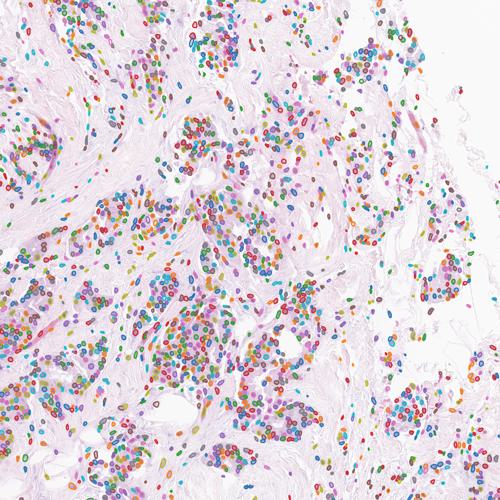}
  \end{subfigure}
\hfill
 \begin{subfigure}[b]{0.18000000000000002\textwidth}
  \centering
  \includegraphics[width=\textwidth]{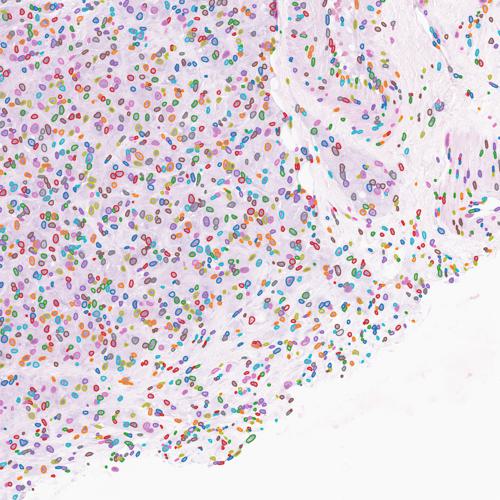}
  \end{subfigure}
\hfill
 \begin{subfigure}[b]{0.18000000000000002\textwidth}
  \centering
  \includegraphics[width=\textwidth]{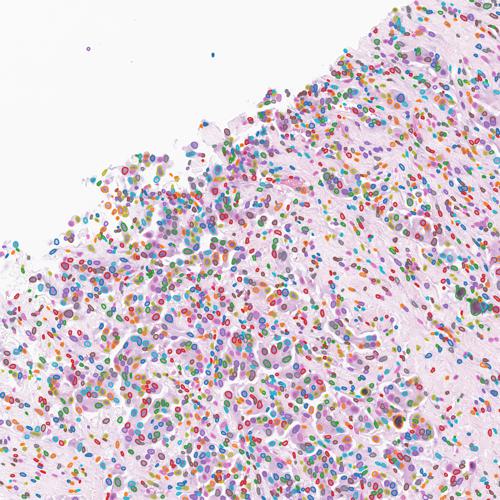}
  \end{subfigure}
\hfill
 \begin{subfigure}[b]{0.18000000000000002\textwidth}
  \centering
  \includegraphics[width=\textwidth]{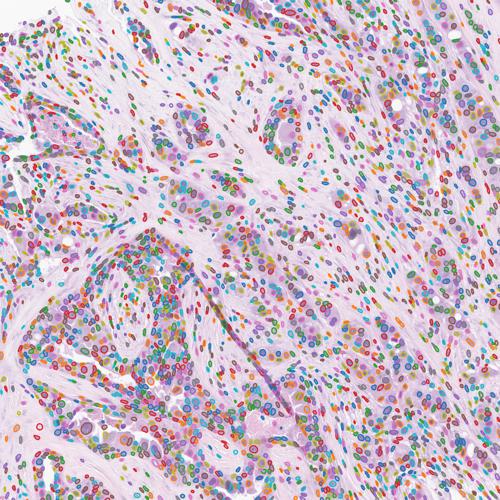}
  \end{subfigure}
\hfill
 \begin{subfigure}[b]{0.18000000000000002\textwidth}
  \centering
  \includegraphics[width=\textwidth]{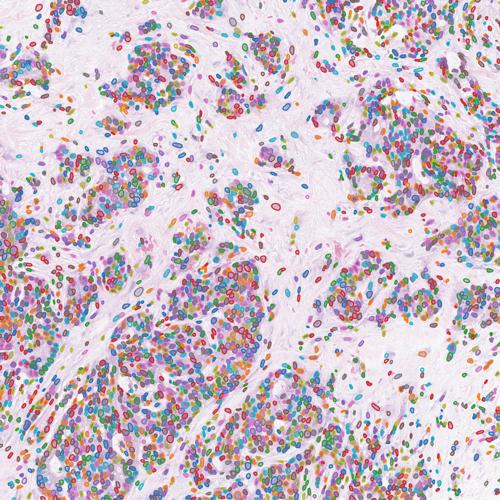}
  \end{subfigure}
\\
 \begin{subfigure}[b]{0.18000000000000002\textwidth}
  \centering
  \includegraphics[width=\textwidth]{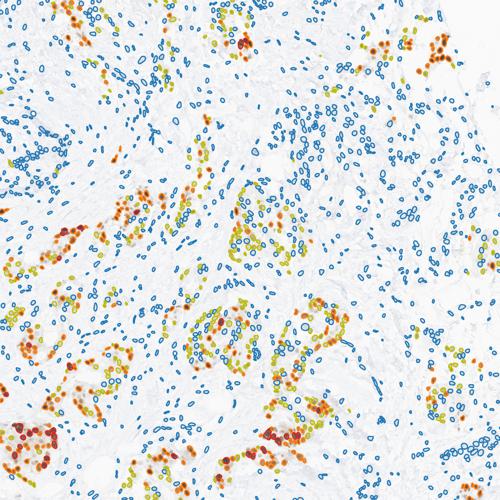}
  \end{subfigure}
\hfill
 \begin{subfigure}[b]{0.18000000000000002\textwidth}
  \centering
  \includegraphics[width=\textwidth]{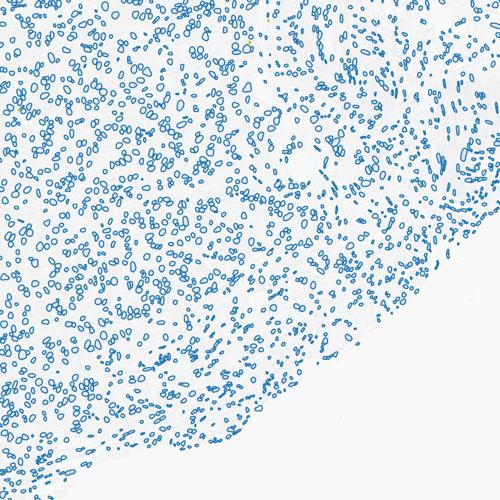}
  \end{subfigure}
\hfill
 \begin{subfigure}[b]{0.18000000000000002\textwidth}
  \centering
  \includegraphics[width=\textwidth]{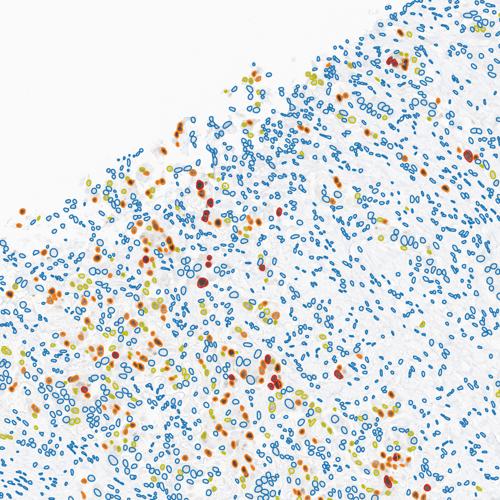}
  \end{subfigure}
\hfill
 \begin{subfigure}[b]{0.18000000000000002\textwidth}
  \centering
  \includegraphics[width=\textwidth]{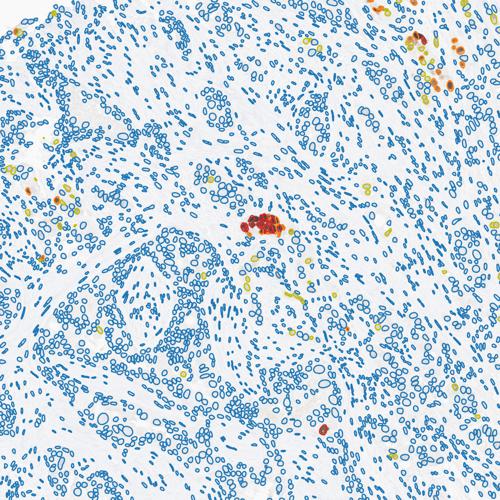}
  \end{subfigure}
\hfill
 \begin{subfigure}[b]{0.18000000000000002\textwidth}
  \centering
  \includegraphics[width=\textwidth]{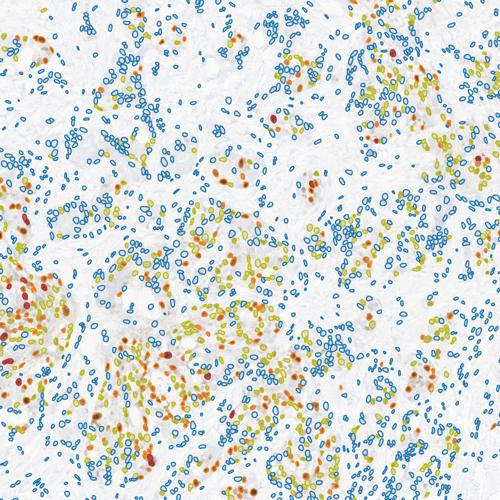}
  \end{subfigure}\\
  \caption{
   Examples of  H\&E-PR pairs in the dataset. Each column corresponds to a H\&E-IHC pair. Row1: H\&E, Row2: IHC, Row3:nuclei sementations on H\&E, Row4: the result of DAB-analysis on IHC where blue, yellow, orange, and red colors mark 0, 1+, 2+, and 3+ nuclei, respectively.
  }
  \label{fig:dsintro_PR}
\end{figure}
\begin{figure}
\centering
 \begin{subfigure}[b]{0.18000000000000002\textwidth}
  \centering
  \includegraphics[width=\textwidth]{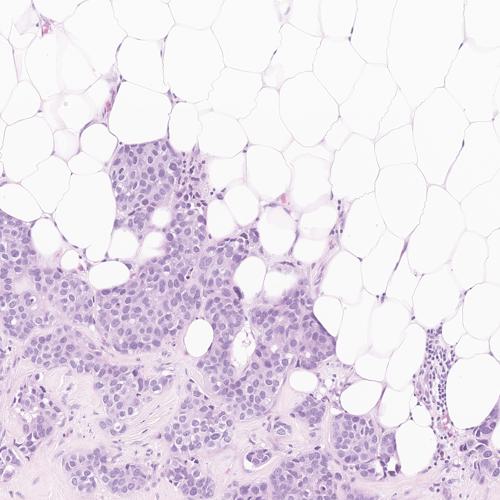}
  \end{subfigure}
\hfill
 \begin{subfigure}[b]{0.18000000000000002\textwidth}
  \centering
  \includegraphics[width=\textwidth]{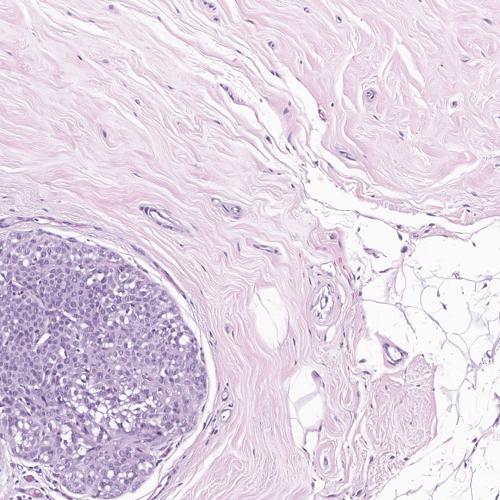}
  \end{subfigure}
\hfill
 \begin{subfigure}[b]{0.18000000000000002\textwidth}
  \centering
  \includegraphics[width=\textwidth]{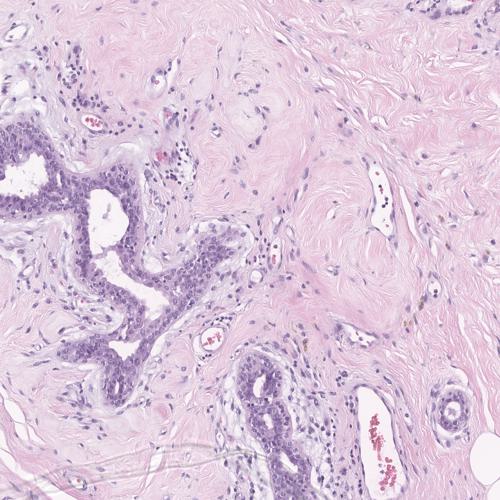}
  \end{subfigure}
\hfill
 \begin{subfigure}[b]{0.18000000000000002\textwidth}
  \centering
  \includegraphics[width=\textwidth]{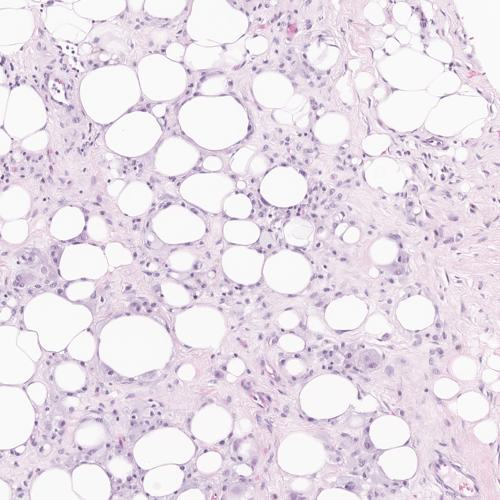}
  \end{subfigure}
\hfill
 \begin{subfigure}[b]{0.18000000000000002\textwidth}
  \centering
  \includegraphics[width=\textwidth]{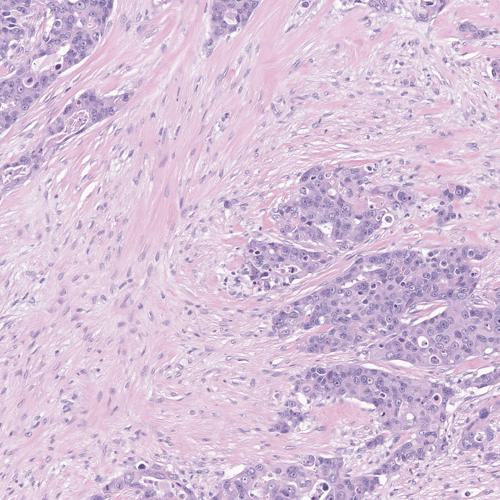}
  \end{subfigure}
\\
 \begin{subfigure}[b]{0.18000000000000002\textwidth}
  \centering
  \includegraphics[width=\textwidth]{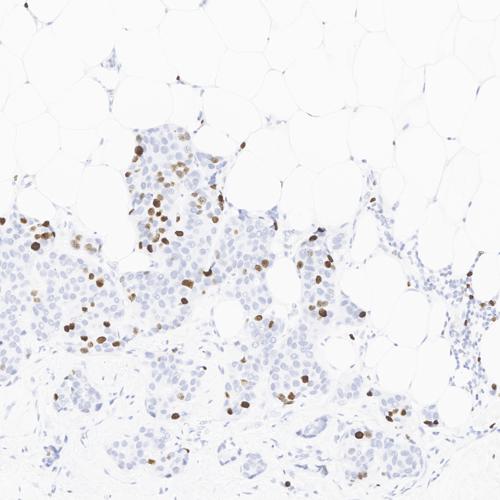}
  \end{subfigure}
\hfill
 \begin{subfigure}[b]{0.18000000000000002\textwidth}
  \centering
  \includegraphics[width=\textwidth]{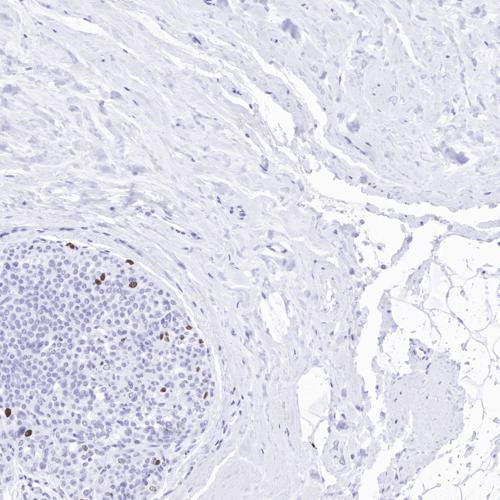}
  \end{subfigure}
\hfill
 \begin{subfigure}[b]{0.18000000000000002\textwidth}
  \centering
  \includegraphics[width=\textwidth]{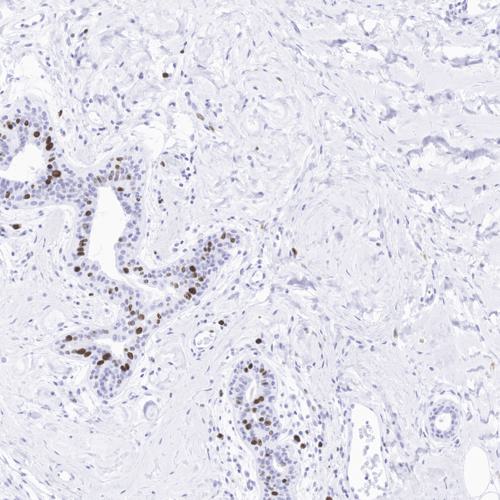}
  \end{subfigure}
\hfill
 \begin{subfigure}[b]{0.18000000000000002\textwidth}
  \centering
  \includegraphics[width=\textwidth]{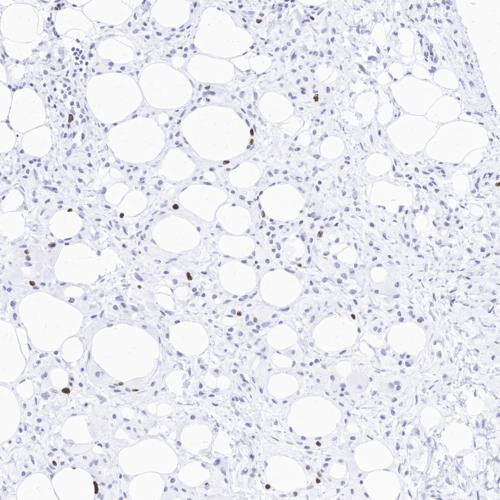}
  \end{subfigure}
\hfill
 \begin{subfigure}[b]{0.18000000000000002\textwidth}
  \centering
  \includegraphics[width=\textwidth]{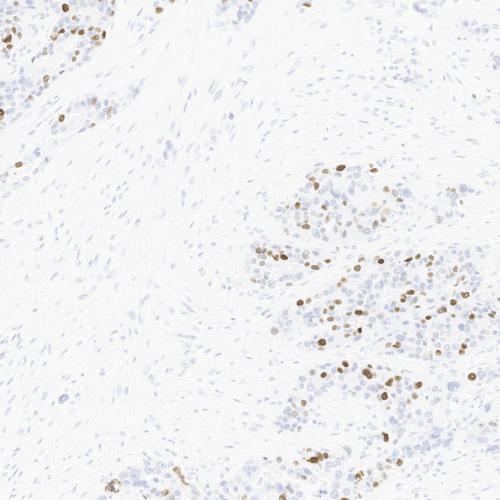}
  \end{subfigure}
\\
 \begin{subfigure}[b]{0.18000000000000002\textwidth}
  \centering
  \includegraphics[width=\textwidth]{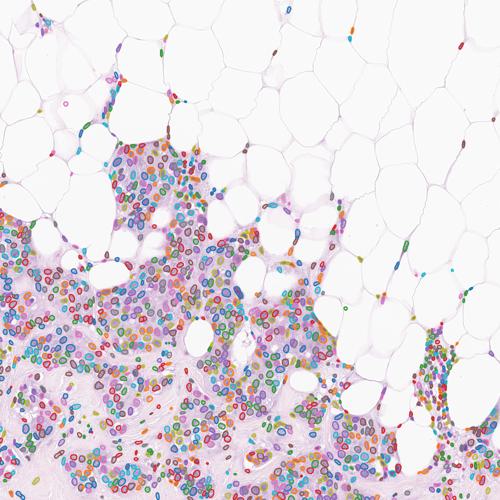}
  \end{subfigure}
\hfill
 \begin{subfigure}[b]{0.18000000000000002\textwidth}
  \centering
  \includegraphics[width=\textwidth]{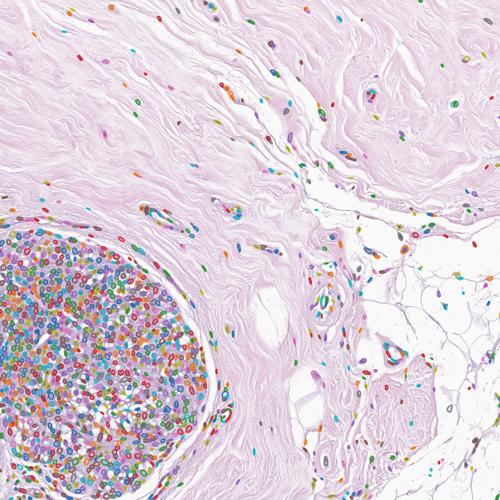}
  \end{subfigure}
\hfill
 \begin{subfigure}[b]{0.18000000000000002\textwidth}
  \centering
  \includegraphics[width=\textwidth]{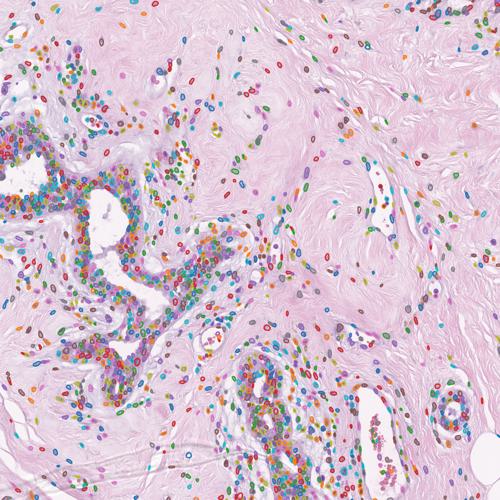}
  \end{subfigure}
\hfill
 \begin{subfigure}[b]{0.18000000000000002\textwidth}
  \centering
  \includegraphics[width=\textwidth]{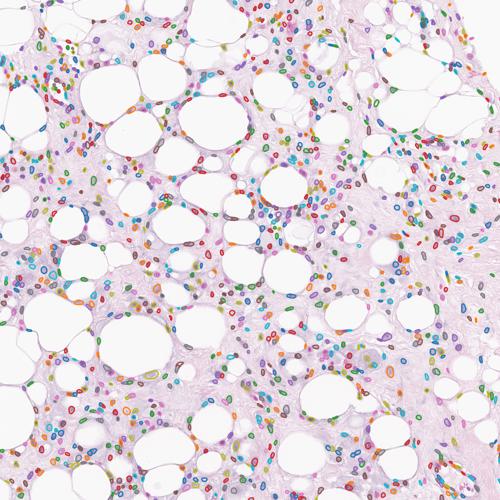}
  \end{subfigure}
\hfill
 \begin{subfigure}[b]{0.18000000000000002\textwidth}
  \centering
  \includegraphics[width=\textwidth]{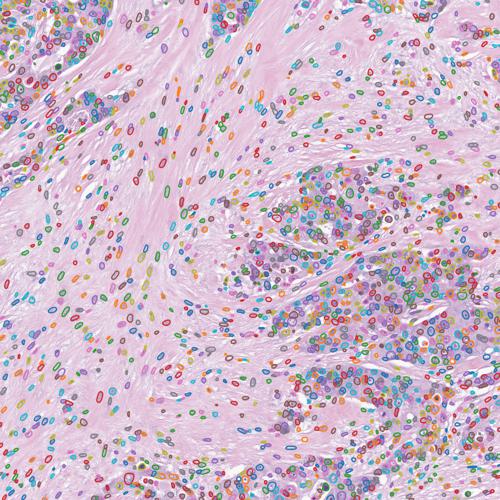}
  \end{subfigure}
\\
 \begin{subfigure}[b]{0.18000000000000002\textwidth}
  \centering
  \includegraphics[width=\textwidth]{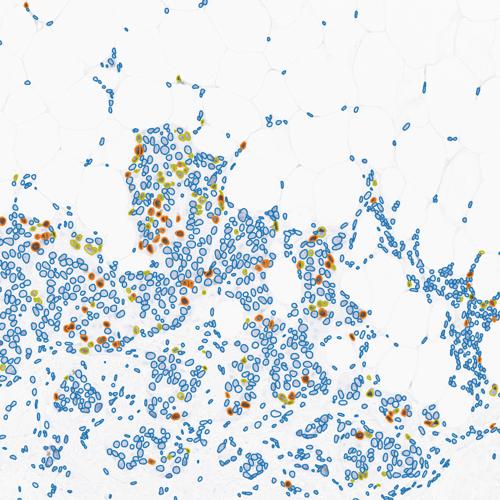}
  \end{subfigure}
\hfill
 \begin{subfigure}[b]{0.18000000000000002\textwidth}
  \centering
  \includegraphics[width=\textwidth]{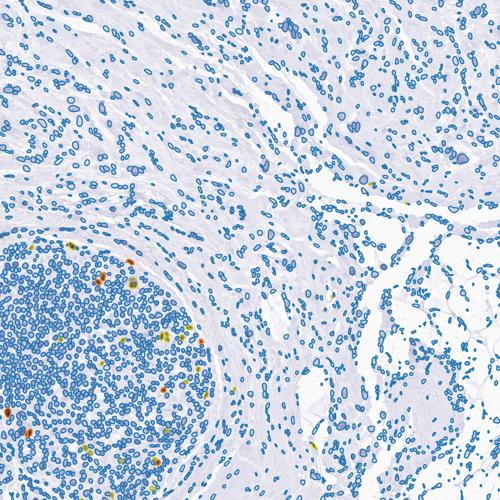}
  \end{subfigure}
\hfill
 \begin{subfigure}[b]{0.18000000000000002\textwidth}
  \centering
  \includegraphics[width=\textwidth]{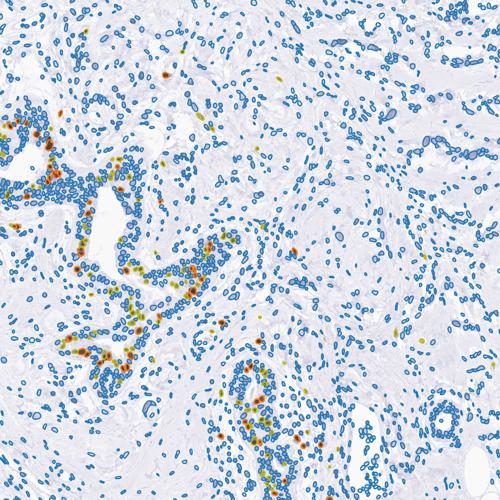}
  \end{subfigure}
\hfill
 \begin{subfigure}[b]{0.18000000000000002\textwidth}
  \centering
  \includegraphics[width=\textwidth]{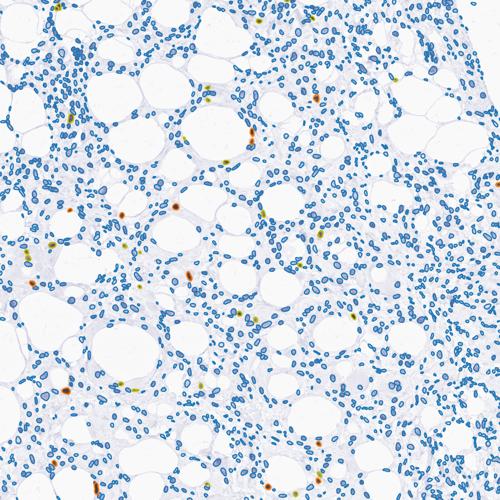}
  \end{subfigure}
\hfill
 \begin{subfigure}[b]{0.18000000000000002\textwidth}
  \centering
  \includegraphics[width=\textwidth]{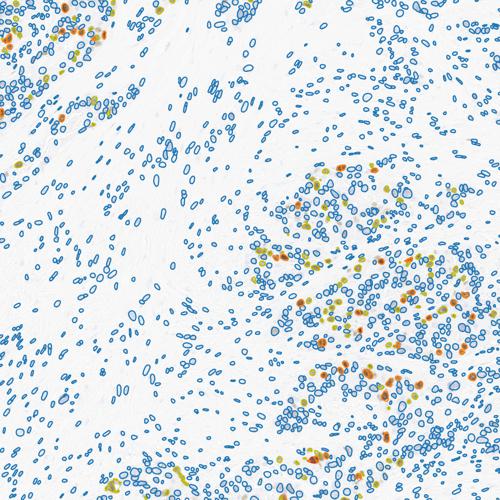}
  \end{subfigure}\\
  \caption{
  Examples of H\&E-Ki67 pairs in the dataset. Each column corresponds to a H\&E-IHC pair. Row1: H\&E, Row2: IHC, Row3:nuclei sementations on H\&E, Row4: the result of DAB-analysis on IHC where blue, yellow, orange, and red colors mark 0, 1+, 2+, and 3+ nuclei, respectively. 
  }
  \label{fig:dsintro_Ki67}
\end{figure}

\begin{figure}
\centering
 \begin{subfigure}[b]{0.18000000000000002\textwidth}
  \centering
  \includegraphics[width=\textwidth]{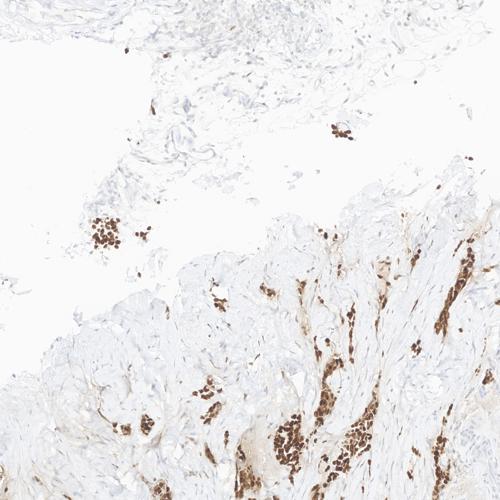}
  \end{subfigure}
\hfill
 \begin{subfigure}[b]{0.18000000000000002\textwidth}
  \centering
  \includegraphics[width=\textwidth]{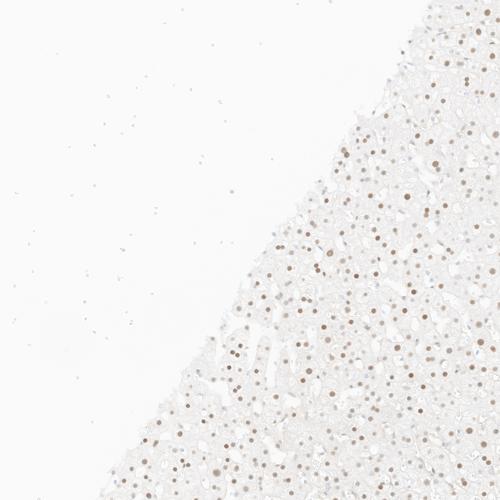}
  \end{subfigure}
\hfill
 \begin{subfigure}[b]{0.18000000000000002\textwidth}
  \centering
  \includegraphics[width=\textwidth]{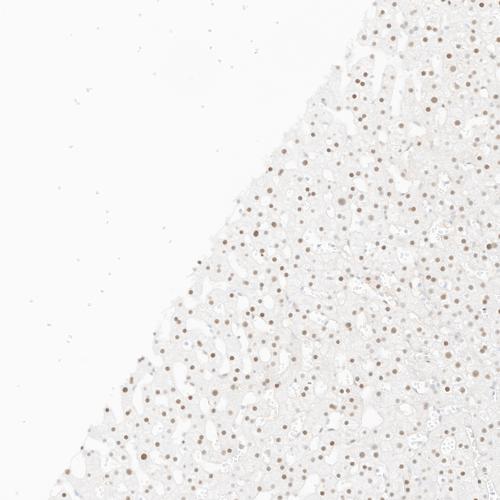}
  \end{subfigure}
\hfill
 \begin{subfigure}[b]{0.18000000000000002\textwidth}
  \centering
  \includegraphics[width=\textwidth]{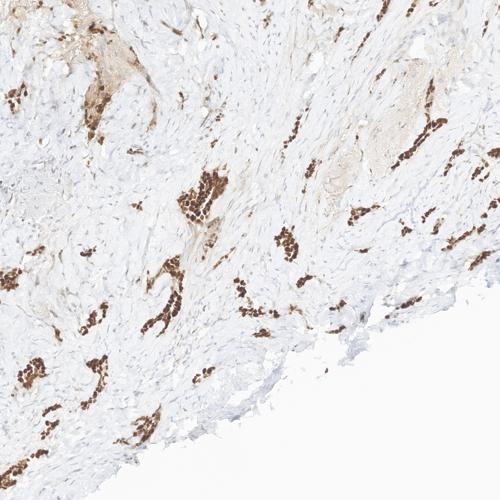}
  \end{subfigure}
\hfill
 \begin{subfigure}[b]{0.18000000000000002\textwidth}
  \centering
  \includegraphics[width=\textwidth]{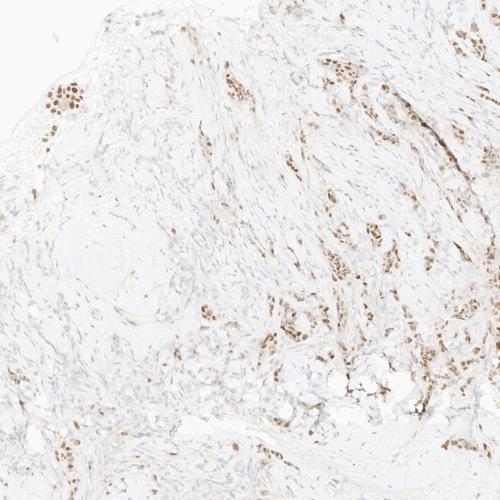}
  \end{subfigure}
\\
 \begin{subfigure}[b]{0.18000000000000002\textwidth}
  \centering
  \includegraphics[width=\textwidth]{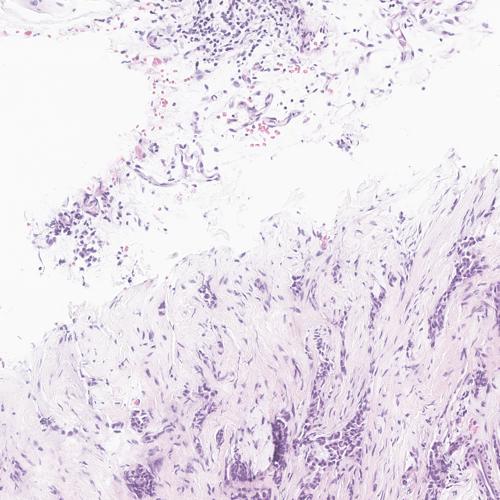}
  \end{subfigure}
\hfill
 \begin{subfigure}[b]{0.18000000000000002\textwidth}
  \centering
  \includegraphics[width=\textwidth]{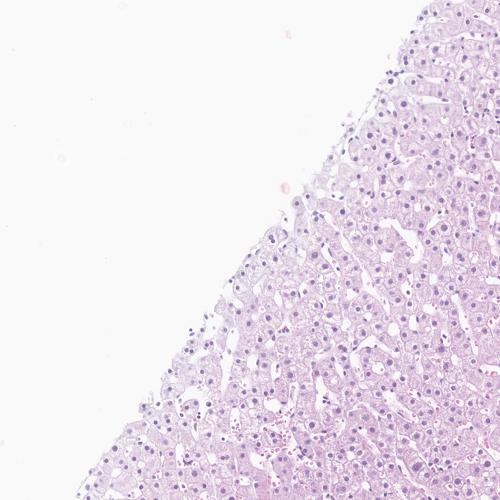}
  \end{subfigure}
\hfill
 \begin{subfigure}[b]{0.18000000000000002\textwidth}
  \centering
  \includegraphics[width=\textwidth]{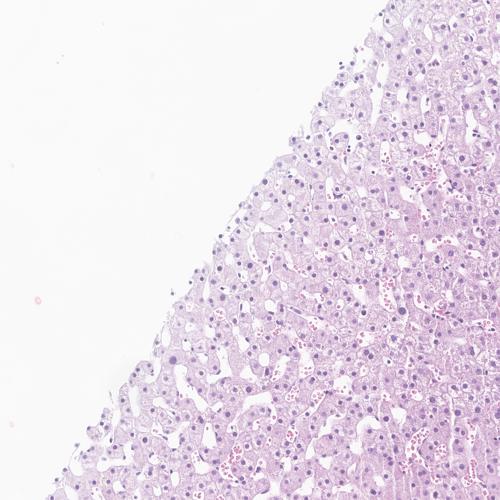}
  \end{subfigure}
\hfill
 \begin{subfigure}[b]{0.18000000000000002\textwidth}
  \centering
  \includegraphics[width=\textwidth]{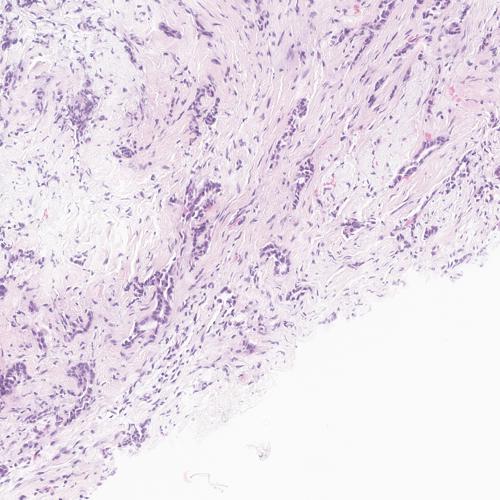}
  \end{subfigure}
\hfill
 \begin{subfigure}[b]{0.18000000000000002\textwidth}
  \centering
  \includegraphics[width=\textwidth]{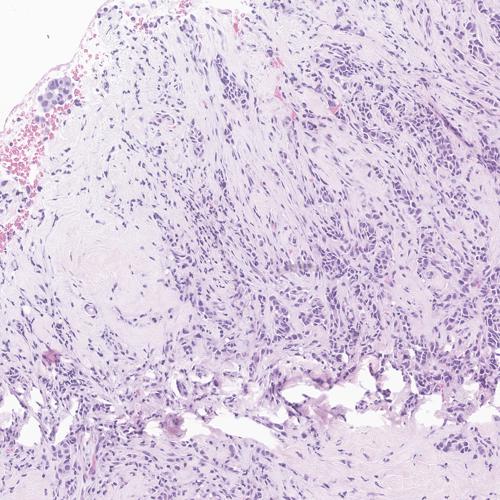}
  \end{subfigure}
\\
 \begin{subfigure}[b]{0.18000000000000002\textwidth}
  \centering
  \includegraphics[width=\textwidth]{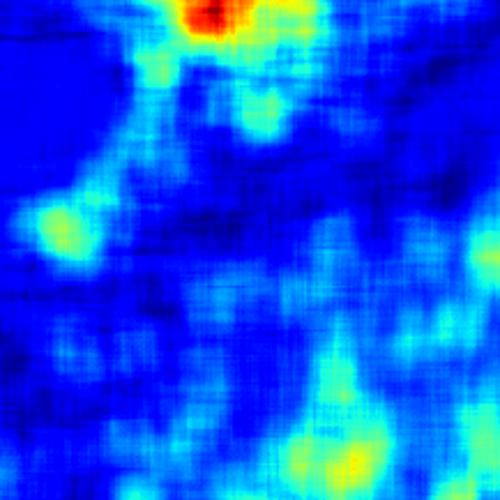}
  \end{subfigure}
\hfill
 \begin{subfigure}[b]{0.18000000000000002\textwidth}
  \centering
  \includegraphics[width=\textwidth]{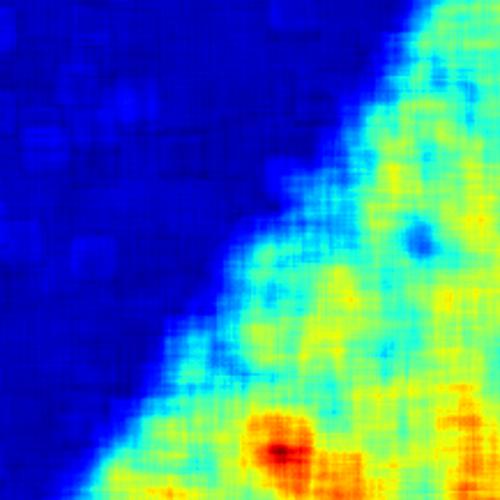}
  \end{subfigure}
\hfill
 \begin{subfigure}[b]{0.18000000000000002\textwidth}
  \centering
  \includegraphics[width=\textwidth]{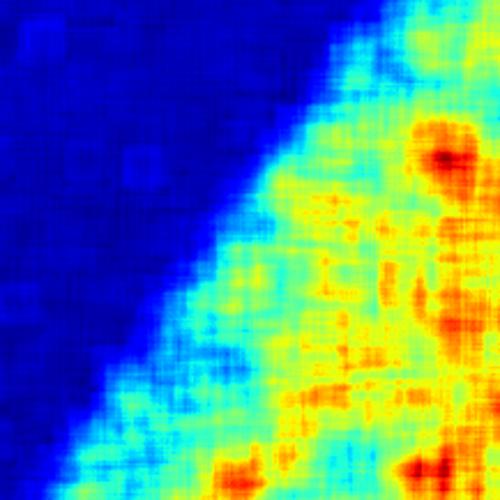}
  \end{subfigure}
\hfill
 \begin{subfigure}[b]{0.18000000000000002\textwidth}
  \centering
  \includegraphics[width=\textwidth]{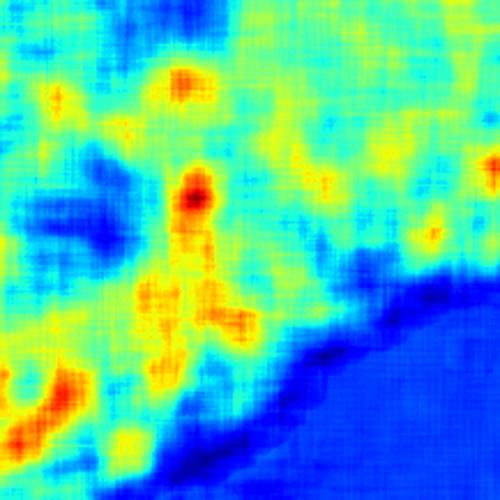}
  \end{subfigure}
\hfill
 \begin{subfigure}[b]{0.18000000000000002\textwidth}
  \centering
  \includegraphics[width=\textwidth]{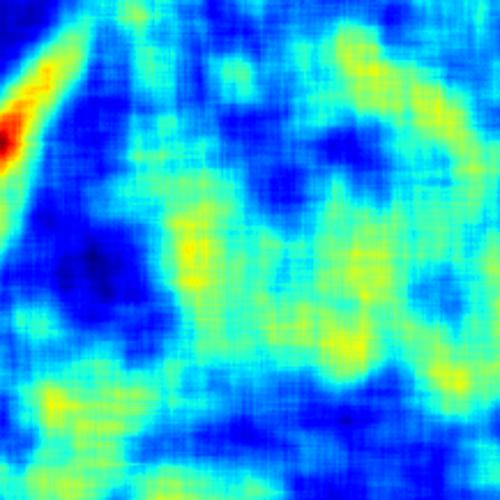}
  \end{subfigure}
\\
 \begin{subfigure}[b]{0.18000000000000002\textwidth}
  \centering
  \includegraphics[width=\textwidth]{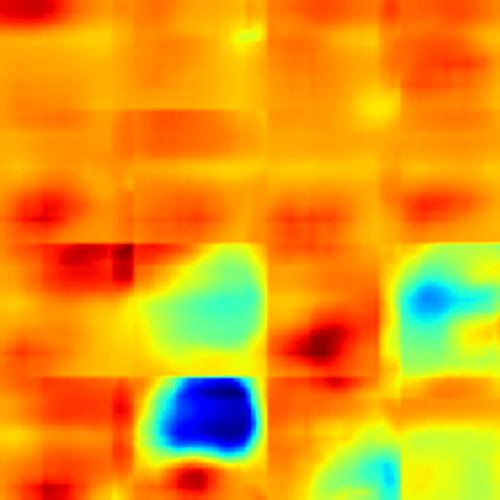}
  \end{subfigure}
\hfill
 \begin{subfigure}[b]{0.18000000000000002\textwidth}
  \centering
  \includegraphics[width=\textwidth]{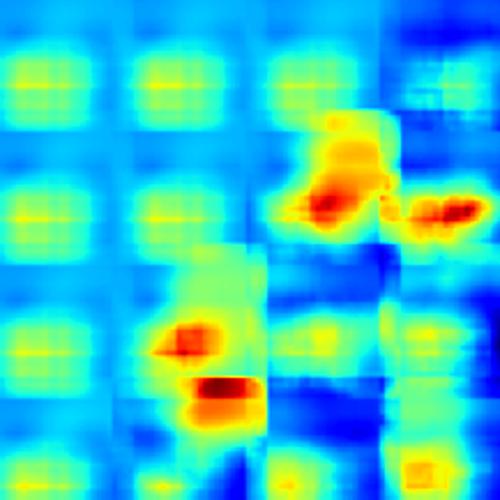}
  \end{subfigure}
\hfill
 \begin{subfigure}[b]{0.18000000000000002\textwidth}
  \centering
  \includegraphics[width=\textwidth]{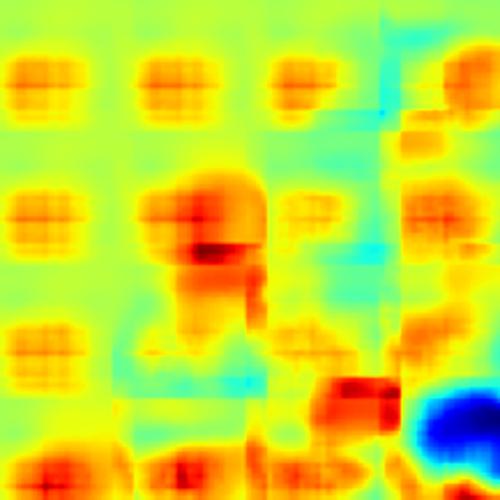}
  \end{subfigure}
\hfill
 \begin{subfigure}[b]{0.18000000000000002\textwidth}
  \centering
  \includegraphics[width=\textwidth]{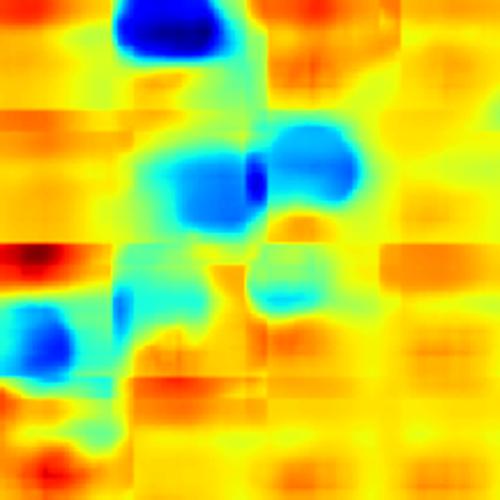}
  \end{subfigure}
\hfill
 \begin{subfigure}[b]{0.18000000000000002\textwidth}
  \centering
  \includegraphics[width=\textwidth]{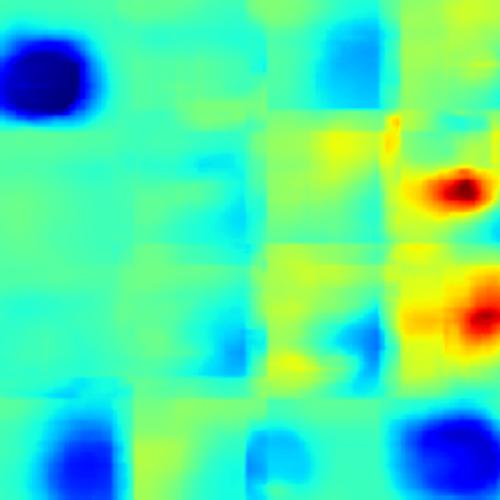}
  \end{subfigure}
\\
 \begin{subfigure}[b]{0.18000000000000002\textwidth}
  \centering
  \includegraphics[width=\textwidth]{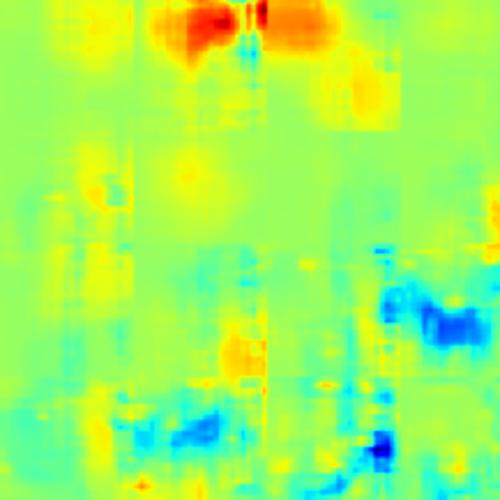}
  \end{subfigure}
\hfill
 \begin{subfigure}[b]{0.18000000000000002\textwidth}
  \centering
  \includegraphics[width=\textwidth]{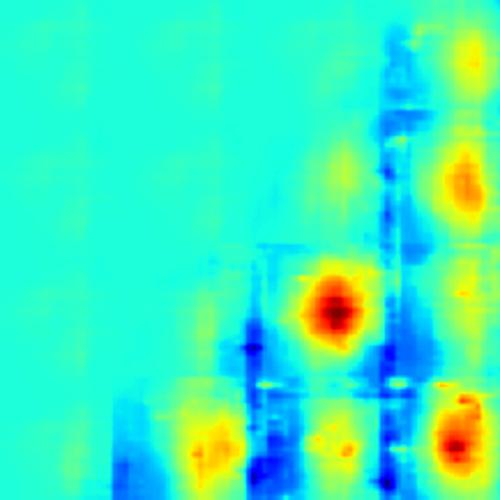}
  \end{subfigure}
\hfill
 \begin{subfigure}[b]{0.18000000000000002\textwidth}
  \centering
  \includegraphics[width=\textwidth]{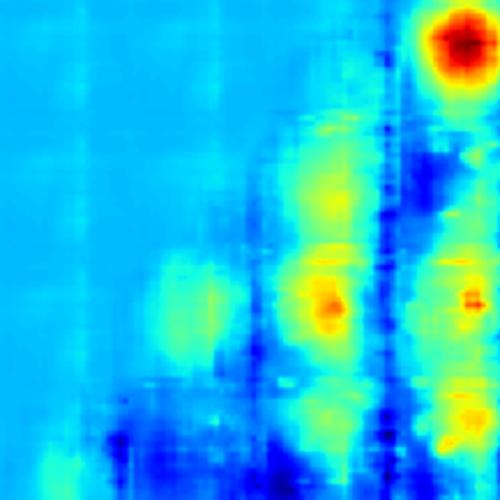}
  \end{subfigure}
\hfill
 \begin{subfigure}[b]{0.18000000000000002\textwidth}
  \centering
  \includegraphics[width=\textwidth]{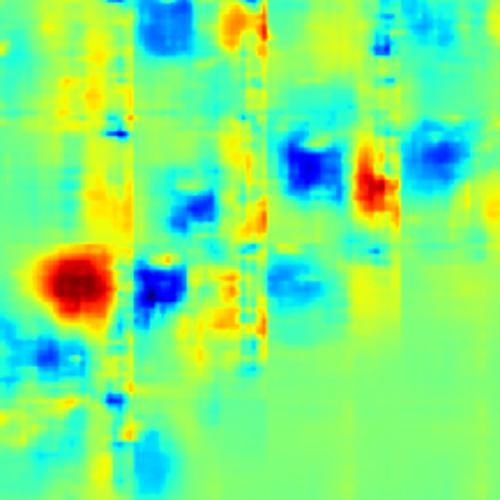}
  \end{subfigure}
\hfill
 \begin{subfigure}[b]{0.18000000000000002\textwidth}
  \centering
  \includegraphics[width=\textwidth]{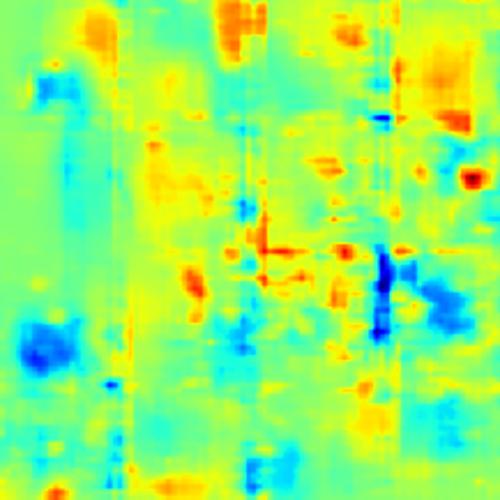}
  \end{subfigure}
\\
 \begin{subfigure}[b]{0.18000000000000002\textwidth}
  \centering
  \includegraphics[width=\textwidth]{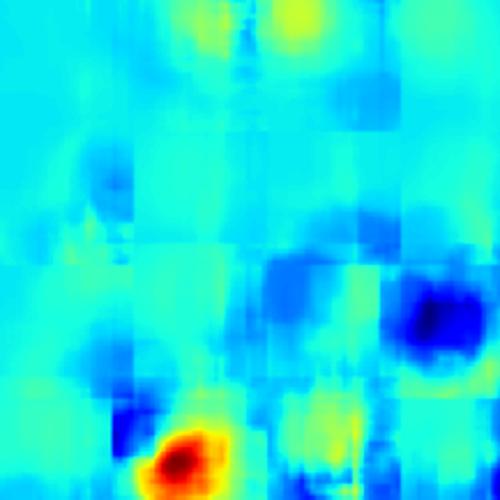}
  \end{subfigure}
\hfill
 \begin{subfigure}[b]{0.18000000000000002\textwidth}
  \centering
  \includegraphics[width=\textwidth]{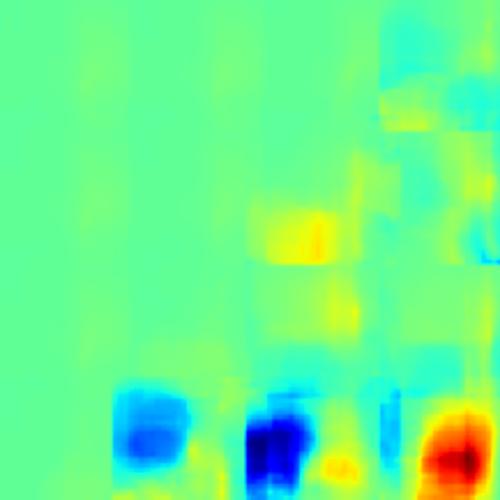}
  \end{subfigure}
\hfill
 \begin{subfigure}[b]{0.18000000000000002\textwidth}
  \centering
  \includegraphics[width=\textwidth]{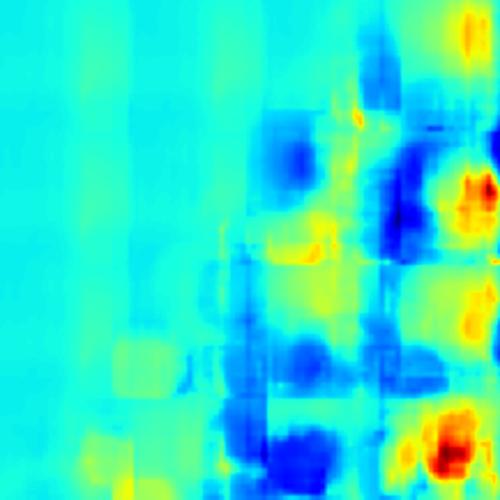}
  \end{subfigure}
\hfill
 \begin{subfigure}[b]{0.18000000000000002\textwidth}
  \centering
  \includegraphics[width=\textwidth]{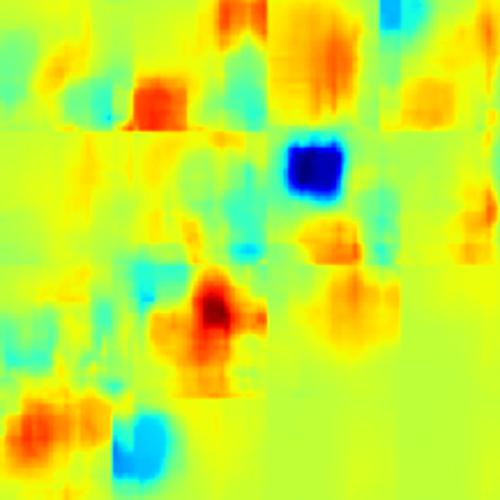}
  \end{subfigure}
\hfill
 \begin{subfigure}[b]{0.18000000000000002\textwidth}
  \centering
  \includegraphics[width=\textwidth]{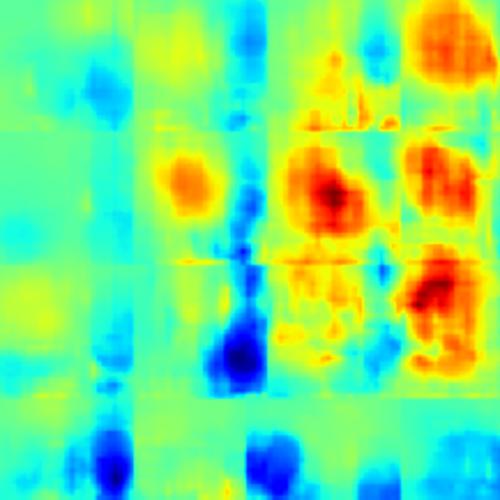}
  \end{subfigure}
  \\
  \label{fig:localizations_er}
  \caption{
   Localizations for ER predictors trained/evaluated on the split shown by the purple circle in Fig. \akfigreferaucs. Each column corresponds to a H\&E-IHC pair. Row 1: IHC, row 2: H\&E, row 3: CLAM \akciteclam's attention mask, row 4: the sensitivity of the classifier labeled as "ViT, high vs. low". rows 5: the average sensitivity of heads of the classifier labeled as "ViT, with G.Z.". 
   rows 6: the average sensitivity of heads of the classifier labeled as "ViT, without G.Z.".
   In all heatmaps we used JET color-map in which high and low values appear in red and blue, respectively. 
  }
\end{figure}
\begin{figure}
\centering
 \begin{subfigure}[b]{0.18000000000000002\textwidth}
  \centering
  \includegraphics[width=\textwidth]{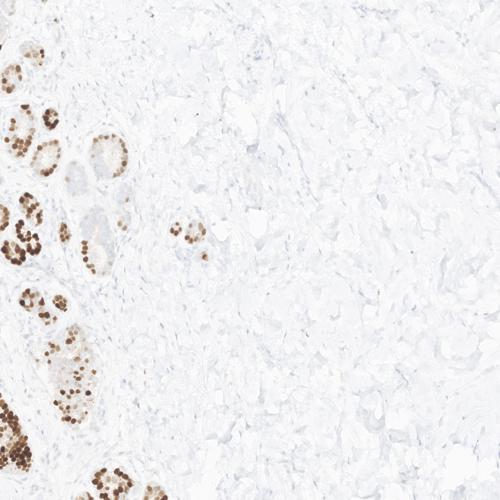}
  \end{subfigure}
\hfill
 \begin{subfigure}[b]{0.18000000000000002\textwidth}
  \centering
  \includegraphics[width=\textwidth]{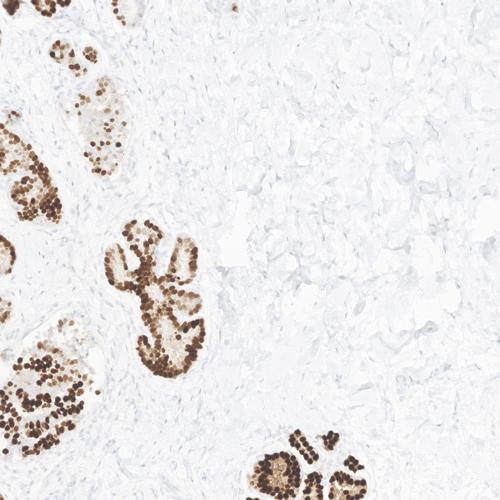}
  \end{subfigure}
\hfill
 \begin{subfigure}[b]{0.18000000000000002\textwidth}
  \centering
  \includegraphics[width=\textwidth]{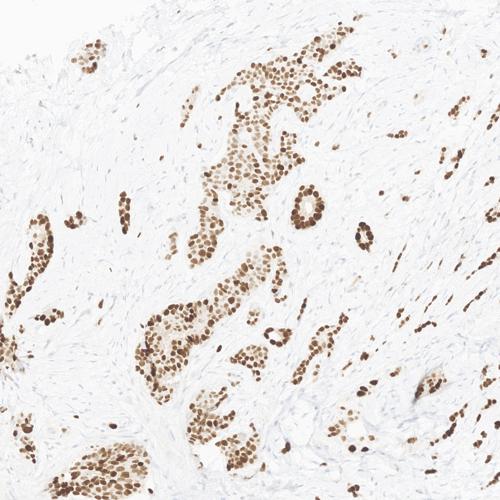}
  \end{subfigure}
\hfill
 \begin{subfigure}[b]{0.18000000000000002\textwidth}
  \centering
  \includegraphics[width=\textwidth]{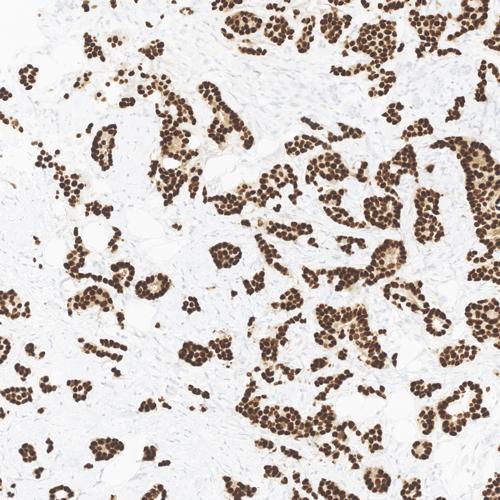}
  \end{subfigure}
\hfill
 \begin{subfigure}[b]{0.18000000000000002\textwidth}
  \centering
  \includegraphics[width=\textwidth]{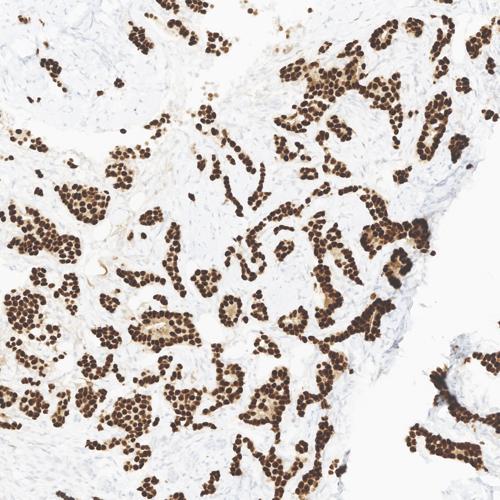}
  \end{subfigure}
\\
 \begin{subfigure}[b]{0.18000000000000002\textwidth}
  \centering
  \includegraphics[width=\textwidth]{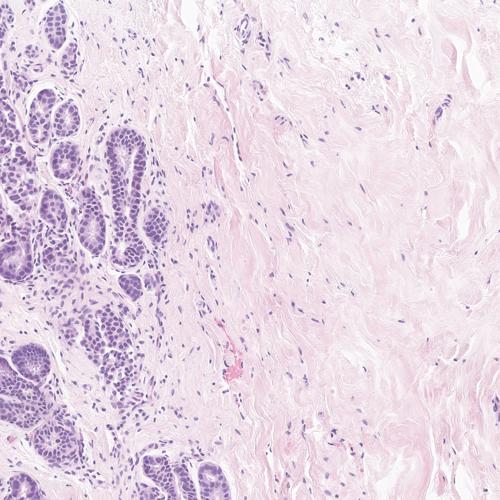}
  \end{subfigure}
\hfill
 \begin{subfigure}[b]{0.18000000000000002\textwidth}
  \centering
  \includegraphics[width=\textwidth]{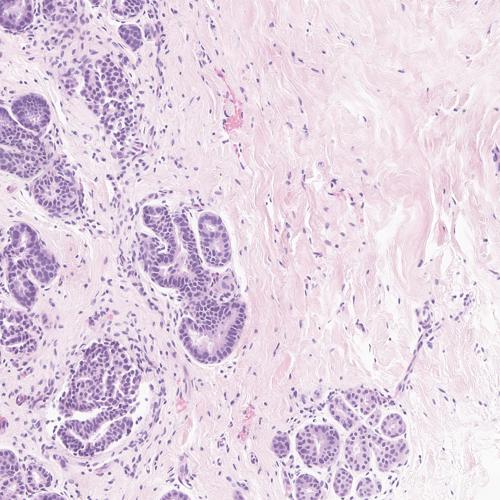}
  \end{subfigure}
\hfill
 \begin{subfigure}[b]{0.18000000000000002\textwidth}
  \centering
  \includegraphics[width=\textwidth]{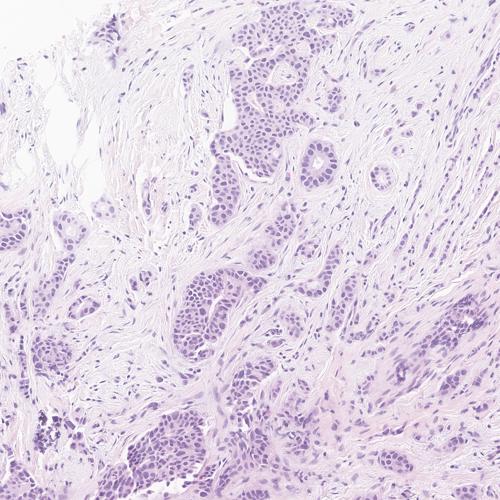}
  \end{subfigure}
\hfill
 \begin{subfigure}[b]{0.18000000000000002\textwidth}
  \centering
  \includegraphics[width=\textwidth]{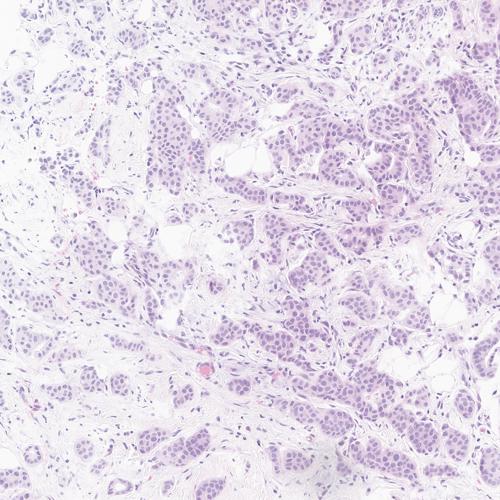}
  \end{subfigure}
\hfill
 \begin{subfigure}[b]{0.18000000000000002\textwidth}
  \centering
  \includegraphics[width=\textwidth]{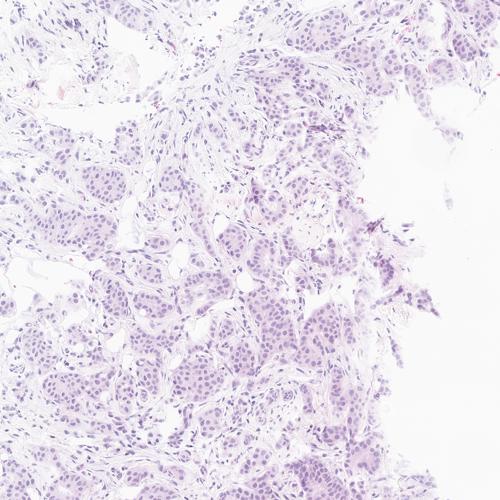}
  \end{subfigure}
\\
 \begin{subfigure}[b]{0.18000000000000002\textwidth}
  \centering
  \includegraphics[width=\textwidth]{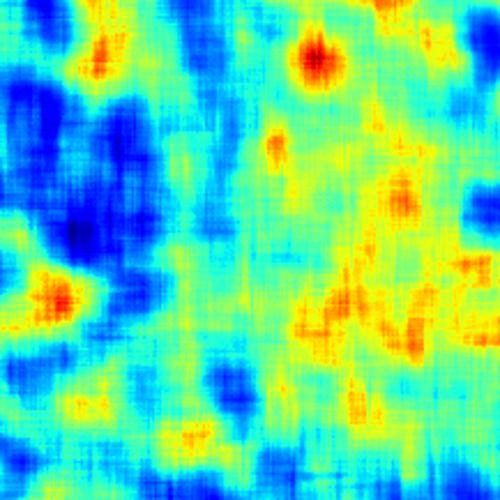}
  \end{subfigure}
\hfill
 \begin{subfigure}[b]{0.18000000000000002\textwidth}
  \centering
  \includegraphics[width=\textwidth]{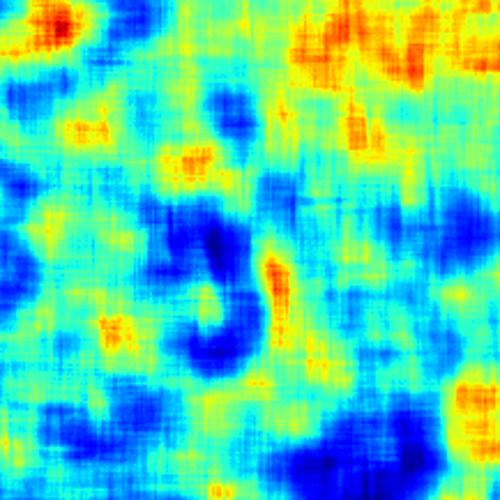}
  \end{subfigure}
\hfill
 \begin{subfigure}[b]{0.18000000000000002\textwidth}
  \centering
  \includegraphics[width=\textwidth]{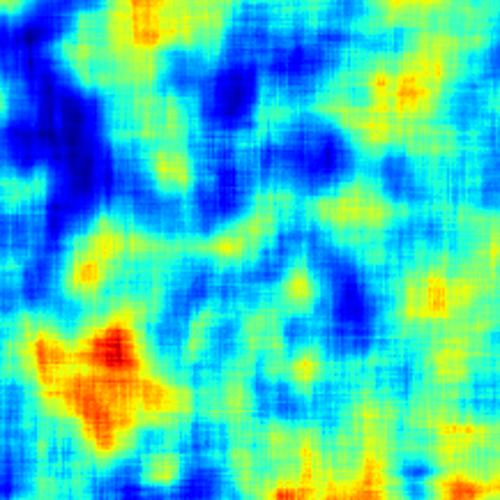}
  \end{subfigure}
\hfill
 \begin{subfigure}[b]{0.18000000000000002\textwidth}
  \centering
  \includegraphics[width=\textwidth]{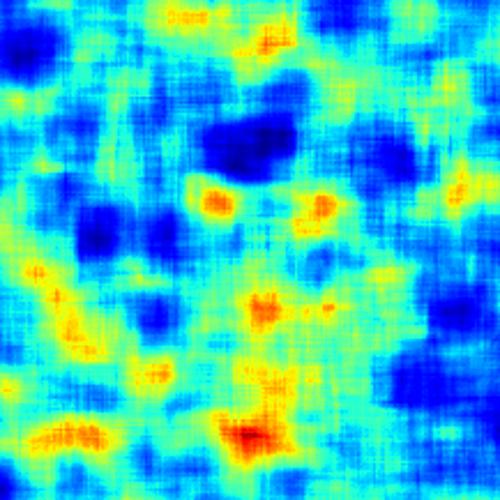}
  \end{subfigure}
\hfill
 \begin{subfigure}[b]{0.18000000000000002\textwidth}
  \centering
  \includegraphics[width=\textwidth]{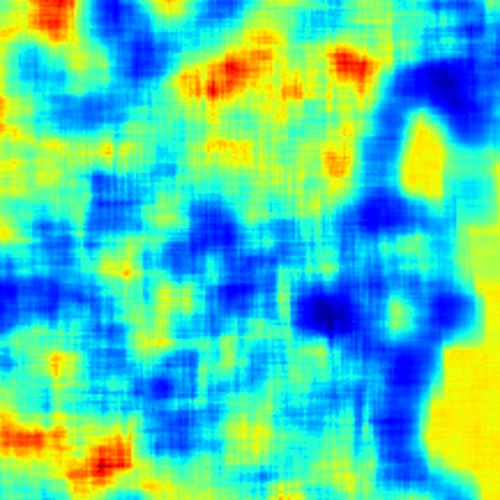}
  \end{subfigure}
\\
 \begin{subfigure}[b]{0.18000000000000002\textwidth}
  \centering
  \includegraphics[width=\textwidth]{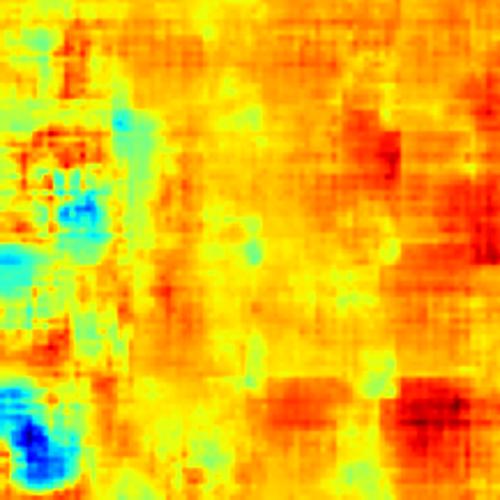}
  \end{subfigure}
\hfill
 \begin{subfigure}[b]{0.18000000000000002\textwidth}
  \centering
  \includegraphics[width=\textwidth]{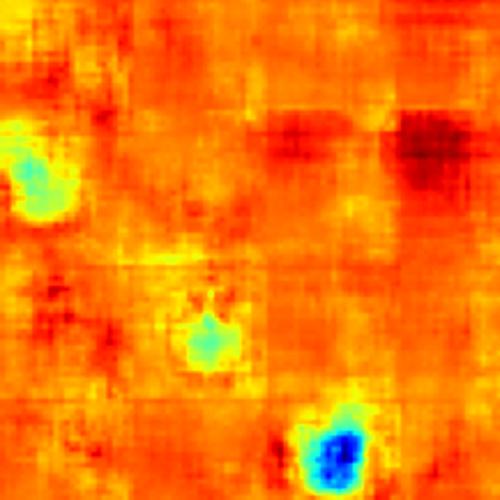}
  \end{subfigure}
\hfill
 \begin{subfigure}[b]{0.18000000000000002\textwidth}
  \centering
  \includegraphics[width=\textwidth]{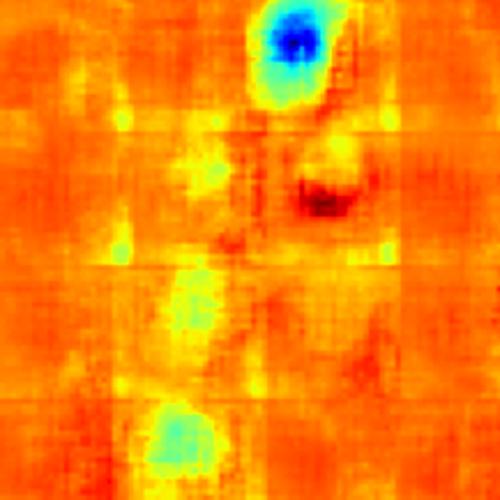}
  \end{subfigure}
\hfill
 \begin{subfigure}[b]{0.18000000000000002\textwidth}
  \centering
  \includegraphics[width=\textwidth]{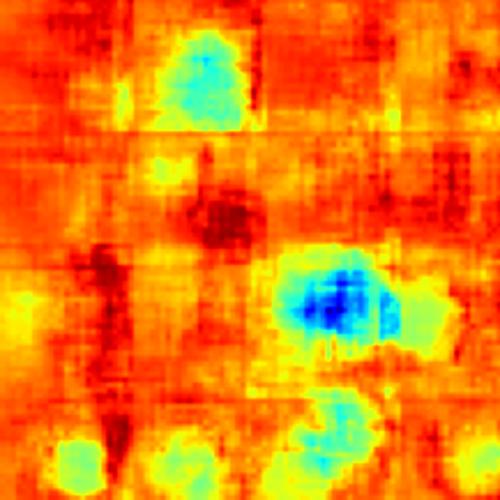}
  \end{subfigure}
\hfill
 \begin{subfigure}[b]{0.18000000000000002\textwidth}
  \centering
  \includegraphics[width=\textwidth]{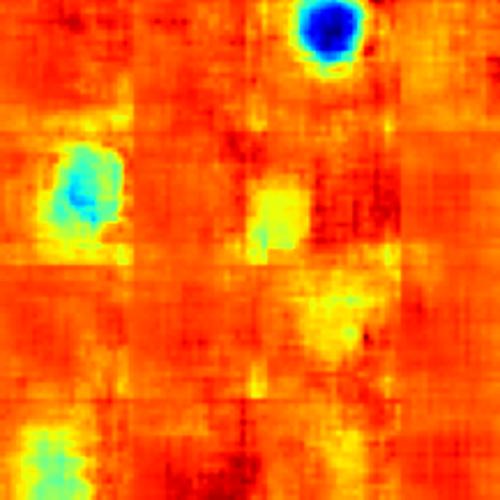}
  \end{subfigure}
\\
 \begin{subfigure}[b]{0.18000000000000002\textwidth}
  \centering
  \includegraphics[width=\textwidth]{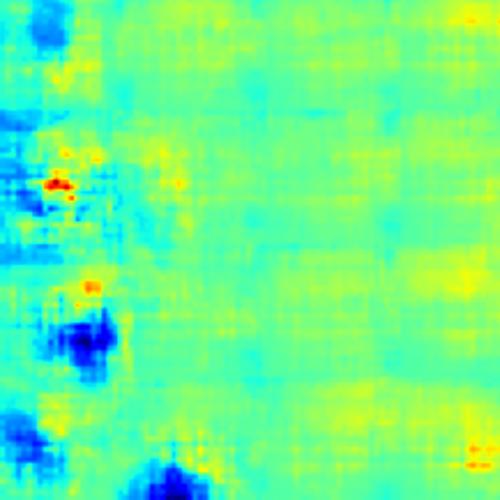}
  \end{subfigure}
\hfill
 \begin{subfigure}[b]{0.18000000000000002\textwidth}
  \centering
  \includegraphics[width=\textwidth]{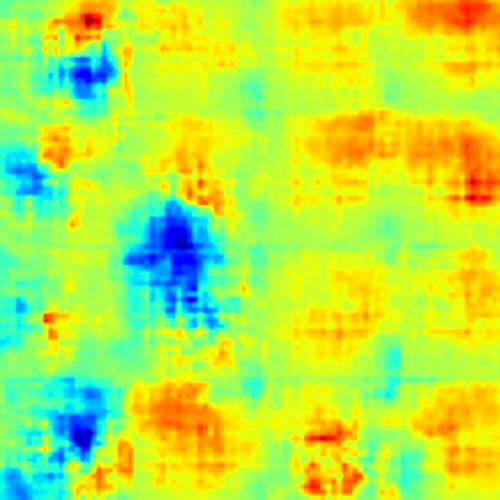}
  \end{subfigure}
\hfill
 \begin{subfigure}[b]{0.18000000000000002\textwidth}
  \centering
  \includegraphics[width=\textwidth]{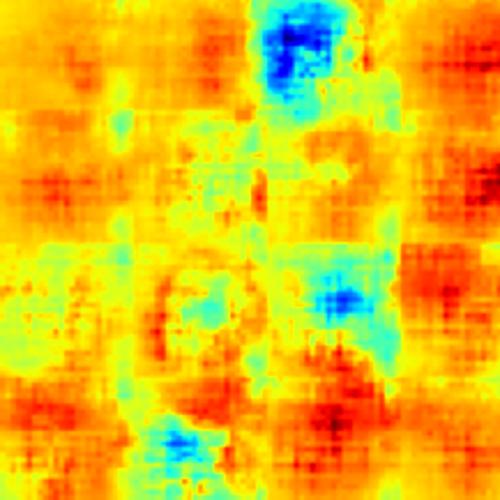}
  \end{subfigure}
\hfill
 \begin{subfigure}[b]{0.18000000000000002\textwidth}
  \centering
  \includegraphics[width=\textwidth]{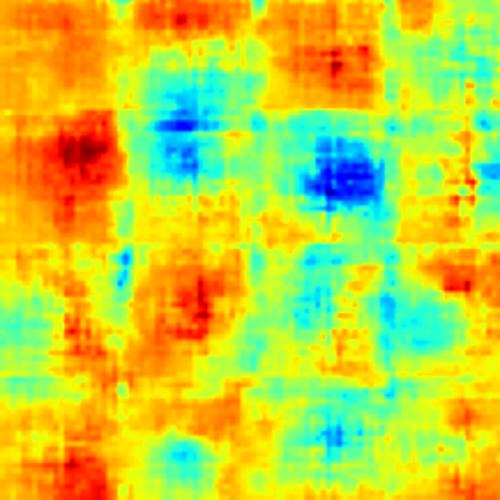}
  \end{subfigure}
\hfill
 \begin{subfigure}[b]{0.18000000000000002\textwidth}
  \centering
  \includegraphics[width=\textwidth]{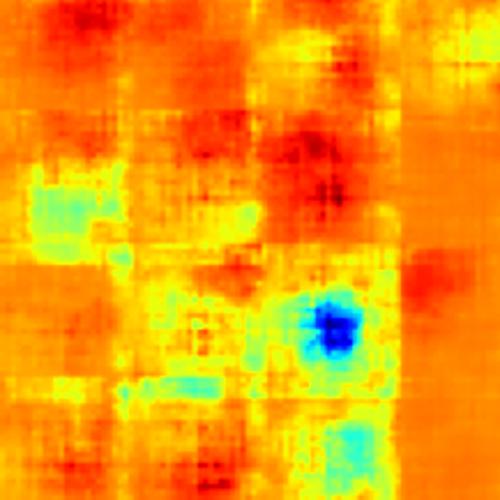}
  \end{subfigure}
\\
 \begin{subfigure}[b]{0.18000000000000002\textwidth}
  \centering
  \includegraphics[width=\textwidth]{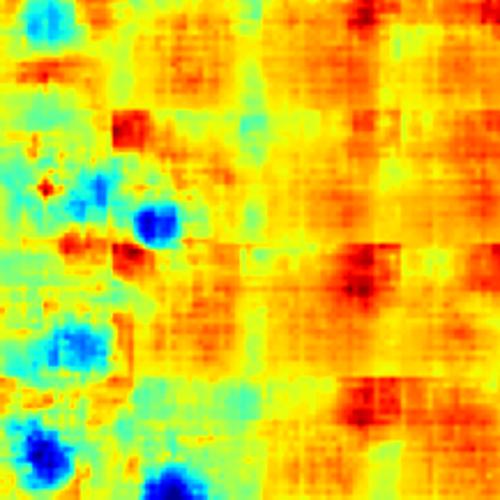}
  \end{subfigure}
\hfill
 \begin{subfigure}[b]{0.18000000000000002\textwidth}
  \centering
  \includegraphics[width=\textwidth]{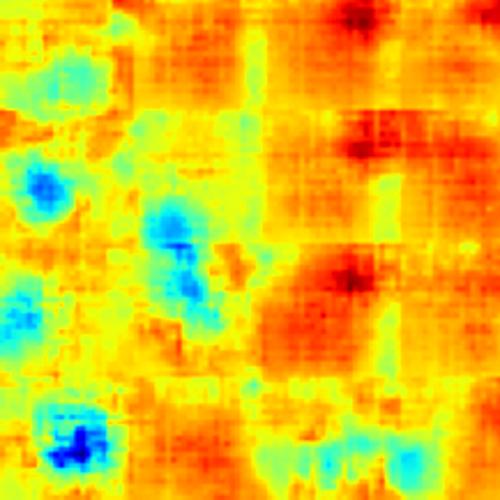}
  \end{subfigure}
\hfill
 \begin{subfigure}[b]{0.18000000000000002\textwidth}
  \centering
  \includegraphics[width=\textwidth]{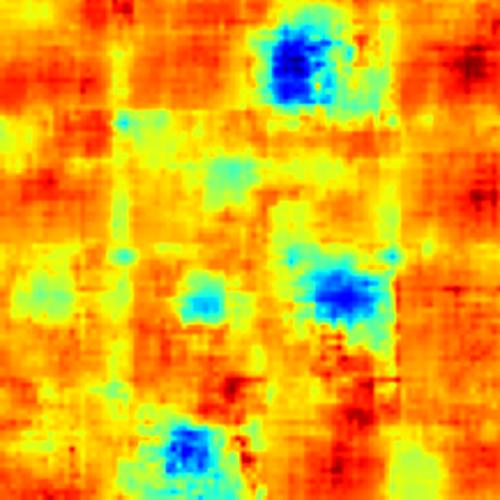}
  \end{subfigure}
\hfill
 \begin{subfigure}[b]{0.18000000000000002\textwidth}
  \centering
  \includegraphics[width=\textwidth]{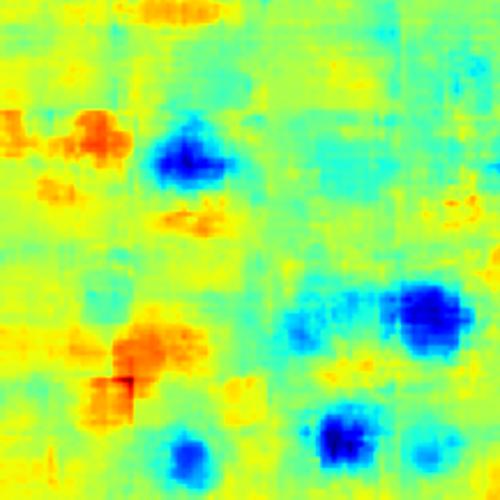}
  \end{subfigure}
\hfill
 \begin{subfigure}[b]{0.18000000000000002\textwidth}
  \centering
  \includegraphics[width=\textwidth]{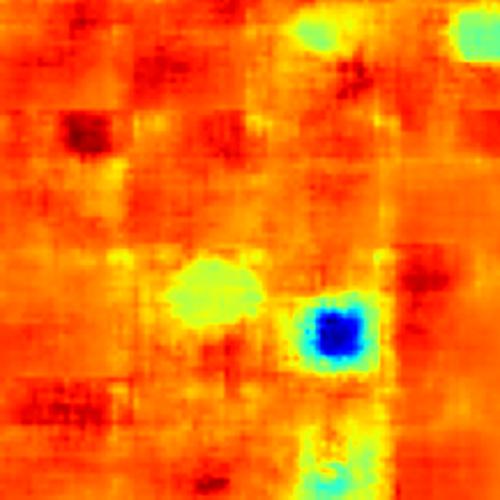}
  \end{subfigure}
  \\
  \label{fig:localizations_pr}
  \caption{
    Localizations for PR predictors trained/evaluated on the split shown by the red circle in Fig. \akfigrefpraucs. Each column corresponds to a H\&E-IHC pair. Row 1: IHC, row 2: H\&E, row 3: CLAM \akciteclam's attention mask, row 4: the sensitivity of the classifier labeled as "ViT, high vs. low". rows 5: the average sensitivity of heads of the classifier labeled as "ViT, with G.Z.". 
   rows 6: the average sensitivity of heads of the classifier labeled as "ViT, without G.Z.".
   In all heatmaps we used JET color-map in which high and low values appear in red and blue, respectively.
  }
\end{figure}

\begin{figure}
\centering
 \begin{subfigure}[b]{0.18000000000000002\textwidth}
  \centering
  \includegraphics[width=\textwidth]{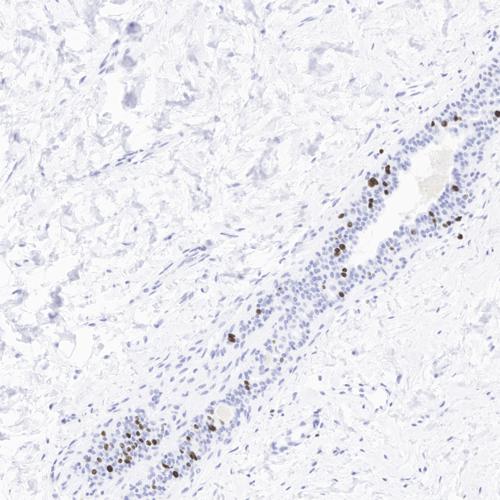}
  \end{subfigure}
\hfill
 \begin{subfigure}[b]{0.18000000000000002\textwidth}
  \centering
  \includegraphics[width=\textwidth]{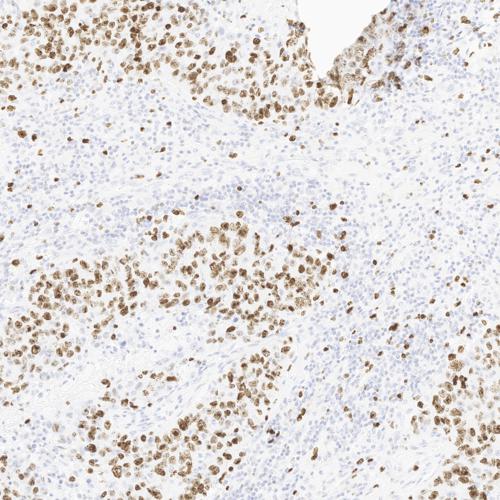}
  \end{subfigure}
\hfill
 \begin{subfigure}[b]{0.18000000000000002\textwidth}
  \centering
  \includegraphics[width=\textwidth]{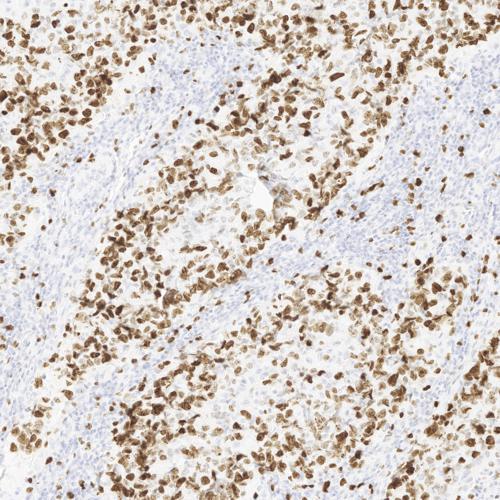}
  \end{subfigure}
\hfill
 \begin{subfigure}[b]{0.18000000000000002\textwidth}
  \centering
  \includegraphics[width=\textwidth]{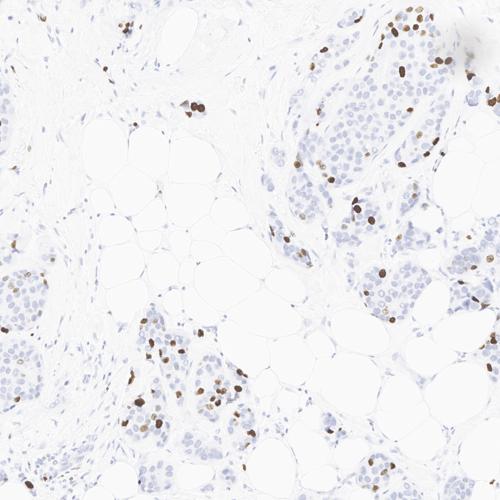}
  \end{subfigure}
\hfill
 \begin{subfigure}[b]{0.18000000000000002\textwidth}
  \centering
  \includegraphics[width=\textwidth]{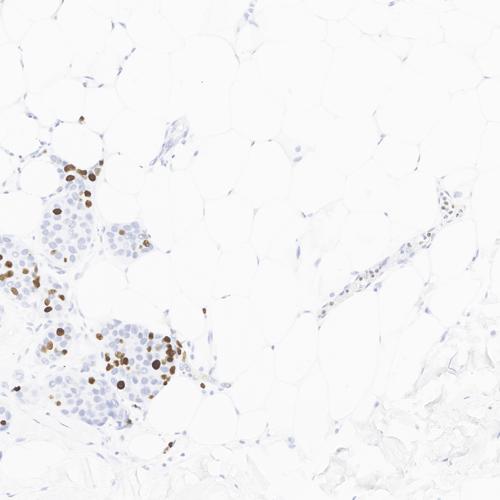}
  \end{subfigure}
\\
 \begin{subfigure}[b]{0.18000000000000002\textwidth}
  \centering
  \includegraphics[width=\textwidth]{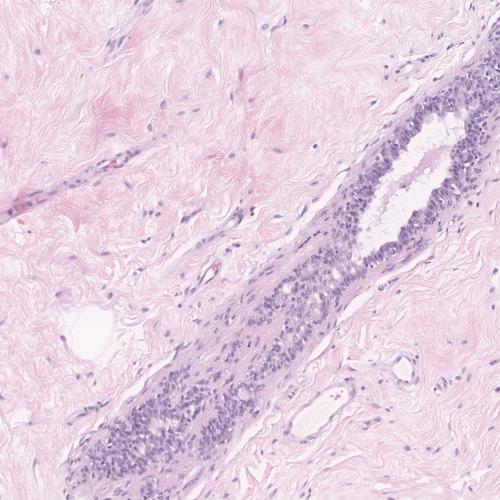}
  \end{subfigure}
\hfill
 \begin{subfigure}[b]{0.18000000000000002\textwidth}
  \centering
  \includegraphics[width=\textwidth]{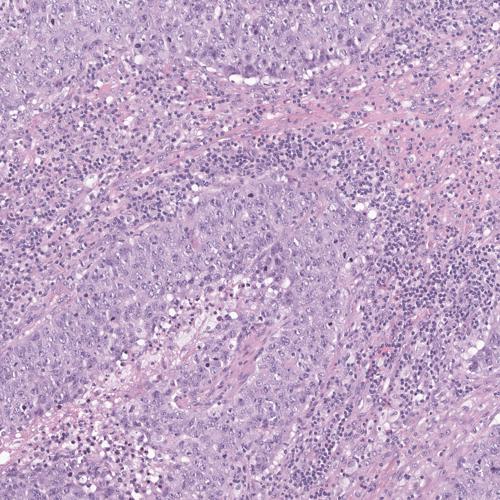}
  \end{subfigure}
\hfill
 \begin{subfigure}[b]{0.18000000000000002\textwidth}
  \centering
  \includegraphics[width=\textwidth]{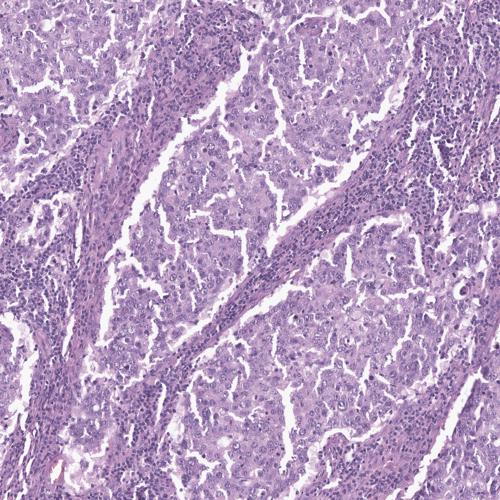}
  \end{subfigure}
\hfill
 \begin{subfigure}[b]{0.18000000000000002\textwidth}
  \centering
  \includegraphics[width=\textwidth]{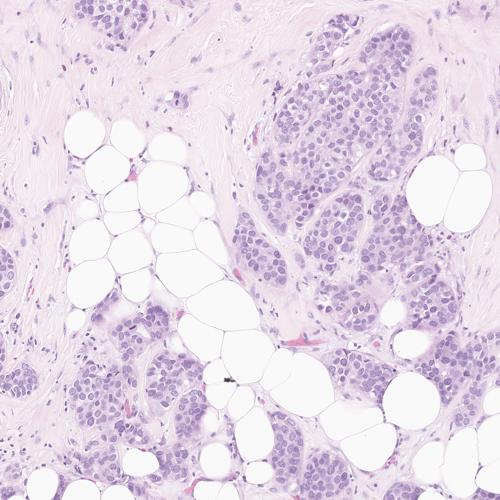}
  \end{subfigure}
\hfill
 \begin{subfigure}[b]{0.18000000000000002\textwidth}
  \centering
  \includegraphics[width=\textwidth]{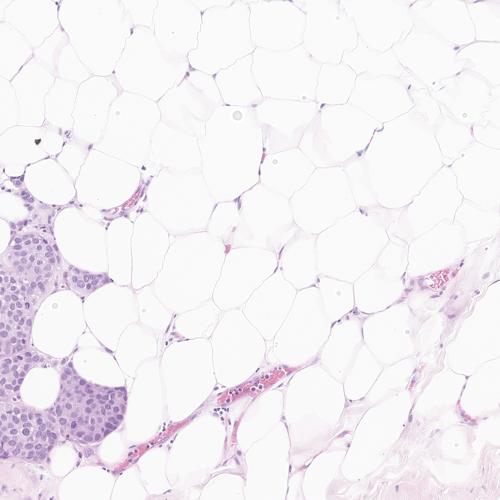}
  \end{subfigure}
\\
 \begin{subfigure}[b]{0.18000000000000002\textwidth}
  \centering
  \includegraphics[width=\textwidth]{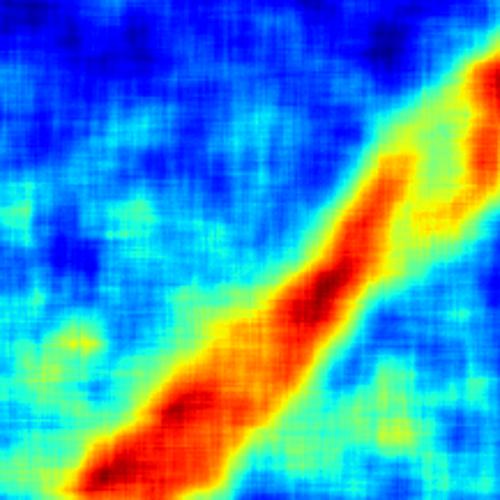}
  \end{subfigure}
\hfill
 \begin{subfigure}[b]{0.18000000000000002\textwidth}
  \centering
  \includegraphics[width=\textwidth]{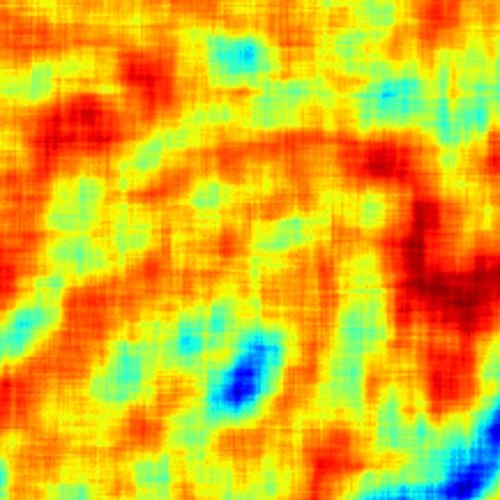}
  \end{subfigure}
\hfill
 \begin{subfigure}[b]{0.18000000000000002\textwidth}
  \centering
  \includegraphics[width=\textwidth]{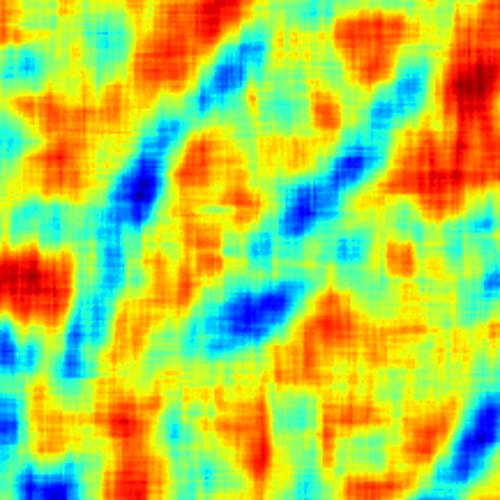}
  \end{subfigure}
\hfill
 \begin{subfigure}[b]{0.18000000000000002\textwidth}
  \centering
  \includegraphics[width=\textwidth]{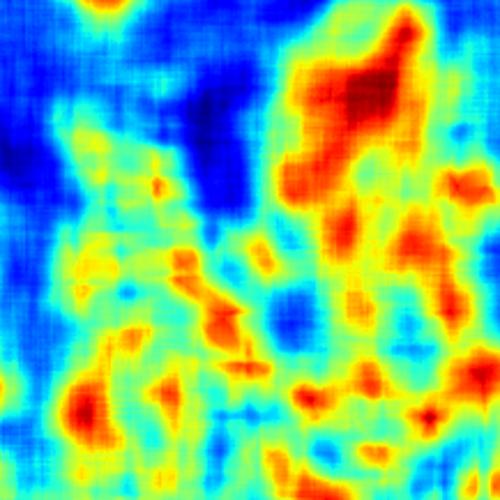}
  \end{subfigure}
\hfill
 \begin{subfigure}[b]{0.18000000000000002\textwidth}
  \centering
  \includegraphics[width=\textwidth]{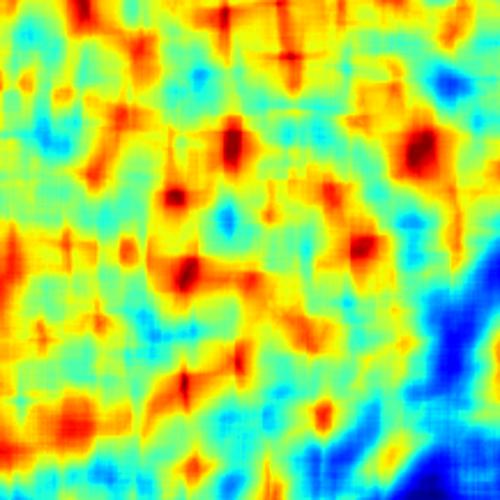}
  \end{subfigure}
\\
 \begin{subfigure}[b]{0.18000000000000002\textwidth}
  \centering
  \includegraphics[width=\textwidth]{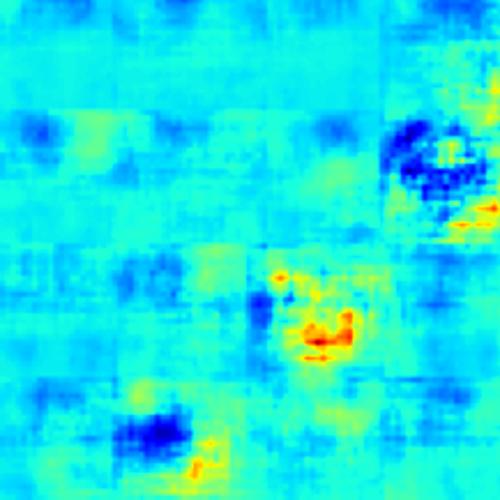}
  \end{subfigure}
\hfill
 \begin{subfigure}[b]{0.18000000000000002\textwidth}
  \centering
  \includegraphics[width=\textwidth]{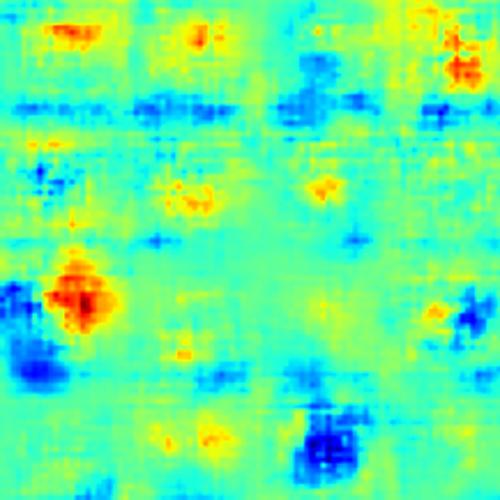}
  \end{subfigure}
\hfill
 \begin{subfigure}[b]{0.18000000000000002\textwidth}
  \centering
  \includegraphics[width=\textwidth]{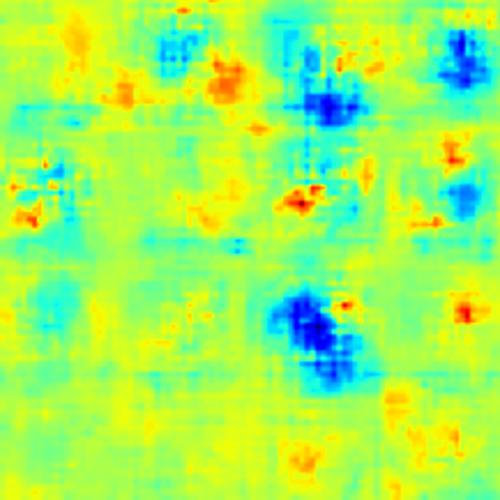}
  \end{subfigure}
\hfill
 \begin{subfigure}[b]{0.18000000000000002\textwidth}
  \centering
  \includegraphics[width=\textwidth]{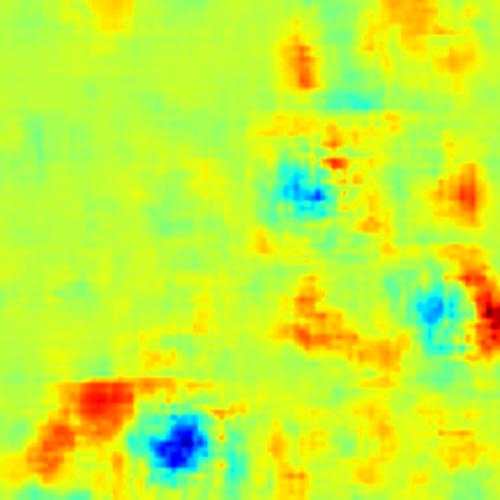}
  \end{subfigure}
\hfill
 \begin{subfigure}[b]{0.18000000000000002\textwidth}
  \centering
  \includegraphics[width=\textwidth]{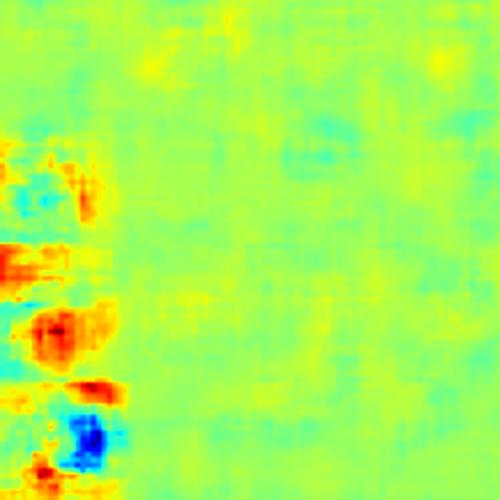}
  \end{subfigure}
\\
 \begin{subfigure}[b]{0.18000000000000002\textwidth}
  \centering
  \includegraphics[width=\textwidth]{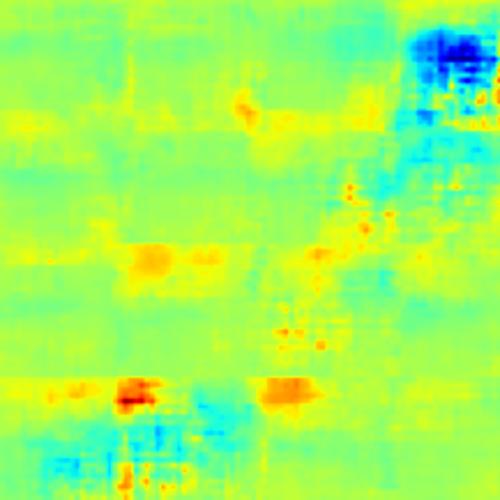}
  \end{subfigure}
\hfill
 \begin{subfigure}[b]{0.18000000000000002\textwidth}
  \centering
  \includegraphics[width=\textwidth]{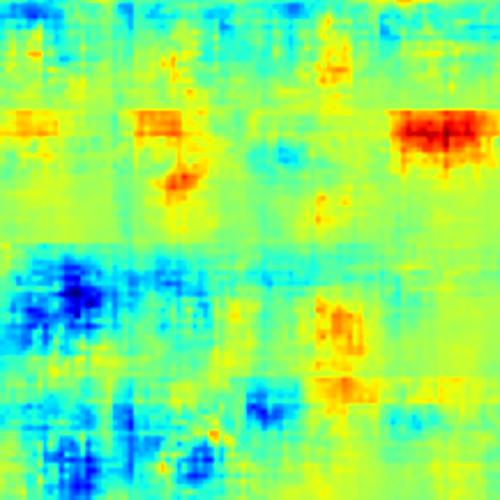}
  \end{subfigure}
\hfill
 \begin{subfigure}[b]{0.18000000000000002\textwidth}
  \centering
  \includegraphics[width=\textwidth]{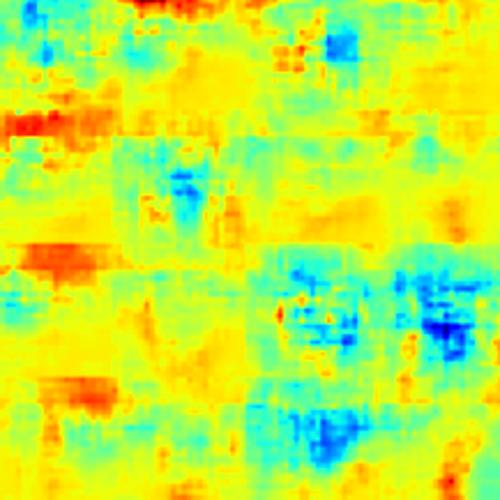}
  \end{subfigure}
\hfill
 \begin{subfigure}[b]{0.18000000000000002\textwidth}
  \centering
  \includegraphics[width=\textwidth]{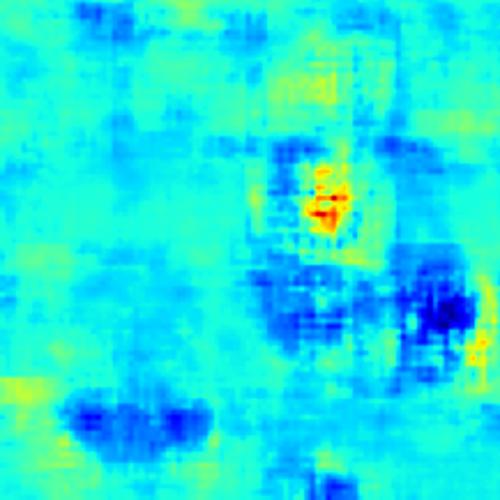}
  \end{subfigure}
\hfill
 \begin{subfigure}[b]{0.18000000000000002\textwidth}
  \centering
  \includegraphics[width=\textwidth]{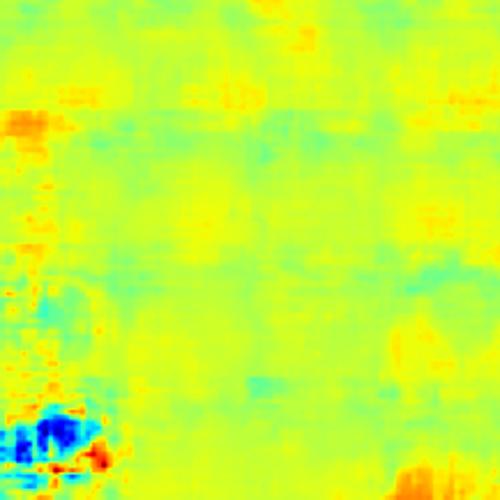}
  \end{subfigure}
\\
 \begin{subfigure}[b]{0.18000000000000002\textwidth}
  \centering
  \includegraphics[width=\textwidth]{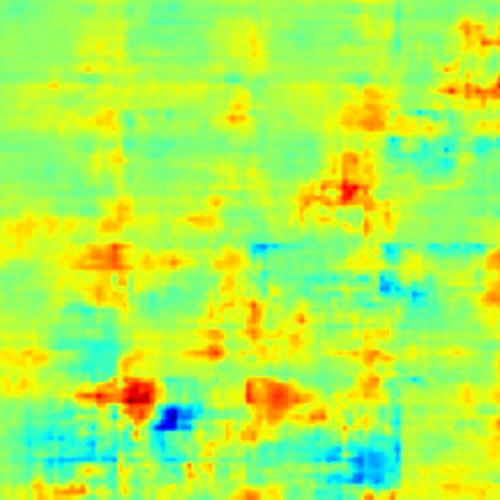}
  \end{subfigure}
\hfill
 \begin{subfigure}[b]{0.18000000000000002\textwidth}
  \centering
  \includegraphics[width=\textwidth]{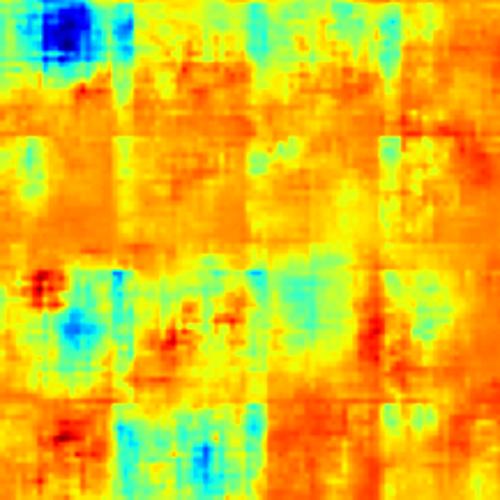}
  \end{subfigure}
\hfill
 \begin{subfigure}[b]{0.18000000000000002\textwidth}
  \centering
  \includegraphics[width=\textwidth]{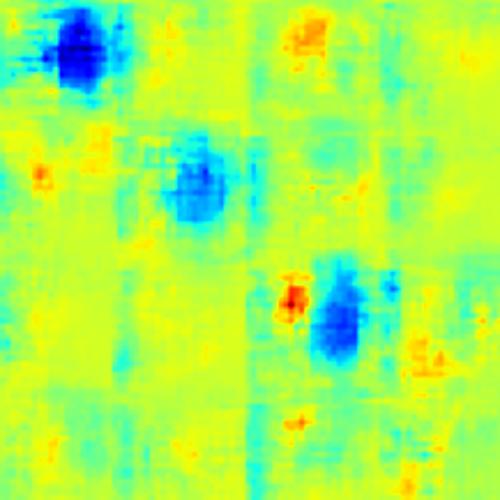}
  \end{subfigure}
\hfill
 \begin{subfigure}[b]{0.18000000000000002\textwidth}
  \centering
  \includegraphics[width=\textwidth]{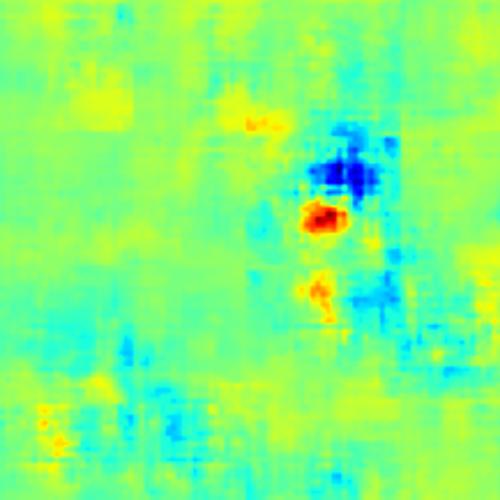}
  \end{subfigure}
\hfill
 \begin{subfigure}[b]{0.18000000000000002\textwidth}
  \centering
  \includegraphics[width=\textwidth]{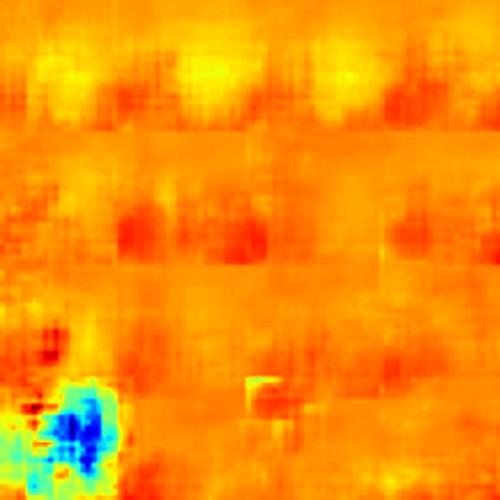}
  \end{subfigure}
  \\
  \label{fig:localizations_ki67}
  \caption{
  Localizations for Ki67 predictors trained/evaluated on the split shown by the red circle in Fig. \akfigrefkiaucs. Each column corresponds to a H\&E-IHC pair. Row 1: IHC, row 2: H\&E, row 3: CLAM \akciteclam's attention mask, row 4: the sensitivity of the classifier labeled as "ViT, high vs. low". rows 5: the average sensitivity of heads of the classifier labeled as "ViT, with G.Z.". 
   rows 6: the average sensitivity of heads of the classifier labeled as "ViT, without G.Z.".
   In all heatmaps we used JET color-map in which high and low values appear in red and blue, respectively.
  }
\end{figure}








\end{document}